\documentclass[a4paper,11pt]{article}
\usepackage[DIV12]{typearea}
\usepackage{longtable}
\usepackage{booktabs,colortbl}
\usepackage{multirow}
\usepackage{amsmath}
\usepackage{amssymb}
\usepackage{bbm}
\usepackage[utf8]{inputenc}
\usepackage{pdflscape}
\usepackage{xspace}
\usepackage[caption=false]{subfig}
\usepackage{nicefrac}
\usepackage{graphicx}
\usepackage{cite}  
\usepackage[small]{caption2}
\usepackage{scalerel} %to scale subscripts in equations
\usepackage{mathtools}

\numberwithin{equation}{section}

\DeclareMathOperator{\re}{Re}
\DeclareMathOperator{\im}{Im}

\newcommand{\rep}[1]{\ensuremath\boldsymbol{#1}}

\newcommand{\Z}[1]{\ensuremath{\mathbbm{Z}_{#1}}} % z_N ->\Z{N}
\newcommand{\SO}[1]{\ensuremath{\mathrm{SO}(#1)}}
\newcommand{\SU}[1]{\ensuremath{\mathrm{SU}(#1)}}
\newcommand{\SL}[1]{\ensuremath{\mathrm{SL}(#1)}}
\newcommand{\Sp}[1]{\ensuremath{\mathrm{Sp}(#1,\mathbbm{Z})}}
\newcommand{\U}[1]{\ensuremath{\mathrm{U}(#1)}}
\newcommand{\E}[1]{\ensuremath{\mathrm{E}_{#1}}}
\newcommand{\e}{\mathrm{e}}
\newcommand{\I}{\mathrm{i}}
\newcommand{\Id}{\mathbbm{1}}
\newcommand{\com}[2]{\lbrack #1, #2\rbrack}
\newcommand{\CP}{\ensuremath{\mathcal{CP}}\xspace}
\newcommand{\x}{\ensuremath{\times}}
\newcommand{\vev}[1]{\ensuremath{\langle{#1}\rangle}}

\usepackage{xcolor}
\advance \headheight by 3.0truept       % for 12pt mandatory...
\usepackage[pdftex]{hyperref}
\hypersetup{
    pdftitle = {Completing the eclectic flavor scheme of the \Z2 orbifold},
    pdfauthor = {Baur, Kade, Nilles, Ramos-Sanchez, Vaudrevange}
}

\begin{document}

\begin{titlepage}

%\vspace*{-3.0cm}
\begin{flushright}
\normalsize{TUM-HEP 1333/21}
\end{flushright}

\vspace*{1.0cm}

\begin{center}
{\Large\textbf{\boldmath Completing the eclectic flavor scheme of the \Z2 orbifold}\unboldmath}

\vspace{1cm}

\textbf{Alexander Baur$^{a,c}$, Moritz Kade$^{a}$,}
\textbf{Hans Peter Nilles$^{b}$, Sa\'ul Ramos--S\'anchez$^{c}$, Patrick K.S. Vaudrevange$^{a}$}
\\[8mm]
\textit{$^a$\small Physik Department T75, Technische Universit\"at M\"unchen,\\ James-Franck-Stra\ss e 1, 85748 Garching, Germany}
\\[2mm]
\textit{$^b$\small Bethe Center for Theoretical Physics and Physikalisches Institut der Universit\"at Bonn,\\ Nussallee 12, 53115 Bonn, Germany}
\\[2mm]
\textit{$^c$\small Instituto de F\'isica, Universidad Nacional Aut\'onoma de M\'exico,\\ POB 20-364, Cd.Mx. 01000, M\'exico}
\end{center}

\vspace{1cm}

\vspace*{1.0cm}

\begin{abstract}
We present a detailed analysis of the eclectic flavor structure of the two-dimensional \Z2 
orbifold with its two unconstrained moduli $T$ and $U$ as well as 
$\SL{2,\Z{}}_T\times \SL{2,\Z{}}_U$ modular symmetry. This provides a thorough understanding of 
mirror symmetry as well as the $R$-symmetries that appear as a consequence of the automorphy 
factors of modular transformations. It leads to a complete picture of local flavor unification in 
the $(T,U)$ modulus landscape. In view of applications towards the flavor structure of 
particle physics models, we are led to top-down constructions with high predictive power. The 
first reason is the very limited availability of flavor representations of twisted matter fields as 
well as their (fixed) modular weights. This is followed by severe restrictions from traditional and 
(finite) modular flavor symmetries, mirror symmetry, $\CP$ and $R$-symmetries on the 
superpotential and K\"ahler potential of the theory.
\end{abstract}

\end{titlepage}

\newpage

%%%%%%%%%%%%%%%%%%%%%%%%%%%%%%%%%%%%%%%%%%%%%%%%%%%%%%%%%%%%%%%%%%%%%%%%%%%%%%%%%%%%%%%%%%%%%%%%%%%%%%%%%%%%%%%%%%%%%%%%%%%%%%%%%%%%%%%%%
\section{Introduction}

In this paper we extend our previous discussion~\cite{Baur:2020jwc} of the $\mathbbm T^2/\Z2$ 
orbifold. $\mathbbm T^2/\Z2$ is the only two-dimensional orbifold with two unconstrained moduli 
$T$, $U$ that transform under $\SL{2,\Z{}}_T\times \SL{2,\Z{}}_U$ and under mirror symmetry, which 
interchanges $T$ and $U$. Hence, it can serve as a building block for the discussion of 
six-dimensional orbifolds.\footnote{For a general discussion of $\mathbbm T^2/\Z{K}$ including 
$K>2$, we refer to refs.~\cite{Nilles:2020tdp,Nilles:2020gvu} and~\cite{Baur:2019kwi,Baur:2019iai,Nilles:2020kgo} 
for $K=3$.} In our previous study, ref.~\cite{Baur:2020jwc}, we had identified the traditional 
flavor symmetries and the finite modular symmetries $\Gamma_N$ for the $\mathbbm T^2/\Z2$ orbifold. 
The groups $\Gamma_N$ (for small $N$) are isomorphic to groups like $S_3$, $A_4$, $S_4$ and $A_5$ 
that could be suitable for a description of discrete flavor symmetries in particle 
physics~\cite{Feruglio:2017spp,Criado:2018thu,Feruglio:2019ktm}. Modular symmetries, however, 
require more than just a discussion of the finite modular groups $\Gamma_N$. In addition, we have 
to include automorphy factors corresponding to the explicit modular weights of matter fields. In 
the present paper, we discuss the implications of these automorphy factors in the case of the 
$\mathbbm T^2/\Z2$ orbifold. Once they are taken into account, we find an extension of the finite 
modular flavor symmetry in form of an $R$-symmetry, which implies further restrictions to the 
superpotential and K\"ahler potential of the theory. This is one of the reasons why a modular 
flavor symmetry has more predictive power than traditional flavor symmetries. In the top-down 
approach (which we adopt here), this extension of the symmetry reflects the symmetries of the 
underlying string theory, which restrict the modular weights to well-defined specific 
values.\footnote{For a previous discussion on the modular weights of the $\mathbbm T^2/\Z3$ 
orbifold, see refs.~\cite{Nilles:2020kgo,Nilles:2020tdp,Nilles:2020gvu}. Although in this case the 
modulus $U$ is fixed, there remains a discrete $R$-symmetry from $\SL{2,\Z{}}_U$.} In the bottom-up 
approach to modular flavor symmetries, the choice of the modular weights of matter fields is part 
of model building and can be used to obtain so-called ``shaping symmetries'' that appear as 
additional accidental symmetries for specific choices of the modular 
weights~\cite{King:2020qaj,Feruglio:2021dte}.

The main results of the paper can be summarized as follows:
\begin{itemize}
\item We identify the full eclectic flavor symmetry~\cite{Nilles:2020nnc} of the $\mathbbm T^2/\Z2$ 
orbifold to be
\begin{equation}
\left[(D_8 \times D_8)/\Z2\right] ~\cup~ \left[(S_3^T \times S_3^U)\rtimes\Z4^{\hat M}\right] ~\cup~ \Z4^R\;.
\end{equation}
It includes a $\Z4^R$ $R$-symmetry that originates from the discussion of the automorphy factors 
and extends the order of the eclectic flavor group from 2304 to 4608. With \CP, the order of the 
eclectic flavor symmetry is further enhanced to a group of order 9216.

\item We provide a discussion of the {\it landscape} of flavor symmetries in $(T,U)$-moduli space 
and identify the local unified flavor groups at specific points and lines in this moduli space. The 
results are given in figure~\ref{fig:LocalFlavorGroups}, accompanied by an explicit discussion of 
the flavor symmetries in the cases of two specific geometrical shapes (the tetrahedron and the squared 
raviolo) as well as $T \leftrightarrow U$ mirror symmetry in section~\ref{sec:flavorUnification}.

\item We observe a specific relation between mirror symmetry and the allowed values of modular 
weights of matter fields (discussed explicitly in section~\ref{sec:RSymmetries}).

\item The additional $R$-symmetry is closely related to the modular symmetry and leads to 
further constraints on the allowed values of modular weights of matter fields. Hence, it further 
restricts the form of superpotential and K\"ahler potential, as explicitly discussed in 
section~\ref{sec:EFT}.

\item We discover the appearance of continuous gauge symmetries for specific configurations in 
moduli space.
\end{itemize}

The paper is structured as follows. In section~\ref{sec:oldSummary}, we recall the results of our 
previous study~\cite{Baur:2020jwc}. Section~\ref{sec:RSymmetries} discusses the automorphy factors 
and modular weights of matter fields. We identify the additional $R$-symmetry and the extended 
eclectic flavor group accordingly. This includes a discussion of the interplay of the modular 
weights with both, $T \leftrightarrow U$ mirror symmetry and the $R$-symmetry. In 
section~\ref{sec:flavorUnification} we analyze the unified local flavor groups that appear at 
specific points, lines and other hyper-surfaces in moduli space. The results including \CP are 
illustrated in figure~\ref{fig:LocalFlavorGroups}. Section~\ref{sec:EFT} is devoted to the 
discussion of the superpotential and K\"ahler potential. We observe the appearance of continuous 
gauge symmetries for certain configurations of the moduli (na\"ively, they might appear as 
accidental symmetries, but they are consequences of underlying symmetries in string theory). In 
section~\ref{sec:conclusions} we give conclusions and outlook. Technical details are relegated to 
several appendices that complete the general discussion of ref.~\cite{Baur:2020jwc}.

%%%%%%%%%%%%%%%%%%%%%%%%%%%%%%%%%%%%%%%%%%%%%%%%%%%%%%%%%%%%%%%%%%%%%%%%%%%%%%%%%%%%%%%%%%%%%%%%%%%%%%%%%%%%%%%%%%%%%%%%%%%%%%%%%%%%%%%%%
\section{What do we know already?}
\label{sec:oldSummary}

Technical details of the eclectic flavor symmetries of $\mathbbm T^2/\Z{K}$ orbifolds ($K=2,3,4,6$) 
have been given in section~2 of ref.~\cite{Nilles:2020gvu}. In the cases $K>2$, the complex 
structure modulus $U$ has to be fixed to allow for the orbifold twist. For $\mathbbm T^2/\Z2$, in 
contrast, we have two unconstrained moduli $T$ and $U$ with the corresponding modular 
transformations $\SL{2,\Z{}}_T\x\SL{2,\Z{}}_U$. For generic values of the moduli, we find the 
traditional flavor symmetry (we use the Small Group notation from \texttt{GAP}~\cite{GAP4})
\begin{equation}
(D_8 \times D_8)/\Z{2} ~\cong~ [32,49]
\end{equation}
as the result of geometry and string selection rules (see refs.~\cite{Kobayashi:2006wq,Olguin-Trejo:2018wpw})
or, equivalently, as a result of the outer automorphisms of the (Narain) space group that describes 
the orbifold~\cite{Baur:2019kwi,Baur:2019iai}. Furthermore, the finite modular symmetry for the 
$\mathbbm T^2/\Z2$ orbifold is shown to be the multiplicative closure of 
$\Gamma_2^T\x \Gamma_2^U = S_3^T\x S_3^U$ and mirror symmetry (which exchanges $T$ and $U$), as 
discussed in ref.~\cite{Baur:2020jwc}. The full mirror symmetry acting on the matter fields turns 
out to be $\Z4^{\hat M}$ (which acts on the moduli as $\Z2$, cf.~\cite{Ding:2020zxw}). This leads 
to the finite modular group $[144,115]$. If we include a \CP-like symmetry acting on the moduli as 
$T\rightarrow-\overline{T}$ and $U\rightarrow-\overline{U}$, the finite modular group enhances to
\begin{equation}
\left[(S_3^T \times S_3^U)\rtimes\Z4^{\hat M}\right] ~\times~ \Z2^{\CP} ~\cong~ [288,880]\;.
\end{equation}
In combination with the generators of the traditional flavor symmetry $[32,49]$, we obtained an 
eclectic flavor group with 4608 elements (2304 without \CP).

Only some of the eclectic flavor symmetries are linearly realized. For generic values of the moduli 
just the traditional flavor group $[32,49]$ remains unbroken. For specific ``geometrical'' 
configurations, this symmetry is enhanced to a larger subgroup of the eclectic flavor group (via 
the so-called stabilizer subgroups). The generators of the unbroken groups are displayed explicitly 
in figure~7 of ref.~\cite{Baur:2020jwc}. Relevant values correspond to the moduli $\vev U=\I$ 
(the squared raviolo) and $\vev U=\exp(\pi\I/3)$ (the tetrahedron) as well as the line 
$\vev T=\vev U$ as a consequence of mirror symmetry. At $\vev T=\vev U$, we find the enhancement of 
$[32,49]$ to $[64,257]$. For the tetrahedron, the group $[32,49]$ is enhanced to $[96,204]$ as 
discussed in section~4.2 of ref.~\cite{Baur:2020jwc}, while for the raviolo, we shall see here in 
section~\ref{sec:flavorUnification} that $[32,49]$ is enhanced to $[128,523]$. If we include the 
\CP-like transformation, we gain a further enhancement of the number of elements by a factor of 
two. The largest linearly realized subgroup of the eclectic flavor group (including \CP) 
was found (in ref.~\cite{Baur:2020jwc}) to be $[1152,157463]$ at $\vev T=\vev U=\exp(\pi\I/3)$.

So far the results are based on the finite modular groups. A full discussion of modular symmetries 
should, however, not only include the finite symmetries $\Gamma_N$ (here 
$\Gamma_2^T\x\Gamma_2^U = S_3^T\x S_3^U$), but also the so-called automorphy factors that arise 
from the non-trivial (fractional) modular weights $(n_T, n_U)$ of $\SL{2,\Z{}}_T\x\SL{2,\Z{}}_U$. 
This leads to further restrictions on the action (given by K\"ahler and superpotential) of the 
theory with an enhancement of the symmetries. As discussed in 
refs.~\cite{Nilles:2020tdp,Nilles:2020gvu}, these automorphy factors lead to discrete phases 
resulting in $R$-symmetries. In our previous paper~\cite{Baur:2020jwc}, for the sake of 
clarity and simplicity, we had not included these automorphy factors in our discussion. We shall 
include them in the following in full detail.

%%%%%%%%%%%%%%%%%%%%%%%%%%%%%%%%%%%%%%%%%%%%%%%%%%%%%%%%%%%%%%%%%%%%%%%%%%%%%%%%%%%%%%%%%%%%%%%%%%%%%%%%%%%%%%%%%%%%%%%%%%%%%%%%%%%%%%%%%
\section[Discrete $R$-symmetries and mirror symmetry]{\boldmath Discrete $R$-symmetries and mirror symmetry \unboldmath}
\label{sec:RSymmetries}

In this section, we show that a $\mathbbm T^2/\Z2$ orbifold sector gives rise to a $\Z{4}^R$ 
symmetry that originates from modular transformations, where the automorphy factors of certain 
modular transformations give rise to the $R$-charges. As the $\mathbbm T^2/\Z2$ orbifold sector is 
equipped with two moduli, $T$ and $U$, there exists a modular group for each of them, $\SL{2,\Z{}}_T$ 
and $\SL{2,\Z{}}_U$, each associated with a modular weight $(n_T, n_U$). Since $R$-charges can be 
defined in terms of both modular groups, these modular weights are highly constrained. Furthermore, 
we give a detailed discussion about the action of mirror symmetry on matter fields and discover a 
new relation between mirror symmetry and the $R$-symmetry.

%%%%%%%%%%%%%%%%%%%%%%%%%%%%%%%%%%%%%%%%%%%%%%%%%%%%
\subsection[Automorphy factors of modular transformations]{\boldmath Automorphy factors of modular transformations \unboldmath}
\label{sec:AutomorphyFactors}

Let us consider a general matter field $\Phi_{(n_T, n_U)}$ originating from string theory with 
modular weights $n_T$ and $n_U$ corresponding to $\SL{2,\Z{}}_T$ and $\SL{2,\Z{}}_U$. Then, under a 
(non \CP-like) modular transformation $\hat\Sigma \in \mathrm{O}_{\hat{\eta}}(2,2,\Z{})$, the field 
transforms as
\begin{equation}\label{eq:generalModTrafoOnFields}
\Phi_{(n_T,n_U)} ~\xmapsto{~\hat{\Sigma}~}~ j^{(n_T,n_U)}(\hat{\Sigma},T,U)\, \rho_{\rep{r}}(\hat{\Sigma})\,\Phi_{(n_T,n_U)}\;.
\end{equation}
Here, $j^{(n_T,n_U)}(\hat{\Sigma},T,U)$ is the automorphy factor of the modular transformation and 
$\rho_{\rep{r}}(\hat{\Sigma})$ is the representation matrix of $\hat{\Sigma}$ that forms a 
representation $\rep{r}$ of the finite modular group, as derived in appendix~\ref{app:TrafoOfTwistedStrings}.

The modular weights of matter fields can be computed in string theory, as reviewed in 
appendix~\ref{app:MirrorForStrings}, and it turns out that, apart from $n_T=n_U$, there is also the
possibility
\begin{equation}
n_T ~\neq~ n_U
\end{equation}
in string theory. In order to determine the automorphy factor $j^{(n_T,n_U)}(\hat{\Sigma},T,U)$, 
we might use as a first step the analogy to Siegel modular forms based on $\Sp{4}$. 
However, Siegel modular forms are defined for parallel weights $n:=n_T=n_U$ only. In 
this case, following refs.~\cite{Ding:2020zxw,Ding:2021iqp}, we have
\begin{equation}\label{eq:jfromSP4Z}
j^{(n)}(M,T,U) = \big(\det(C\,\Omega+D)\big)^{n} \quad\mathrm{for}~ M = \begin{pmatrix}A&B\\ C&D\end{pmatrix} \in \Sp{4} ~\mathrm{and}~ 
        \Omega = \begin{pmatrix}U&0\\ 0&T\end{pmatrix}.
\end{equation}
\Sp{4} contains $\SL{2,\Z{}}_T$ and $\SL{2,\Z{}}_U$ via the element 
$M_{(\gamma_T,\gamma_U)}\in\Sp{4}$, where 
\begin{equation}
\gamma_T ~:=~ \begin{pmatrix}a_T&b_T\\ c_T&d_T\end{pmatrix} ~\in~\SL{2,\Z{}}_T \quad\mathrm{and}\quad \gamma_U ~:=~ \begin{pmatrix}a_U&b_U\\ c_U&d_U\end{pmatrix} ~\in~\SL{2,\Z{}}_U\;,
\end{equation}
as defined in ref.~\cite{Baur:2020yjl}. Then, eq.~\eqref{eq:jfromSP4Z} yields
\begin{equation}\label{eq:AutomorphyFactorSLxSLParallelWeights}
j^{(n)}(M_{(\gamma_T,\gamma_U)},T,U) ~=~ \left(c_T\,T+d_T\right)^{n} \left(c_U\,U+d_U\right)^{n}\;.
\end{equation}
Using the dictionary~\cite{Baur:2020yjl} that relates $\Sp{4}$ with the modular group 
$\mathrm{O}_{\hat\eta}(2,2+16,\Z{})$ of our string setting, the \Sp{4} element 
$M_{(\gamma_T,\gamma_U)}$ is equivalent to $\hat{\Sigma}_{(\gamma_T,\gamma_U)}\in\mathrm{O}_{\hat\eta}(2,2+16,\Z{})$ 
defined in appendix~\ref{app:OriginOfModular}. We take this as a motivation to define
\begin{equation}\label{eq:AutomorphyFactorSLxSL}
j^{(n_T,n_U)}(\hat{\Sigma}_{(\gamma_T,\gamma_U)},T,U) ~:=~ \left(c_T\,T+d_T\right)^{n_T} \left(c_U\,U+d_U\right)^{n_U}\;,
\end{equation}
also for the case $n_T \neq n_U$. It turns out that for the other cases 
$\hat{\Sigma}\neq\hat{\Sigma}_{(\gamma_T,\gamma_U)}$ one can use the automorphy factor 
eq.~\eqref{eq:jfromSP4Z} for the element $M\in\Sp{4}$ that corresponds to $\hat\Sigma$ using again 
the dictionary of ref.~\cite{Baur:2020yjl}. It is important to note that the resulting automorphy 
factor will be independent of the specific choice $n=n_T$ or $n=n_U$, since $n_T=n_U\ \mathrm{mod}\ 2$. 
In the following, we will see this explicitly at some examples.

%%%%%%%%%%%%%%%%%%%%%%%%%%%%%%%%%%%%%%%%%%%%%%%%%%%%
\subsection[Discrete R-symmetry]{\boldmath Discrete $R$-symmetry \unboldmath}
\label{sec:RSymmery}

In the $\mathbbm T^2/\Z2$ orbifold sector, a \Z2 sublattice rotation is given by 
\begin{equation}
\hat\Theta_{(2)} ~:=~ \hat{C}_\mathrm{S}^2 ~=~ \hat{K}_\mathrm{S}^2 ~=~ -\Id_4 ~\in~\mathrm{O}_{\hat{\eta}}(2,2,\Z{})\;,
\end{equation}
i.e.\ in the Narain formulation, $\hat\Theta_{(2)}=-\Id_4$ is a left-right symmetric $180^\circ$ 
rotation in the $\mathbbm T^2/\Z2$ orbifold sector that leaves the orthogonal compact dimensions 
invariant, see refs.~\cite{Nilles:2020tdp,Nilles:2020gvu}. It is an outer automorphism of the full 
Narain space group of the six-dimensional orbifold and, hence, it is a symmetry of the theory. Using the 
definition of $\hat{\Sigma}_{(\gamma_T,\gamma_U)}$ given in eq.~\eqref{eq:SigmaTU} of 
appendix~\ref{app:OriginOfModular}, this $\mathrm{O}_{\hat{\eta}}(2,2,\Z{})$ transformation can be 
expressed in two ways as 
\begin{equation}\label{eq:RfromSL2ZU}
\hat\Theta_{(2)}~=~\hat{\Sigma}_{(\mathrm{R},\Id_2)}~=~\hat{\Sigma}_{(\Id_2,\mathrm{R})}\;, \quad\mathrm{where}\quad \mathrm{R} ~=~ \mathrm{S}^2 ~=~ \left(\begin{array}{cc} -1 & 0 \\ 0 & -1 \end{array} \right) ~\in~ \SL{2,\Z{}}\,.
\end{equation}
As the generalized metric $\mathcal{H}(T,U)$ is invariant under a 
transformation~\eqref{eq:TrafoOfGeneralizedMetric} with $\hat\Theta_{(2)}$, this \Z2 sublattice 
rotation leaves $T$ and $U$ invariant. Hence, the sublattice rotation $\hat\Theta_{(2)}$ 
corresponds to a traditional flavor symmetry. Still, the transformation with $\hat\Theta_{(2)}$ 
originates from a modular transformation, $\hat\Theta_{(2)}=\hat{C}_\mathrm{S}^2=\hat{K}_\mathrm{S}^2$. 
So, we expect the appearance of an automorphy factor. Since $\mathrm{R}\in\SL{2,\Z{}}_T$ and 
$\mathrm{R}\in\SL{2,\Z{}}_U$ are identified in $\mathrm{O}_{\hat{\eta}}(2,2,\Z{})$, we have to 
ensure that we compute the automorphy factor correctly: we can use either the factor 
$(c_T T+d_T)^{n_T}=(-1)^{n_T}$ or $(c_U U+d_U)^{n_U}=(-1)^{n_U}$ for the transformation 
$\mathrm{R}$. Yet, the resulting automorphy factor must coincide in both cases, 
$(-1)^{n_T}=(-1)^{n_U}$. Hence, we see that
\begin{equation}\label{eq:RelationBetweenModularWeights}
n_U ~\stackrel{!}{=}~ n_T \ \mathrm{mod}\ 2\;.
\end{equation}
This relation is satisfied for all (massless) matter from the $\mathbbm T^2/\Z2$ orbifold sector, 
as one can see from table~\ref{tab:Z2modularWeights}. Moreover, eq.~\eqref{eq:RelationBetweenModularWeights} 
also holds for all massive strings, as shown in appendix~\ref{app:MirrorForStrings}. Consequently, 
having control over the automorphy factor, we can choose $\mathrm{R}\in\SL{2,\Z{}}_U$ and the modular 
weight $n_U$ in the following. 

\begin{table}[t!]
\centering
\resizebox{\textwidth}{!}{
\begin{tabular}{lllclllcccc}
\toprule
\phantom{traditionalabcd}& \multicolumn{2}{c}{bulk matter} & \phantom{ab} & \multicolumn{3}{c}{twisted matter} & \phantom{ab} &  & &   \\\cmidrule{2-3}\cmidrule{5-7}
		& $\Phi_{(0,0)}$ & $\Phi_{(-1,-1)}$ & & $\Phi_{(\nicefrac{-1}{2},\nicefrac{-1}{2})}$ & $\Phi_{(\nicefrac{-3}{2},\,\nicefrac{1}{2})}$ & $\Phi_{(\,\nicefrac{1}{2},\nicefrac{-3}{2})}$ & & $Y_{\rep{4}_3}^{(2)}$ & & $\mathcal{W}$ \\\midrule
\arrayrulecolor{lightgray}
traditional& $\phantom{-}\rep{1}_0$ & $\phantom{-}\rep{1}_0$ & & $\phantom{-}\rep{4}_{\phantom{1}}$ & $\phantom{-}\rep{4}_{\phantom{1}}$ & $\phantom{-}\rep{4}_{\phantom{1}}$ & & $\rep{1}_0$ & & $\phantom{-}\rep{1}_0$ \\\cmidrule{1-11}
modular    & $\phantom{-}\rep{1}_0$ & $\phantom{-}\rep{1}_0$ & & $\phantom{-}\rep{4}_{1}$ & \multicolumn{2}{c}{$\left(\rep{4}_{1}\oplus \rep{4}_{1}\right)\quad~~$} & & $\rep{4}_3$ & & $\phantom{-}\rep{1}_0$ \\\cmidrule{1-11}
$n_T$      & $\phantom{-}0$ &           $-1$ & & $\nicefrac{-1}{2}$ & $\nicefrac{-3}{2}$ & $\phantom{-}\nicefrac{1}{2}$ & & $2_{\phantom{1}}$ & & $-1_{\phantom{1}}$ \\\cmidrule{1-11}
$n_U$      & $\phantom{-}0$ &           $-1$ & & $\nicefrac{-1}{2}$ & $\phantom{-}\nicefrac{1}{2}$ & $\nicefrac{-3}{2}$ & & $2_{\phantom{1}}$ & & $-1_{\phantom{1}}$ \\\cmidrule{1-11}
$R$-charge & $\phantom{-}0$ & $\phantom{-}2$ & &     $\phantom{-}3$ &               $\phantom{-}1$ &     $\phantom{-}1$ & & $0_{\phantom{1}}$ & &  $2_{\phantom{1}}\ \mathrm{mod}\ 4$ \\
\arrayrulecolor{black}
\bottomrule
\end{tabular}}

\caption{All admissible modular weights of massless matter fields $\Phi_{(n_T,n_U)}$ as well as 
their representations under the flavor symmetries and $\Z{4}^R$ of a $\mathbbm T^2/\Z2$ orbifold 
sector, see appendix~\ref{app:TrafoOfTwistedStrings}. The $\Z{4}^R$ $R$-charges are normalized 
to be integer for matter superfields $\Phi_{(n_T,n_U)}$. The traditional flavor group is $[64,266]$ 
and the modular flavor group is $[144,115]$.}
\label{tab:Z2modularWeights}
\end{table}

The superpotential $\mathcal{W}$ transforms under $\hat\Theta_{(2)}$ as
\begin{equation}
\mathcal{W} ~\xmapsto{\hat\Theta_{(2)}}~ -\mathcal{W}\;,
\end{equation}
due to the automorphy factor $(c_U U+d_U)^{-1} = -1$ evaluated for $\mathrm{R}$ given 
in eq.~\eqref{eq:RfromSL2ZU}. Thus, the transformation $\hat\Theta_{(2)}$ generates a discrete 
$R$-symmetry~\cite{Nilles:2020tdp,Nilles:2020gvu}.

The action of $\hat\Theta_{(2)}$ on matter fields $\Phi_{(n_T, n_U)}$ with $\SL{2,\Z{}}_U$ modular weights 
$n_U \in\{ 0,-1,\nicefrac{-1}{2},\nicefrac{1}{2},\nicefrac{-3}{2}\}$, as listed in 
table~\ref{tab:Z2modularWeights}, is given by
\begin{equation}\label{eq:RTrafoOnMatterField}
\Phi_{(n_T, n_U)} ~\xmapsto{\hat\Theta_{(2)}}~ (-1)^{n_U}\, \rho_{\rep{r}}(\hat\Theta_{(2)})\, \Phi_{(n_T, n_U)} ~=~ (-1)^{n_U}\, \Phi_{(n_T, n_U)} ~=:~ \exp\left(\nicefrac{2\pi\I\,R}{4}\right)\, \Phi_{(n_T, n_U)}\;.
\end{equation}
Here, we used that $\rho_{\rep{r}}(\hat\Theta_{(2)})=\rho_{\rep{r}}(\hat{K}_\mathrm{S})^2=\rho_{\rep{r}}(\hat{C}_\mathrm{S})^2=\Id$.
For the allowed modular weights $n_U \in\{ 0,-1,\nicefrac{-1}{2},\nicefrac{1}{2},\nicefrac{-3}{2}\}$ 
the multivalued phase factor gives rise to $\Z{4}^R$ $R$-charges $R\in\{0,2,3,1,1\}$, respectively. 
Thus, for the $\mathbbm T^2/\Z2$ orbifold sector we find that the $R$-charge $R$ is given in terms 
of the modular weight $n_U$ (or $n_T$) as
\begin{equation}
R ~=~ 2\,n_U \ \mathrm{mod}\ 4 ~=~ 2\,n_T \ \mathrm{mod}\ 4\;,
\end{equation}
cf.\ ref.~\cite{Nilles:2013lda}. Note that due to the fractional modular weights $n_U$, 
$(\hat\Theta_{(2)})^2$ gives a non-trivial transformation with charges $2R = 4n_U\ \mathrm{mod}\ 4$ 
that turns out to be equivalent to the point group selection rule of eq.~\eqref{eq:PG}. Since the
$R$-symmetry transformation acts trivially on all moduli, it belongs to the traditional flavor 
symmetry, which gets enhanced to 
\begin{equation}
\label{eq:Z2traditionalFlavorWithR}
  \frac{\left(D_8\x D_8\right)/\Z2\,\x\,\Z4^R}{\Z2} ~\cong~ [64,266]\,,
\end{equation}
where the \Z2 in the latter quotient identifies the point group selection rule of $\mathbbm T^2/\Z2$
contained in both the $\Z4^R$ and the traditional symmetry $\left(D_8\x D_8\right)/\Z2$. 
In string theory, modular symmetries are anomaly-free (see e.g.~\cite{Ibanez:1992hc,Araki:2008ek}
for details on anomaly cancellation for modular symmetries). Hence, since
the $\Z4^R$ $R$-symmetry arises from modular symmetries, it is anomaly-free.

Due to the $\Z4^R$ $R$-symmetry, the eclectic flavor group of a $\mathbbm T^2/\Z2$ orbifold sector 
gets extended to
\begin{equation}
\left[(D_8\x D_8)/\Z2\right] ~\cup~ \left[(S_3^T\x S_3^U)\rtimes\Z4^{\hat M}\right] ~\cup~ \Z4^R \;,
\end{equation}
which results in a group of order 4608. Including a \CP-like transformation, the order of the 
eclectic flavor group is further enhanced to a group of order 9216.

%%%%%%%%%%%%%%%%%%%%%%%%%%%%%%%%%%%%%%%%%%%%%%%%%%%%
\subsection{The action of mirror symmetry on matter fields}
\label{sec:mirrorOnMatter}

In order to analyze the action of mirror symmetry $\hat{M} \in \mathrm{O}_{\hat\eta}(2,2,\Z{})$ on 
matter fields in our string setup, we have to determine the automorphy factor first. Using the 
results of section~\ref{sec:AutomorphyFactors}, we consider the mirror element in \Sp{4}, which 
reads
\begin{equation}
M_{\times} ~=~ \begin{pmatrix}0&1&0&0\\ 1&0&0&0\\ 0&0&0&1\\ 0&0&1&0\end{pmatrix} ~\in~\Sp{4}\;.
\end{equation}
Thus, the automorphy factor~\eqref{eq:jfromSP4Z} of a mirror transformation is given by 
$(-1)^n$. Since $n_T=n_U \, \mathrm{mod}\, 2$ as derived in eq.~\eqref{eq:RelationBetweenModularWeights}, 
one can assign $n=n_U$ and the automorphy factor of a mirror transformation $\hat{M}$ is given as 
$(-1)^{n_U}$, without loss of generality. Moreover, note that the $\Z4^R$ $R$-charges given in 
eq.~\eqref{eq:RTrafoOnMatterField} are analogously defined. Thus, the automorphy factor of 
$\hat{M}$ can be removed using the $\Z{4}^R$ $R$-symmetry, as we will do in the following.

Now, let us assume that under a mirror transformation $\hat{M}$ we have
\begin{equation}\label{eq:hatMpreliminary}
\Phi_{(n_T, n_U)} ~\xmapsto{~\hat{M}~}~ \rho_{\rep{r}}(\hat{M})\,\Phi_{(n_T, n_U)}\;,
\end{equation}
for a matter field $\Phi_{(n_T, n_U)}$ that transforms in the representation $\rep{r}$ of the 
finite modular group $(S_3^T\x S_3^U) \rtimes\Z4^{\hat M}\cong [144,115]$ (where we have absorbed 
the automorphy factor $(-1)^{n_U}$ using $\Z{4}^R$). In the following, we will see that the 
transformation~\eqref{eq:hatMpreliminary} is correct for $n_T=n_U$, but must be modified in the 
case $n_T \neq n_U$. To do so, let us consider the following chain of transformations
\begin{subequations}\label{eq:MgammaUMinv}
\begin{eqnarray}
\Phi_{(n_T, n_U)}&\xmapsto{~~\hat{M}~~}                      &\rho_{\rep{r}}(\hat{M})\,\Phi_{(n_T, n_U)}\\
                 &\xmapsto{\hat{\Sigma}_{(\Id_2,\gamma_U)}}  &(c_U\,U+d_U)^{n_U}\,\rho_{\rep{r}}(\hat{M})\,\rho_{\rep{r}}(\hat{\Sigma}_{(\Id_2,\gamma_U)})\,\Phi_{(n_T, n_U)}\label{eq:MgammaUMinvb}\\
                 &\xmapsto{~\hat{M}^{-1}~}                   &(c_U\,T+d_U)^{n_U}\,\rho_{\rep{r}}(\hat{M})\,\rho_{\rep{r}}(\hat{\Sigma}_{(\Id_2,\gamma_U)})\,\rho_{\rep{r}}(\hat{M})^{-1}\,\Phi_{(n_T, n_U)}\;,\label{eq:MgammaUMinvc}
\end{eqnarray}
\end{subequations}
under the assumption that eq.~\eqref{eq:hatMpreliminary} were correct. However, a mirror 
transformation maps an element $\hat{\Sigma}_{(\Id_2,\gamma_U)} \in \SL{2,\Z{}}_U$ to an element 
$\hat{\Sigma}_{(\gamma_U,\Id_2)} \in \SL{2,\Z{}}_T$, see eq.~\eqref{eq:MirrorOnNarainSigma}. Thus, 
eq.~\eqref{eq:MgammaUMinvc} must be equal to
\begin{equation}\label{eq:gammaT}
\Phi_{(n_T, n_U)} ~\xmapsto{\hat{\Sigma}_{(\gamma_U,\Id_2)}}~ (c_U\,T+d_U)^{n_T}\,\rho_{\rep{r}}(\hat{\Sigma}_{(\gamma_U,\Id_2)})\,\Phi_{(n_T, n_U)}\;,
\end{equation}
where the $2\times 2$ matrix $\gamma_T\in\SL{2,\Z{}}_T$ in $\hat{\Sigma}_{(\gamma_T,\Id_2)}$ has 
to be equal to the matrix $\gamma_U$ used in eq.~\eqref{eq:MgammaUMinvb}. Now, we have to distinguish 
between two cases: First, in the case of so-called parallel weights (i.e.\ if $n_T=n_U$) the 
representation matrices $\rho_{\rep{r}}(\hat{\Sigma}_{(\gamma_U,\Id_2)})$ and 
$\rho_{\rep{r}}(\hat{\Sigma}_{(\Id_2,\gamma_U)})$ have to be related as follows
\begin{equation}\label{eq:MirrorOnSL2Zs}
\rho_{\rep{r}}(\hat{\Sigma}_{(\gamma_U,\Id_2)}) ~=~ \rho_{\rep{r}}(\hat{M})\,\rho_{\rep{r}}(\hat{\Sigma}_{(\Id_2,\gamma_U)})\,\rho_{\rep{r}}(\hat{M})^{-1}\;,
\end{equation}
and eq.~\eqref{eq:MgammaUMinvc} coincides with eq.~\eqref{eq:gammaT}. In the second case (i.e.\ 
if $n_T \neq n_U$), the (preliminary) chain of transformations given in eq.~\eqref{eq:MgammaUMinv} 
has to be modified
\begin{subequations}\label{eq:MgammaUMinv2}
\begin{eqnarray}
\Phi_{(n_T, n_U)}&\xmapsto{~~\hat{M}~~}                    &\rho_{\rep{r}}(\hat{M})\,\Phi_{(n_U, n_T)}\\
                 &\xmapsto{\hat{\Sigma}_{(\Id_2,\gamma_U)}}&(c_U\,U+d_U)^{n_T}\,\rho_{\rep{r}}(\hat{M})\,\rho_{\rep{r}}(\hat{\Sigma}_{(\Id_2,\gamma_U)})\,\Phi_{(n_U, n_T)}\\
                 &\xmapsto{~\hat{M}^{-1}~}                 &(c_U\,T+d_U)^{n_T}\,\rho_{\rep{r}}(\hat{M})\,\rho_{\rep{r}}(\hat{\Sigma}_{(\Id_2,\gamma_U)})\,\rho_{\rep{r}}(\hat{M})^{-1}\,\Phi_{(n_T, n_U)}\label{eq:MgammaUMinv2Last}\;.
\end{eqnarray}
\end{subequations}
Then, we have to impose condition~\eqref{eq:MirrorOnSL2Zs} and, consequently, 
eq.~\eqref{eq:MgammaUMinv2Last} coincides with eq.~\eqref{eq:gammaT} using 
$\hat{\Sigma}_{(\gamma_T,\Id_2)}$ with $\gamma_T$ equal to $\gamma_U$.

In other words, for each matter field $\Phi_{(n_T, n_U)}$ with $n_T \neq n_U$ (satisfying the 
constraint~\eqref{eq:RelationBetweenModularWeights}) there must exist a partner field, denoted by 
$\Phi_{(n_U, n_T)}$, which coincides in all properties with $\Phi_{(n_T, n_U)}$ but has 
interchanged modular weights. Then, a mirror transformation has to act on matter fields 
$(\Phi_{(n_T, n_U)}, \Phi_{(n_U, n_T)})$ as 
\begin{equation}\label{eq:MirrorOnTwistedMatter}
\begin{pmatrix}\Phi_{(n_T, n_U)}\\\Phi_{(n_U, n_T)}\end{pmatrix} ~\xmapsto{~\hat{M}~}~ \begin{pmatrix}0&\rho_{\rep{r}}(\hat{M})\\\rho_{\rep{r}}(\hat{M})& 0\end{pmatrix}\,\begin{pmatrix}\Phi_{(n_T, n_U)}\\\Phi_{(n_U, n_T)}\end{pmatrix} \quad\mathrm{if}\quad n_T ~\neq~ n_U\;,
\end{equation}
and eq.~\eqref{eq:MirrorOnSL2Zs} has to hold. On the other hand, the transformations 
$\hat{K}_\mathrm{S}$, $\hat{K}_\mathrm{T}$, $\hat{C}_\mathrm{S}$ and $\hat{C}_\mathrm{T}$ act 
diagonally on $(\Phi_{(n_T, n_U)}, \Phi_{(n_U, n_T)})$. For example, after an appropriate basis 
change one can check using the character table~\ref{tab:144115char} that the twisted matter 
fields $(\Phi_{(\nicefrac{-3}{2},\,\nicefrac{1}{2})}, \Phi_{(\,\nicefrac{1}{2},\nicefrac{-3}{2})})$ 
transform in the representations $\rep{4}_1 \oplus \rep{4}_1$ of the finite modular group 
$(S_3^T \times S_3^U) \rtimes \Z4^{\hat M} \cong [144,115]$ (see table~\ref{tab:Z2modularWeights}).
Finally, in appendix~\ref{app:MirrorForStrings} 
we show in string theory why a mirror transformation interchanges $\Phi_{(n_T, n_U)}$ and 
$\Phi_{(n_U, n_T)}$ if $n_T \neq n_U$, as we have seen in this bottom-up discussion that has lead 
to eq.~\eqref{eq:MirrorOnTwistedMatter}.

%%%%%%%%%%%%%%%%%%%%%%%%%%%%%%%%%%%%%%%%%%%%%%%%%%%%%%%%%%%%%%%%%%%%%%%%%%%%%%%%%%%%%%%%%%%%%%%%%%%%%%%%%%%%%%%%%%%%%%%%%%%%%%%%%%%%%%%%%
\section[Local flavor unification]{\boldmath Local flavor unification \unboldmath}
\label{sec:flavorUnification}

The full eclectic flavor group of the $\mathbbm T^2/\Z2$ orbifold sector is a group of order 4608 
that consists of the enhanced traditional flavor symmetry $(D_8\x D_8)/\Z2\cup\Z4^{R}\cong [64,266]$, 
and the finite modular symmetry $(S_3^T \times S_3^U) \rtimes \Z4^{\hat M} \cong [144,115]$. 
Adding the \CP-like generator $\hat\Sigma_*$ (see eq.~\eqref{eq:GeneratorsMirrorAndCP}) enhances 
the finite modular symmetry to $[288,880]$ and the eclectic flavor group with \CP has order 9216. 
However, the full eclectic flavor group gets broken spontaneously by non-vanishing vevs of the 
moduli $(T,U)$. In this section, we complete the analysis of ref.~\cite{Baur:2020jwc} of the 
unbroken groups at various points $(\vev T,\vev U)$ in moduli space.

The couplings of interest among matter fields are governed by the modular forms of (parallel) 
weight $(2,2)$. They can be spanned by
\begin{equation}
\label{eq:ModularFormWeight22}
\hat{Y}^{(2)}_{\rep{4}_3}(T,U) ~=~ \begin{pmatrix}\hat{Y}_1(T,U)\\ \hat{Y}_2(T,U)\\ \hat{Y}_3(T,U)\\ \hat{Y}_4(T,U)\end{pmatrix} ~:=~ 
\begin{pmatrix}\hat{Y}_1(T) \, \hat{Y}_1(U)\\ \hat{Y}_2(T) \, \hat{Y}_1(U)\\ \hat{Y}_1(T) \, \hat{Y}_2(U)\\ \hat{Y}_2(T) \, \hat{Y}_2(U)\end{pmatrix}\;,
\end{equation}
where $\hat{Y}_1(\tau)$ and $\hat{Y}_2(\tau)$ are the $S_3$ modular forms of weight 2, see for 
example refs.~\cite{Kobayashi:2018vbk,Ding:2020zxw}. $\hat{Y}_1(\tau)$ and $\hat{Y}_2(\tau)$ can be 
written as
\begin{subequations}
\begin{eqnarray}
\hat{Y}_1(\tau) &:=& \frac{1}{16}       \left(\left(\vartheta_{00}(0; \tau)\right)^4 + \left(\vartheta_{01}(0; \tau)\right)^4\right)\;,\\
\hat{Y}_2(\tau) &:=& \frac{\sqrt{3}}{16}\left(\left(\vartheta_{00}(0; \tau)\right)^4 - \left(\vartheta_{01}(0; \tau)\right)^4\right)\;,
\end{eqnarray}
\end{subequations}
in terms of the Jacobi theta function 
\begin{equation}\label{eq:vartheta}
\vartheta(z, \tau) ~:=~ \sum_{m\in\Z{}} \exp\left(\pi\I\tau m^2 + 2\pi\I m z\right)\;,
\end{equation}
where we have defined $\vartheta_{00}(z,\tau):=\vartheta(z,\tau)$ and 
$\vartheta_{01}(z,\tau):=\vartheta(z+\nicefrac{1}{2},\tau)$. The modular form 
$\hat{Y}^{(2)}_{\rep{4}_3}(T,U)$ transforms under modular transformations according to 
eq.~\eqref{eq:YTrafo} as a $\rep{4}_3$ of the finite modular group $[144,115]$, but is invariant 
under the traditional flavor group. Further details are given in appendix~\ref{app:FMGIrreps}, see 
also table~\ref{tab:Z2modularWeights}.

At special points in moduli space, i.e.\ for fixed vacuum expectation values $(\vev{T},\vev{U})$, 
some of the modular transformations are left unbroken in the vacuum, i.e.\ they leave invariant the 
moduli, building the so-called stabilizer subgroup,
\begin{equation}
\label{eq:stabilizer}
H_{(\vev T,\vev U)}~:=~
\big\langle ~\gamma ~\big|~ \gamma ~\in~ \mathrm{O}_{\hat\eta}(2,2,\Z{}) \quad
     \text{with}\quad \gamma(\vev{T}) ~=~ \vev{T} \;\;\mathrm{and}\;\; \gamma(\vev{U}) ~=~ \vev{U}~\big\rangle\,.
\end{equation}
Here, $\mathrm{O}_{\hat\eta}(2,2,\Z{})$, generated by $\{\hat K_{\mathrm S},\hat K_{\mathrm T},\hat C_{\mathrm S},
\hat C_{\mathrm T},\hat \Sigma_*,\hat{M}\}$, is the modular group of the $\mathbbm T^2/\Z2$ 
orbifold sector, see appendix~\ref{app:OriginOfModular}.

At these points, the couplings $\hat{Y}^{(2)}_{\rep{4}_3}(\vev T,\vev U)$ are left invariant under 
stabilizer elements $\gamma\in H_{(\vev T,\vev U)}$. If $\gamma$ takes the form given 
in eq.~\eqref{eq:SigmaTU}, i.e.\ if it is an $\SL{2,\Z{}}_T\x\SL{2,\Z{}}_U$ element,
its action on the couplings, eq.~\eqref{eq:YTrafo}, becomes the eigenvalue equation
\begin{equation}
\label{eq:YAlignment}
\rho_{\rep{4}_3}(\gamma)\,\hat{Y}^{(2)}_{\rep{4}_3}(\vev T,\vev U) ~\stackrel!=~ 
(c_T\vev T+d_T)^{-2}\,(c_U\vev U+d_U)^{-2}\,\hat{Y}^{(2)}_{\rep{4}_3}(\vev T,\vev U)\,,
\end{equation}
where $c_T,d_T,c_U,d_U$ are integers that define $\gamma$. This relation can be straighforwardly 
extended to include stabilizer elements that involve the mirror symmetry $\hat M$ and the \CP-like 
$\hat\Sigma_*$ generators, as they do not induce automorphy factors, see the discussion in 
section~\ref{sec:mirrorOnMatter} and appendix~\ref{sec:appDModularGroup}. Since the 
representation $\rho_{\rep{4}_3}(\gamma)$ of the finite modular group is unitary, its eigenvalues 
and hence the automorphy factors must be phases, see also ref.~\cite[section 6]{Nilles:2020gvu}. 
Note that eq.~\eqref{eq:YAlignment} corresponds to the mechanism of flavon alignment in the context 
of modular flavor symmetries, as also discussed in e.g.\ 
refs.~\cite{Novichkov:2018ovf,Novichkov:2018yse,Gui-JunDing:2019wap}.

A consequence of the automorphy factors being phases at $(\vev{T},\vev{U})$ is that the modular 
transformations from $H_{(\vev T,\vev U)}$ act linearly on matter fields. Hence, the stabilizer 
enhances the traditional flavor symmetry to the multiplicative closure of the traditional flavor 
group and the stabilizer modular subgroup, i.e.\ to the so-called unified flavor group
\begin{equation}
  \big[(D_8\x D_8)/\Z2\x\Z4^{R}\big]/\Z2 ~\cup~ H_{(\vev T,\vev U)}\,.
\end{equation}
Explicitly, from eqs.~\eqref{eq:generalModTrafoOnFields} and~\eqref{eq:AutomorphyFactorSLxSL}, 
the action of a (non-\CP-like) stabilizer element $\gamma\in H_{(\vev T,\vev U)}$ on a field 
$\Phi_{(n_T,n_U)}$ with modular weights $(n_T,n_U)$ is given by
\begin{equation}
\label{eq:unifiedTrafo}
  \Phi_{(n_T,n_U)} ~\xmapsto{~\gamma~}~ \rho_{\rep{t},(\vev T,\vev U)} \Phi_{(n_T,n_U)}
  ~:=~ (c_T\vev{T}+d_T)^{n_T}(c_U\vev{U}+d_U)^{n_U} \rho_{\rep{r}}(\gamma) \Phi_{(n_T,n_U)}\,,
\end{equation}
where $\rho_{\rep t,(\vev T,\vev U)}$ is a $t$-dimensional representation $\rep t$ of the unified 
flavor group, whereas $\rep r$ is a representation of the finite modular group $[144,115]$. We 
stress here the presence of the automorphy factors in the transformation~\eqref{eq:unifiedTrafo}, 
which can enhance the order of the unbroken transformations due to the possibility of 
fractional weights of matter fields.

%%%%%%%%%%%%%%%%%%%%%%%%%%%%%
\subsection{Unified flavor groups at generic points in moduli space}

Even for generic values of the moduli, the traditional flavor group is enhanced
for $\vev T = \vev U$.
In this case, the mirror transformation $\hat M$ is left unbroken. Considering its 
$\Z4^{\hat M}$ action on matter fields, as given by eqs.~\eqref{eq:hatMpreliminary} 
or~\eqref{eq:MirrorOnTwistedMatter}, we find that the unified flavor group in this case 
is given by $[128,2316]$. 

\begin{figure}[t!]
    \centering
    \begin{minipage}{0.48\textwidth}%
        \subfloat[(a)][$U$ fixed at a generic value $\vev U\neq\I,e^{\nicefrac{\pi\I}{3}}$.]{
            \label{fig:genericU}
            \centering \includegraphics[page=1,scale=1.8]{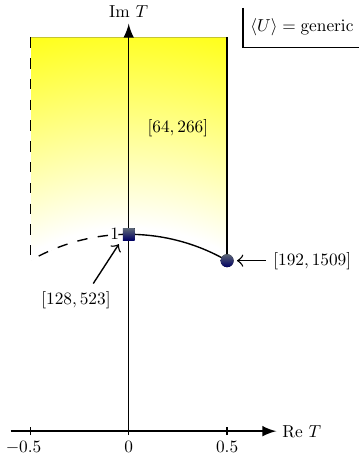}
            }
    \end{minipage}%
    \begin{minipage}{0.48\textwidth}%
        \subfloat[(b)][$T$ fixed at a generic value $\vev T\neq\I,e^{\nicefrac{\pi\I}{3}}$.]{
            \label{fig:genericT}
            \centering \includegraphics[page=1,scale=1.8]{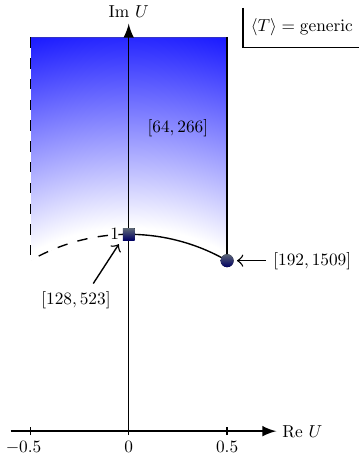}
        }
    \end{minipage}%
    \caption{Unified flavor groups at special points in moduli space, including the $\Z4^R$ symmetry. 
    (a) For generic $\vev U$, only at $\vev T=\I$ (square) and $\vev T=e^{\nicefrac{\pi\I}{3}}$ 
    (bullet) the traditional flavor symmetry is enhanced by $H_{(\I,\vev U)}=\langle \hat{K}_\mathrm{S}\rangle$ 
    and $H_{(e^{\nicefrac{\pi\I}{3}},\vev U)}=\langle \hat{K}_\mathrm{T} \hat{K}_\mathrm{S}\rangle$, respectively.
    (b) For generic $\vev T$, the results are equivalent due to mirror symmetry $\hat M$, which
    exchanges $T\leftrightarrow U$, $\hat{K}_\mathrm{T}\leftrightarrow\hat{C}_\mathrm{T}$ and
    $\hat{K}_\mathrm{S}\leftrightarrow\hat{C}_\mathrm{S}$.
    \label{fig:UTmodulispaces}}
\end{figure}

We can consider also the case that one of the moduli has a generic value while the second modulus
is fixed at one of the special points, $\I$ or $e^{\nicefrac{\pi\I}{3}}$. In figure~\ref{fig:UTmodulispaces}, 
we display the different unified flavor groups achieved by incorporating the unbroken modular 
transformations at those special points in moduli space. Let us consider the results for 
generic $\vev U$, as presented in figure~\ref{fig:genericU}. At $\vev T =\I$, we know 
(cf.~ref.~\cite{Baur:2020jwc}) that the stabilizer subgroup is generated by $\hat K_{\mathrm S}$, 
which becomes a \Z8 generator, considering the admissible automorphy factors 
$(-\I)^{n_T}=e^{\nicefrac{-\pi\I\,n_{\scaleto{T}{3pt}}}{2}}$ 
for matter fields with $n_T\in\{0,-1,\nicefrac{\pm 1}{2}, \nicefrac{-3}{2}\}$. 
However, the traditional flavor group $[64,266]$ is only enhanced to $[128,523]$ because 
$(\hat K_{\mathrm S})^2=(\hat C_{\mathrm S})^2$ amounts to the $\Z4^R$ symmetry already 
contained in the traditional symmetry. Further, at $\vev T = e^{\nicefrac{\pi\I}{3}}$, 
the stabilizer is generated by $\hat K_{\mathrm T}\hat K_{\mathrm S}$. Taking the 
automorphy factor $(-e^{\nicefrac{\pi\I}{3}})^{n_T}=e^{\nicefrac{-2\pi\I\,n_{\scaleto{T}{3pt}}}{3}}$,
it corresponds to a \Z6 symmetry, such that $(\hat K_{\mathrm T}\hat K_{\mathrm S})^3$ 
is equivalent to the point group of $\mathbbm T^2/\Z2$. Thus, the order of the traditional flavor group is only 
enhanced by a factor of three, leading to the unified flavor group $[192,1509]$.
Because of the mirror symmetry $\hat M$, it suffices to consider generic $\vev U$ to 
learn what happens also for generic $\vev T$, as we notice by comparing 
figures~\ref{fig:genericT} and~\ref{fig:genericU}. For example, for generic value of
the K\"ahler modulus and $\vev U=\I$, the stabilizer is generated by $\hat C_{\mathrm S}$,
which enhances the traditional flavor group $[64,266]$ to the unified flavor group $[128,523]$.

%%%%%%%%%%%%%%%%%%%%%%%%%%
\subsection{Unified flavor groups of the raviolo}
\label{sec:raviolo}

At $\vev U=\I$ (cf.~ref.~\cite{Feruglio:2021dte}), the geometry of the $\mathbbm T^2/\Z2$ 
orbifold adopts the form of a square raviolo, where the corners correspond to the singularities of 
the orbifold and the edges are perpendicular and have the same length. As just mentioned, in this 
case the traditional flavor group is enhanced by $\hat C_{\mathrm S}$ to the unified flavor group 
$[128,523]$, considering the \Z8 phases $(-\I)^{n_U}=e^{\nicefrac{-\pi\I\,n_{\scaleto{U}{3pt}}}{2}}$ 
associated with the automorphy factors of the matter fields. Since $\hat C_{\mathrm S}$ acts on the 
compact space as a $\pi/2$ rotation~\cite[eq.~(63a)]{Nilles:2020gvu}, the unified flavor group 
contains a remnant of the Lorentz symmetry in higher dimensions and is hence a discrete 
$R$-symmetry.

\begin{figure}[t!]
    \centering
    \begin{minipage}{0.48\textwidth}%
        \subfloat[a][$\mathbbm T^2/\Z2$ with $\vev U=\I$]{
            \label{fig:TplaneForU=i}
            \centering \includegraphics[page=2,scale=1.8]{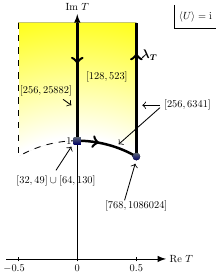}
        }
    \end{minipage}%
    \begin{minipage}{0.48\textwidth}%
        \subfloat[b][$\mathbbm T^2/\Z2$ with $\vev U=e^{\nicefrac{\pi\I}{3}}$]{
            \label{fig:TplaneForU=e^pii/3}
            \centering \includegraphics[page=3,scale=1.8]{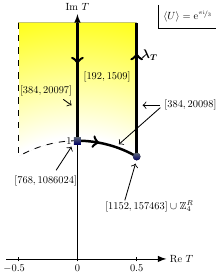}
        }
    \end{minipage}%
    \caption{Unified flavor groups at different points \vev{T} and the boundary $\lambda_T$ 
    of the fundamental domain of $\SL{2,\Z{}}_T$ for two special vevs \vev{U} of the complex structure 
    modulus. We use the shorthands $[32,49]\cup[64,130]\cong\big[[32,49]\x[64,130]\big]/\Z2$ and
    $[1152,157463]\cup\Z4^R\cong \left[[1152,157463]\x\Z4^R\right]/\Z2$, where the \Z2 identifies 
    the point group selection rule of $\mathbbm T^2/\Z2$ contained in the multiplied groups.
    \label{fig:TplaneRavioloTetrahedron}}
\end{figure}

There are two special points in moduli space for the raviolo, where further enhancements occur if 
the \CP-like modular transformation $\hat\Sigma_*$ is considered. First, at $\vev T=\I$ we find the 
stabilizer $H_{(\I,\I)}=\langle \hat C_{\mathrm S},\hat M, \hat\Sigma_*\rangle$, which enhances the 
traditional flavor symmetry to
\begin{equation}
\label{eq:largestUFG}
  \frac{D_8\x D_8}{\Z2}~\cup~ \left(D_8\rtimes D_8\right) ~\cong~ \frac{[32,49]\,\x\,[64,130]}{\Z2}\,,
\end{equation}
where $(D_8\x D_8)/\Z2\cong[32,49]$ is the traditional flavor group without $\Z4^R$, 
$D_8\rtimes D_8\cong[64,130]$ is the modular group corresponding to $H_{(\I,\I)}$ including the 
automorphy factors, and the \Z2 in the quotient on the right-hand side identifies the point group 
selection rule of the $\mathbbm T^2/\Z2$ orbifold sector present in both groups. The unified flavor 
group~\eqref{eq:largestUFG} has thus order 1024, which corresponds to the maximal enhancement of 
the traditional flavor symmetry for $\vev U=\I$. Secondly, at $\vev T=e^{\nicefrac{\pi\I}{3}}$, the 
traditional flavor group is enhanced by the subgroup of modular transformations 
$H_{(e^{\nicefrac{\pi\I}{3}},\I)}=\langle\hat C_{\mathrm S},\hat K_{\mathrm T}\hat K_{\mathrm S},\hat K_{\mathrm S}\hat\Sigma_*\rangle$ 
to the group $[768, 1086024]$. These enhancements are illustrated in figure~\ref{fig:TplaneForU=i}.

Additionally, on the line in moduli space described by the boundary $\lambda_T$ of the fundamental 
domain, depicted in figure~\ref{fig:TplaneForU=i}, we find two more enhancements. We see that the 
K\"ahler modulus fixed at $\vev T=\I\im\vev T>\I$ is left invariant by 
$\langle\hat C_{\mathrm S},\hat\Sigma_*\rangle$. Acting on matter fields along with their 
corresponding automorphy factors, this yields the unified flavor group $[256, 25882]$. 
Furthermore, the traditional flavor symmetry is enhanced to $[256, 6341]$ along the regions of the 
locus $\lambda_T$ where $\vev T=e^{\I\varphi}$ with $\nicefrac\pi3<\varphi<\nicefrac\pi2$, and 
$\vev T=\nicefrac12+\I\im\vev T$ with $\im\vev T>\nicefrac{\sqrt3}{2}$. In these regions, the 
stabilizers are given by $\langle\hat C_{\mathrm S}, \hat K_{\mathrm S}\hat\Sigma_*\rangle$ and 
$\langle\hat C_{\mathrm S}, \hat K_{\mathrm T}\hat\Sigma_*\rangle$, respectively.

%%%%%%%%%%%%%%%%%%%%%%%%%%%%%%%%%%%%%%%%%%%%%%%%%%%%
\subsection{Unified flavor groups of the tetrahedron}
\label{sec:tetrahedron}

When the complex structure is stabilized at $\vev U=e^{\nicefrac{\pi\I}{3}}$, the $\mathbbm T^2/\Z2$ 
orbifold sector has the shape of a tetrahedron, cf.~ref.~\cite[figure~5]{Baur:2020jwc}. As can be 
read off from figure~\ref{fig:genericT}, in the tetrahedron with a generic value for $T$, the 
$\hat C_{\mathrm T}\hat C_{\mathrm S}$ modular transformation leaves the moduli invariant and 
enhances the traditional flavor group to the unified flavor group $[192,1509]$, the generic flavor 
symmetry of the tetrahedron. This contains discrete $R$-symmetries due to the inclusion of the 
discrete rotation in the compact space generated by $\hat C_{\mathrm T}\hat C_{\mathrm S}$.

The symmetry of the tetrahedron can be further enlarged if the K\"ahler modulus is fixed e.g.\ at 
the special values $\vev T=e^{\nicefrac{\pi\I}{3}}$ or $\vev T=\I$. At these points, the respective 
stabilizer subgroups are 
$H_{(e^{\nicefrac{\pi\I}{3}},e^{\nicefrac{\pi\I}{3}})}=\langle \hat C_{\mathrm T}\hat C_{\mathrm S},\hat C_{\mathrm T}\hat K_{\mathrm T}\hat\Sigma_*,\hat M\rangle$ and 
$H_{(\I,e^{\nicefrac{\pi\I}{3}})}=\langle \hat C_{\mathrm T}\hat C_{\mathrm S},\hat C_{\mathrm S}\hat\Sigma_*,\hat K_{\mathrm S}\rangle$, 
which include the \CP-like modular transformation $\hat\Sigma_*$. Considering the action of the 
stabilizer elements in each case, including their automorphy factors, we find the enhancements 
shown in figure~\ref{fig:TplaneForU=e^pii/3}. Note that, like  
in the case of the raviolo, the point at which 
both moduli values coincide, i.e.\ at $(\vev T,\vev U)=(e^{\nicefrac{\pi\I}{3}},e^{\nicefrac{\pi\I}{3}})$, 
is endowed with the largest possible linear enhancement of the traditional flavor symmetry in the 
$\mathbbm T^2/\Z2$ orbifold sector, the group
\begin{equation}
\label{eq:largestUnifiedFlavorGroup}
\frac{[1152,157463]\,\x\,\Z4^R}{\Z2}\,,
\end{equation}
which is of  order 2304. Here, as before, the \Z2 corresponds to the identification of the point group 
selection rule of this orbifold sector, which appears in both groups, $[1152,157463]$ and $\Z4^R$.

Along the locus $\lambda_T$, there are two more enhancements of the traditional flavor group. For 
$\vev T=\I\im\vev T>\I$, not only $\hat C_{\mathrm T}\hat C_{\mathrm S}$ leaves the moduli 
invariant but also $\hat C_{\mathrm T}\hat\Sigma_*$. This implies that the flavor symmetry of the 
tetrahedron $[192,1509]$ is enhanced to the unified flavor group $[384,20097]$. Besides, for 
$\vev T = e^{\I\varphi}$ with $\nicefrac\pi3<\varphi<\nicefrac\pi2$, the stabilizer is 
$\langle\hat C_{\mathrm T}\hat C_{\mathrm S}, \hat K_{\mathrm S}\hat C_{\mathrm S}\hat\Sigma_*\rangle$ 
and leads to the unified flavor group $[384,20098]$. The same flavor enhancement is obtained if the 
K\"ahler modulus sits at $\vev T=\nicefrac12+\I\im\vev T$ with $\im\vev T>\nicefrac{\sqrt3}{2}$, 
where the stabilizer is generated by $\hat C_{\mathrm T}\hat C_{\mathrm S}$ and 
$\hat K_{\mathrm T}\hat C_{\mathrm S}\hat\Sigma_*$.

%%%%%%%%%%%%%%%%%%%%%%%%
\subsubsection[A4 flavor symmetry from the tetrahedron]{\boldmath$A_4$ flavor symmetry from the tetrahedron\unboldmath}
\label{sec:A4symmetry}

Let us turn back to the case of the tetrahedron, $\vev U=e^{\nicefrac{\pi\I}{3}}$, with a generic 
value for the K\"ahler modulus. In this case, according to eq.~\eqref{eq:unifiedTrafo}, the
stabilizer modular generator $\hat{C}_\mathrm{T}\hat{C}_\mathrm{S}$ acts on matter fields as
\begin{equation}\label{eq:CTCSTrafo}
\Phi_{(n_T,n_U)} \xmapsto{\hat{C}_\mathrm{T}\,\hat{C}_\mathrm{S}} 
\rho_{\rep{r},(\vev T,e^{\nicefrac{\pi\I}{3}})}(\hat{C}_\mathrm{T}\,\hat{C}_\mathrm{S})\,
\Phi_{(n_T,n_U)}\,,
\end{equation}
where due to the automorphy factor $(c_U\,\vev U+d_U)^{n_U}$ we get
\begin{equation}
\rho_{\rep{r},(\vev T,e^{\nicefrac{\pi\I}{3}})}(\hat{C}_\mathrm{T}\,\hat{C}_\mathrm{S}) 
~:=~ \left(\exp\left(\nicefrac{-2\pi\I}{3}\right)\right)^{n_U}\!\rho_{\rep{r}}(\hat{C}_\mathrm{T})\rho_{\rep{r}}(\hat{C}_\mathrm{S})\;.
\end{equation}
The admissible modular weights of massless matter fields are $n_U\in\{0,-1\}$ for bulk matter, and 
$n_U\in\{\nicefrac{-3}{2},\nicefrac{-1}{2},\nicefrac{1}{2}\}$ for twisted matter. Hence, 
eq.~\eqref{eq:CTCSTrafo} describes a \Z6 transformation that we can write as $\Z{2}\times\Z{3}$, 
generated by
\begin{subequations}
\begin{align}
&\Z{2}^{\mathrm{(PG)}}\;\;: & \left(\rho_{\rep{r},(\vev T,e^{\nicefrac{\pi\I}{3}})}(\hat{C}_\mathrm{T}\,\hat{C}_\mathrm{S}) \right)^3 &=~ \exp\left(-2\pi\I n_U\right)\,\Id\;,\label{eq:TetraZ2}\\
&\Z{3}^R\;\;\;\;\;\;:       & \left(\rho_{\rep{r},(\vev T,e^{\nicefrac{\pi\I}{3}})}(\hat{C}_\mathrm{T}\,\hat{C}_\mathrm{S}) \right)^2 &=~ \exp\left(\nicefrac{-4\pi\I n_U}{3}\right)\,\left(\rho_{\rep{r}}(\hat{C}_\mathrm{T})\rho_{\rep{r}}(\hat{C}_\mathrm{S})\right)^2\;, \label{eq:TetraZ3}
\end{align}
\end{subequations}
respectively. Using the admissible modular weights, we note that the \Z2 factor in 
eq.~\eqref{eq:TetraZ2} corresponds to the $\Z2^{\mathrm{(PG)}}$ point group selection rule of the 
$\mathbbm T^2/\Z2$ orbifold sector. Moreover, the $\Z{3}$ factor eq.~\eqref{eq:TetraZ3} acts on the 
superpotential $\mathcal{W}$ as 
\begin{equation}\label{eq:CTCSonW}
\mathcal{W} ~\xmapsto{\left(\hat{C}_\mathrm{T}\,\hat{C}_\mathrm{S}\right)^2}~ \omega^2\,\mathcal{W}\;,
\end{equation}
which is a discrete $\Z3^R$ $R$-symmetry using the definition 
$\omega := \exp\left(\nicefrac{2\pi\I}{3}\right)$.

The group generated by the traditional flavor group elements $\rho_{\rep{4}}(h_1)$, 
$\rho_{\rep{4}}(h_2)$ from eq.~\eqref{eq:TraditionalRhoOfTwistedStrings} together with the 
$\Z{2}^{\mathrm{(PG)}}$ factor from eq.~\eqref{eq:TetraZ2} and the $\Z3^R$ factor from 
eq.~\eqref{eq:TetraZ3} turns out to be
\begin{equation}
A_4^R ~\times~ \Z{2}^{\mathrm{(PG)}} ~\cong~ [ 24, 13 ]\;.
\end{equation}
Here, the alternating group $A_4^R$ is a non-Abelian $R$-symmetry as it arises from 
$\rho_{\rep{4}}(h_1)$, $\rho_{\rep{4}}(h_2)$ and the $\Z3^R$ $R$-symmetry, cf.\ 
ref~\cite{Chen:2013dpa} for a general discussion on non-Abelian $R$-symmetries. The matter fields 
and the superpotential $\mathcal{W}$ build the following representations of 
$A_4^R \times \Z{2}^{\mathrm{(PG)}}$
\begin{subequations}
\begin{eqnarray}
\Phi_{(0,0)}                               & : & (\rep{1},\rep{1}_0)\;,\\
\Phi_{(-1,-1)}                             & : & (\rep{1}',\rep{1}_0)\;,\\
\Phi_{(\nicefrac{-1}{2},\nicefrac{-1}{2})} & : & (\rep{3},\rep{1}_1) ~\oplus~ (\rep{1}'',\rep{1}_1)\;,\\
\mathcal{W}                                & : & (\rep{1}',\rep{1}_0)\;,
\end{eqnarray}
\end{subequations}
where we denote the irreducible representations of $\Z{2}$ by $\rep{1}_0$ and $\rep{1}_1$ and the 
ones of $A_4^R$ by $\rep{3}$, $\rep{1}$, $\rep{1}'$ and $\rep{1}''$, see appendix~\ref{app:A4}.

Combined with the $\Z4^R$ symmetry associated with the sublattice rotation $\hat\Theta_{(2)}$ and 
the $\Z2\x\Z2$ generators $\rho_{\rep{4}}(h_3)$ and $\rho_{\rep{4}}(h_4)$ of the traditional flavor 
group, associated with the space group selection rule of the $\mathbbm T^2/\Z2$ orbifold sector, 
the group $A_4^R\x\Z2$ enhances to the unified flavor group of the tetrahedron, $[192,1509]$, see 
figure~\ref{fig:TplaneForU=e^pii/3}.

Compared to the literature (see e.g.\ refs.~\cite{Altarelli:2006kg,Adulpravitchai:2009id,deAnda:2019jxw}), 
we see that in the consistent string approach the na\"ive $A_4$ symmetry obtained by the compactification 
of two extra dimensions on a tetrahedron has to be extended in two ways: First, the $\Z3$ generator 
of $A_4^R$ turns out to be an $R$-symmetry. This can be understood equivalently either as a 
discrete remnant of the extra-dimensional Lorentz symmetry or as a discrete remnant of an 
$\SL{2,\Z{}}_U$ modular symmetry. In addition, $A_4^R$ is enhanced in the full string approach by 
stringy selection rules to $[192, 1509]$ of order 192, which still contrasts with previous 
results~\cite{Kobayashi:2006wq}.

%%%%%%%%%%%%%%%%%%%%%%%%%
\subsection[Other CP-enhanced unified flavor groups]{\boldmath Other \CP-enhanced unified flavor groups \unboldmath}

\begin{figure}[t!]
\centering
\centering{\includegraphics[width=0.9\linewidth]{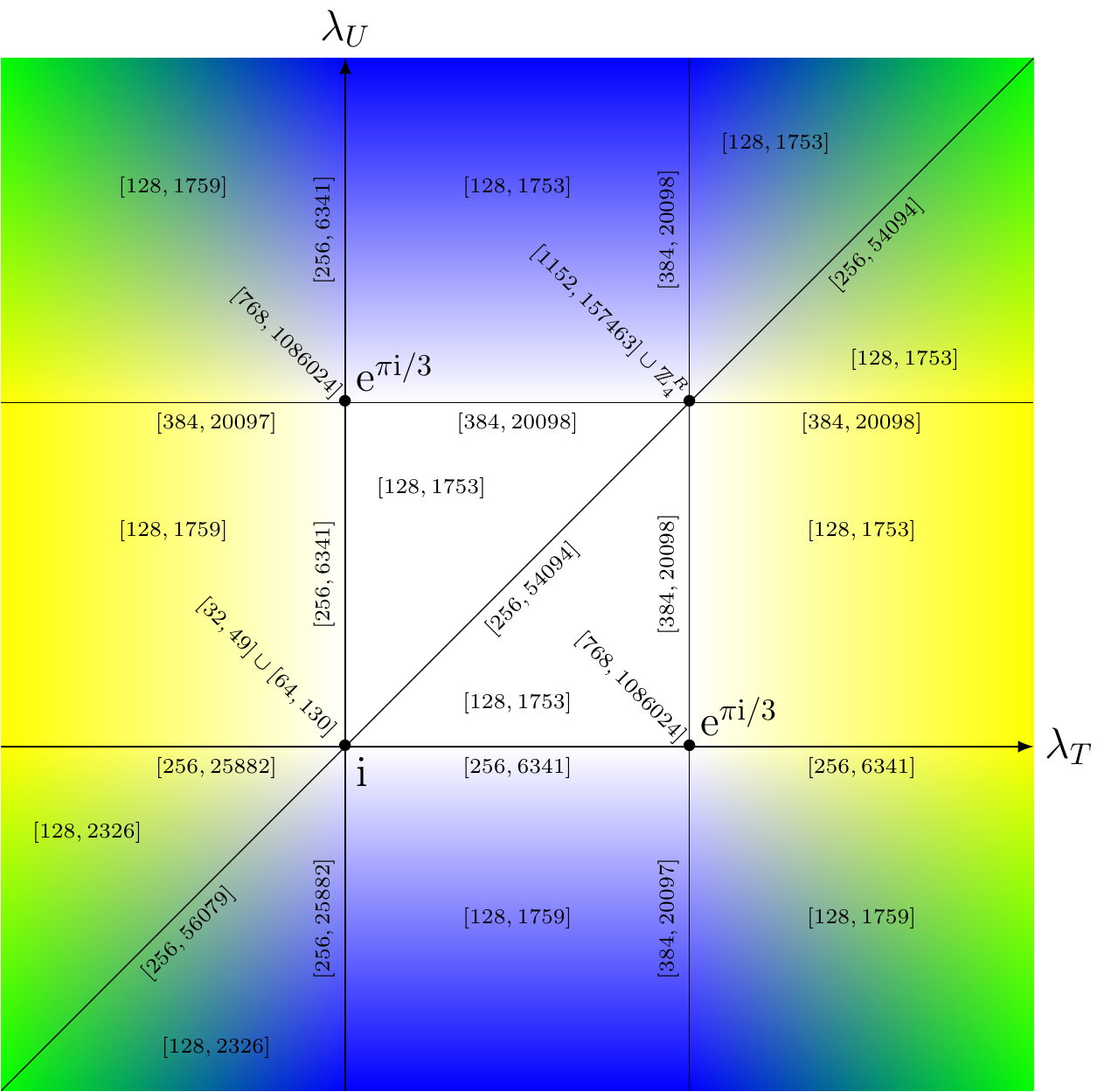}}
\caption{Unified flavor groups with \CP at $\vev{T}\in\lambda_T$ and $\vev{U}\in\lambda_U$. We use the 
shorthands $[32,49]\cup[64,130]\cong\big[[32,49]\x[64,130]\big]/\Z2$ and 
$[1152,157463]\cup\Z4^R\cong \left[[1152,157463]\x\Z4^R\right]/\Z2$, where the \Z2 corresponds
to the point group selection rule of the $\mathbbm T^2/\Z2$ orbifold sector, contained in both 
groups in the product. The axes $\lambda_T$ and $\lambda_U$ describe the boundaries of the 
fundamental domains of $T$ and $U$, see e.g.\ figure~\ref{fig:TplaneRavioloTetrahedron} for $\lambda_T$. 
The diagonal depicts the hypersurface where $\vev U=\vev T$ on the curves $\lambda_T$ and $\lambda_U$. 
The unified flavor groups above and below the diagonal are related by mirror symmetry $\hat M$.
\label{fig:LocalFlavorGroups}}
\end{figure}

Figure~\ref{fig:LocalFlavorGroups} displays all unified flavor groups of the $\mathbbm T^2/\Z2$ 
orbifold sector that include i) the $\Z4^R$ symmetry arising from embedding the orbifold sector in 
higher dimensions, ii) the \CP-like modular transformation $\hat\Sigma_*$, and iii) the phases 
associated with the automorphy factors of modular transformations acting on matter fields 
$\Phi_{(n_T,n_U)}$. We use as reference axes the straightened lines describing the boundaries 
$\lambda_T$ and $\lambda_U$ of the $T$ and $U$ moduli spaces, respectively. For example, 
$\lambda_T$ is illustrated in figure~\ref{fig:TplaneRavioloTetrahedron}. Note that the flavor 
enhancements along the horizontal lines in figure~\ref{fig:LocalFlavorGroups} have already been 
discussed in sections~\ref{sec:raviolo} (bottom line with $\vev U=\I$) 
and~\ref{sec:tetrahedron} (upper line with $\vev U=e^{\nicefrac{\pi\I}{3}}$). 

Mirror symmetry $\hat M$ acts on the moduli and the modular generators as $U\leftrightarrow T$, 
$\hat C_\mathrm{S}\leftrightarrow\hat K_\mathrm{S}$, and 
$\hat C_\mathrm{T}\leftrightarrow\hat K_\mathrm{T}$. As a first consequence, the points along the 
diagonal in figure~\ref{fig:LocalFlavorGroups}, defined by $\vev T=\vev U$, are left invariant by 
$\hat M$. Furthermore, the points below and above this diagonal are connected by a mirror 
transformation. It then follows that $\hat M$ identifies the unified flavor groups in these two 
sectors of moduli space.

Focusing on the lower half of the plane, below the diagonal of figure~\ref{fig:LocalFlavorGroups}, 
we see that the sole enhancements that have not been discussed in the preceding subsections are 
those that lie at the diagonal, and those that are valid in the squared and triangular regions of 
figure~\ref{fig:LocalFlavorGroups}. Let us consider two examples. In the lowest part of the 
diagonal, the stabilizer modular subgroup is $H_{(\I x,\I x)} = \langle\hat M,\hat\Sigma_*\rangle$, 
with $x>1$. Considering the associated transformations on matter fields with their corresponding 
automorphy factors, as given in appendix~\ref{sec:appDModularGroup}, we find that the traditional 
flavor group $[64,266]$ is in this case enhanced to $[256,56079]$. Similarly, in the bottom 
triangle of the figure, the stabilizer is given just by 
$H_{(\I x,\I y)} = \langle\hat\Sigma_*\rangle$, with $x,y>1$ and $y>x$, which yields the unified 
flavor group $[128,2326]$. Other cases can be easily determined by using the proper stabilizer 
subgroups provided in our previous work~\cite[figure~7]{Baur:2020jwc}.

%%%%%%%%%%%%%%%%%%%%%%%%%%%%%%%%%%%%%%%%%%%%%%%%%%%%%%%%%%%%%%%%%%%%%%%%%%%%%%%%%%%%%%%%%%%%%%%%%%%%%%%%%%%%%%%%%%%%%%%%%%%%%%%%%%%%%%%%%
\section[Effective field theory of the Z2 orbifold]{\boldmath Effective field theory of the \Z2 orbifold\unboldmath}
\label{sec:EFT}

In this section, we focus on four $\rep{4}$-plets of twisted matter fields 
$\Phi^i_{(n_T,n_U)}$, for $i\in\{1,2,3,4\}$, with modular weights $n_T=n_U=\nicefrac{-1}{2}$, see 
table~\ref{tab:Z2modularWeights}. Each $\rep{4}$-plet contains four fields $\phi^i_{(n_1,n_2)}$ 
that we label by the winding numbers $(n_1,n_2)\in\{(0,0),(1,0),(0,1),(1,1)\}$ and \emph{not} by 
the modular weights $n_T$ and $n_U$. Hence, each twisted matter field $\phi^i_{(n_1,n_2)}$ is 
localized at one of the four fixed points of the $\mathbbm{T}^2/\Z{2}$ orbifold sector, see 
appendix~\ref{app:StringsOnOrbifolds} and ref.~\cite[figure~1]{Baur:2020jwc}. In other words, we 
consider four $\rep{4}$-plets $\Phi^{i}_{(\nicefrac{-1}{2},\nicefrac{-1}{2})} := (\phi^i_{(0,0)},\phi^i_{(1,0)},\phi^i_{(0,1)},\phi^i_{(1,1)})^\mathrm{T}$. 
We assume that the $\rep{4}$-plets differ in some additional charges, for example with respect to 
the unbroken gauge group from $\E8\times\E8$ (or $\SO{32}$). Then, we use the eclectic flavor 
symmetry to write down the most general K\"ahler and superpotential to lowest order in these fields.

%%%%%%%%%%%%%%%%%%%%%%%%%%%%%%%%%%%%%%%%%%%%%%%%%%%%
\subsection[The K\"ahler potential]{The K\"ahler potential}

The Hermitian K\"ahler potential $K$ of a single twisted matter field 
$\Phi_{(\nicefrac{-1}{2},\nicefrac{-1}{2})}$ reads to leading order~\cite{Dixon:1989fj}
\begin{equation}
K ~\supset~ (-\I T + \I \bar{T})^{\nicefrac{-1}{2}}\, (-\I U + \I \bar{U})^{\nicefrac{-1}{2}} \, \tilde{K}\;,
\end{equation}
where $\tilde{K}$ is an Hermitian bilinear polynomial of the form 
$\tilde{K} = c_{n_1 n_2 n'_1 n'_2} \phi_{(n_1,n_2)}\,\bar{\phi}_{(n'_1,n'_2)}$. In the following, 
we constrain $\tilde{K}$ by imposing the traditional flavor symmetry step by step. First, the space 
and point group selection rules~\eqref{eq:SGandPG} enforce $n_1'=n_1$ and $n_2'=n_2$, resulting in 
$\tilde{K} = c_{n_1 n_2} \phi_{(n_1,n_2)}\,\bar{\phi}_{(n_1,n_2)}$. Then, invariance under 
eq.~\eqref{eq:h1andh2} forces all coefficients $c_{n_1 n_2}$ to be equal (and we normalize them to 
1). Hence,
\begin{equation}
\tilde{K} ~=~ \left( |\phi_{(0,0)}|^2 + |\phi_{(1,0)}|^2 + |\phi_{(0,1)}|^2 + |\phi_{(1,1)}|^2\right)\;.
\end{equation}
Now, we can generalize this easily to four $\rep{4}$-plets of twisted matter fields 
$\Phi^i_{(\nicefrac{-1}{2},\nicefrac{-1}{2})}$, for $i\in\{1,2,3,4\}$. Due to our assumption of 
additional (gauge) charges that distinguish between $\phi^i_{(n_1,n_2)}$ and $\phi^j_{(n_1,n_2)}$ 
for $i\neq j$, we obtain
\begin{equation}
\label{eq:generalK}
\tilde{K} ~=~ \sum_{i=1}^4 \left( |\phi^i_{(0,0)}|^2 + |\phi^i_{(1,0)}|^2 + |\phi^i_{(0,1)}|^2 + |\phi^i_{(1,1)}|^2\right)\;.
\end{equation}
Consequently, for the $\mathbbm T^2/\Z2$ orbifold sector the traditional flavor symmetry already 
enforces the K\"ahler potential to be diagonal in twisted matter fields. Hence, this diagonal 
structure can not be changed by the full eclectic flavor symmetry: Since additional terms involving 
modular forms $\hat{Y}(T,U)$ (as suggested by ref.~\cite{Chen:2019ewa}) are singlets of the 
traditional flavor group, the K\"ahler potential must remain diagonal, cf.\ 
ref.~\cite{Nilles:2020kgo}. Yet, additional corrections to the K\"ahler potential that involve 
flavons are still possible, cf.\ ref.~\cite{Chen:2013aya}.

%%%%%%%%%%%%%%%%%%%%%%%%%%%%%%%%%%%%%%%%%%%%%%%%%%%%
\subsection[The superpotential]{The superpotential}

To lowest order in twisted matter fields $\Phi^{i}_{(\nicefrac{-1}{2},\nicefrac{-1}{2})}$, the 
superpotential reads schematically
\begin{subequations}\label{eq:PreSuperpot}
\begin{eqnarray}
\mathcal{W} & \supset & \hat{Y}^{(0)}(T,U)\, \Phi_{(0,0)}\, \Phi^{i}_{(\nicefrac{-1}{2},\nicefrac{-1}{2})}\, \Phi^{j}_{(\nicefrac{-1}{2},\nicefrac{-1}{2})} \label{eq:PreSuperpot1}\\
            &+& \hat{Y}^{(2)}(T,U)\,\Phi_{(-1,-1)}\, \Phi^1_{(\nicefrac{-1}{2},\nicefrac{-1}{2})}\, \Phi^2_{(\nicefrac{-1}{2},\nicefrac{-1}{2})}\, \Phi^3_{(\nicefrac{-1}{2},\nicefrac{-1}{2})}\, \Phi^4_{(\nicefrac{-1}{2},\nicefrac{-1}{2})}\;.\label{eq:PreSuperpot2}
\end{eqnarray}
\end{subequations}
Here, we have imposed the $\Z{4}^R$ $R$-symmetry and the fact that the modular weights of matter 
fields and couplings have to add up to $(-1,-1)$ for the superpotential, see 
table~\ref{tab:Z2modularWeights}. Thus, $\hat{Y}^{(0)}(T,U)$ has to carry modular weights $(0,0)$, 
while the modular form $\hat{Y}^{(2)}(T,U)$ has modular weights $(2,2)$. There exists a unique 
modular form of weight $(2,2)$, which we denote by $\hat{Y}^{(2)}_{\rep{4}_3}(T,U)$ in the 
following (see eq.~\eqref{eq:ModularFormWeight22}). In addition, the superpotential has to be 
covariant under the full eclectic flavor symmetry, i.e.\ it has to be invariant simultaneously 
under the traditional non-$R$ symmetries and the finite modular flavor symmetry but transform with 
the appropriate phases (automorphy factors) under the $R$-symmetry (modular symmetry).

%%%%%%%%%%%%%%%%%%%%
\subsubsection[Constraints  from the traditional flavor symmetry]{Constraints from the traditional flavor symmetry}

Let us start with invariance under the traditional flavor symmetry 
$\left(D_8 \times D_8\right)/\Z{2} \cong [32,49]$. First, we consider the product of two twisted 
matter fields $\Phi^i_{(\nicefrac{-1}{2},\nicefrac{-1}{2})}\,\Phi^j_{(\nicefrac{-1}{2},\nicefrac{-1}{2})}$ 
needed for the terms~\eqref{eq:PreSuperpot1} in the superpotential $\mathcal{W}$. The fields 
$\Phi^i_{(\nicefrac{-1}{2},\nicefrac{-1}{2})}$ transform as irreducible $\rep{4}$-plets of the 
traditional flavor group $[32,49]$. Hence, we need to consider the tensor product
\begin{equation}\label{eq:TFG4times4}
\rep{4} ~\otimes~ \rep{4} ~=~ \bigoplus_{\alpha,\beta,\gamma,\delta~\in~\{-,+\}} \rep{1}_{\alpha\beta\gamma\delta}\;.
\end{equation}
This tensor product contains one trivial singlet $\rep{1}_{++++}$, which corresponds to the 
terms
\begin{equation}\label{eq:InvariantI0}
\mathcal{I}^{ij}_0 ~=~ \phi^i_{(0,0)} \phi^j_{(0,0)} + \phi^i_{(1,0)} \phi^j_{(1,0)} + \phi^i_{(0,1)} \phi^j_{(0,1)} + \phi^i_{(1,1)} \phi^j_{(1,1)}\;,
\end{equation}
for $i,j\in\{1,2,3,4\}$. The total $R$-charge is $2$ as one can see easily from 
table~\ref{tab:Z2modularWeights}. As a remark, one can check the invariance of the terms 
$\mathcal{I}^{ij}_0$ explicitly using the orthogonality of the representation matrices given in 
eq.~\eqref{eq:TraditionalRhoOfTwistedStrings}. Since $\Phi_{(0,0)}$ is a trivial singlet 
$\rep{1}_{++++}$ of $[32,49]$ with $R$-charge $0$, the terms 
$\Phi_{(0,0)}\, \mathcal{I}^{ij}_0 \subset \mathcal{W}$ are allowed by both, the traditional flavor 
symmetry and $\Z{4}^R$.

Next, we study the product of four twisted matter fields in order to construct the superpotential 
terms in eq.~\eqref{eq:PreSuperpot2}. Since 
\begin{equation}
\rep{1}_{\alpha\beta\gamma\delta} ~\otimes~ \rep{1}_{\alpha\beta\gamma\delta} ~=~ \rep{1}_{++++}\;,
\end{equation}
we know from eq.~\eqref{eq:TFG4times4} that there are 16 invariant combinations $\mathcal{I}_i$, 
$i \in\{1,\ldots,16\}$. We list them in appendix~\ref{app:superpotential}. Consequently, out of the 
$4^4 = 256$ possible terms from 
$\Phi^1_{(\nicefrac{-1}{2},\nicefrac{-1}{2})}\,\Phi^2_{(\nicefrac{-1}{2},\nicefrac{-1}{2})}\,\Phi^3_{(\nicefrac{-1}{2},\nicefrac{-1}{2})}\,\Phi^4_{(\nicefrac{-1}{2},\nicefrac{-1}{2})}$, 
invariance under the traditional flavor symmetry $[32,49]$ allows only 16.

%%%%%%%%%%%%%%%%%%%%
\subsubsection[Constraints from the modular symmetry]{Constraints from the modular symmetry}

As explained in ref.~\cite{Baur:2020jwc} and derived in appendix~\ref{app:TrafoOfTwistedStrings}, 
the modular symmetry $(\SL{2,\Z{}}_T \times \SL{2,\Z{}}_U) \rtimes\Z2$ of the K\"ahler and complex 
structure modulus $T$ and $U$, respectively, is realized as finite modular group 
$(S_3^T\times S_3^U)\rtimes\Z4^{\hat{M}}\cong [144,115]$ on twisted matter fields.

Using the matrix representation eq.~\eqref{eq:ModularRhoOfTwistedStrings} of the twisted matter 
fields $\Phi^i_{(\nicefrac{-1}{2},\nicefrac{-1}{2})}$, we see that the terms 
$\hat{Y}^{(0)}(T,U)\, \Phi_{(0,0)}\, \mathcal{I}^{ij}_0$ in eq.~\eqref{eq:PreSuperpot1} are 
invariant under the finite modular symmetry. 
In contrast, the terms $\mathcal{I}_i$, $i \in\{1,\ldots,16\}$, defined in appendix~\ref{app:superpotential},
do not transform trivially under the generators $\hat{C}_{\mathrm{S}}$, $\hat{C}_{\mathrm{T}}$ and 
$\hat{M}$ of the finite modular symmetry,
\begin{equation}
\begin{pmatrix}
\mathcal{I}_1 \\
\vdots \\
\mathcal{I}_{16}
\end{pmatrix}
~\xmapsto{~\hat{\Sigma}~}~
\left(j^{(\nicefrac{-1}{2})}(\hat{\Sigma},T,U)\right)^4\,R(\hat{\Sigma})\,
\begin{pmatrix}
\mathcal{I}_1 \\
\vdots \\
\mathcal{I}_{16}
\end{pmatrix}\;.
\label{eq:ITrafosUnderFMG}
\end{equation}
We list the $16\times 16$ matrices $R(\hat{\Sigma})$ in appendix~\ref{app:superpotential}. 
Comparing the traces of $R(\hat{\Sigma})$ to the character table of $[144,115]$ given in 
appendix~\ref{app:FMGCharacterTable}, we find that 
$\left(\mathcal{I}_1, ..., \mathcal{I}_{16}\right)^\mathrm{T}$ decomposes into the irreducible 
representations
\begin{equation}\label{eq:tensorproduct44}
\rep{1}_0 \oplus \rep{1}_0 \oplus \rep{1}_0 \oplus \rep{1}_1 \oplus \rep{4}_3 \oplus \rep{4}_3 \oplus \rep{4}_5
\end{equation}
of the finite modular group $[144,115]$. Only the two $\rep{4}_3$ representations yield invariant 
terms when they are combined with the modular form $\hat{Y}^{(2)}_{\rep{4}_3}(T,U)$ defined in 
eq.~\eqref{eq:ModularFormWeight22}. Hence, we identify the two $\rep{4}_3$ representations from the 
tensor product eq.~\eqref{eq:tensorproduct44}. We obtain
\begin{subequations}\label{eq:WQuartic}
\begin{align}
Q_1 ~:=&~ 
\begin{pmatrix}
2\,\mathcal{I}_2 - 2\,\mathcal{I}_3 + \mathcal{I}_6 + \mathcal{I}_9 - \mathcal{I}_5 - \mathcal{I}_8 \\
\sqrt{3}\, \left(\, \mathcal{I}_{12} +\mathcal{I}_{14} - \mathcal{I}_{11} - \mathcal{I}_{13} \,\right)\\
\sqrt{3}\, \left(\, \mathcal{I}_6 + \mathcal{I}_8 - \mathcal{I}_5 - \mathcal{I}_9 \,\right) \\
2\,\mathcal{I}_{15} - 2\,\mathcal{I}_{16} + \mathcal{I}_{12} + \mathcal{I}_{13} - \mathcal{I}_{11} - \mathcal{I}_{14}
\end{pmatrix}\;,\\[4pt]
Q_2 ~:=&~ 
\begin{pmatrix}
2\,\mathcal{I}_3 - 2\,\mathcal{I}_4 + \mathcal{I}_7 + \mathcal{I}_{10} - \mathcal{I}_6 - \mathcal{I}_9 \\
\sqrt{3}\, \left(\, \mathcal{I}_{15} + \mathcal{I}_{16} - \mathcal{I}_{12} - \mathcal{I}_{14} \,\right) \\
\sqrt{3}\, \left(\, \mathcal{I}_7 + \mathcal{I}_9 - \mathcal{I}_6 - \mathcal{I}_{10} \,\right) \\
2\,\mathcal{I}_{11} - 2\,\mathcal{I}_{13} + \mathcal{I}_{12} + \mathcal{I}_{16} - \mathcal{I}_{15} - \mathcal{I}_{14}
\end{pmatrix}\;.
\end{align}
\end{subequations}
Note that the quartic polynomial $Q_1$ is antisymmetric when 
$\Phi^1_{(\nicefrac{-1}{2},\nicefrac{-1}{2})} \leftrightarrow \Phi^4_{(\nicefrac{-1}{2},\nicefrac{-1}{2})}$ or
$\Phi^2_{(\nicefrac{-1}{2},\nicefrac{-1}{2})} \leftrightarrow \Phi^3_{(\nicefrac{-1}{2},\nicefrac{-1}{2})}$ 
are interchanged. Analogously, $Q_2$ is antisymmetric under 
$\Phi^1_{(\nicefrac{-1}{2},\nicefrac{-1}{2})} \leftrightarrow \Phi^2_{(\nicefrac{-1}{2},\nicefrac{-1}{2})}$ or
$\Phi^3_{(\nicefrac{-1}{2},\nicefrac{-1}{2})} \leftrightarrow \Phi^4_{(\nicefrac{-1}{2},\nicefrac{-1}{2})}$.
The invariant terms in the superpotential then read 
\begin{equation}
\mathcal{W}(T,U,\phi) ~\supset~ c_1 \left(\hat{Y}^{(2)}_{\rep{4}_3}(T,U) \cdot Q_1\right) \Phi_{(-1,-1)} \,+\, c_2 \left(\hat{Y}^{(2)}_{\rep{4}_3}(T,U) \cdot Q_2\right) \Phi_{(-1,-1)}\,.
\end{equation}
Hence, the number of unfixed superpotential parameters in eq.~\eqref{eq:PreSuperpot2} is reduced 
from 16 in the case of imposing only the traditional flavor symmetry to 2 (i.e.\ $c_1$ and $c_2$) 
when we include the constraints from the full eclectic flavor symmetry. This is in contrast to the 
leading order K\"ahler potential, see eq.~\eqref{eq:generalK}, where the finite modular symmetry 
did not yield additional constraints compared to the traditional flavor symmetry. Finally, the 
superpotential of eq.~\eqref{eq:PreSuperpot} is thus explicitly given by
\begin{subequations}\label{eq:Superpot}
\begin{eqnarray}
\mathcal{W}(T,U,\phi) & = & c_0^{ij}\, \hat{Y}^{(0)}_{\rep{1}_0}(T,U)\, \mathcal{I}^{ij}_0\, \Phi_{(0,0)} \\
                      & + & c_1        \left(\hat{Y}^{(2)}_{\rep{4}_3}(T,U) \cdot Q_1\right) \Phi_{(-1,-1)} \,+\, c_2 \left(\hat{Y}^{(2)}_{\rep{4}_3}(T,U) \cdot Q_2\right) \Phi_{(-1,-1)}\label{eq:Superpot2}\;,
\end{eqnarray}
\end{subequations}
where $Q_1$ and $Q_2$ are the quartic polynomials in the twisted matter fields 
$\Phi^i_{(\nicefrac{-1}{2},\nicefrac{-1}{2})}$, given in eq.~\eqref{eq:WQuartic}.

%%%%%%%%%%%%%%%%%%%%%%%%%%%%%%%%%%%%%%%%%%%%%%%%%%%%
\subsection[Gauge symmetry enhancement in moduli space]{Gauge symmetry enhancement in moduli space}
\label{sec:Accidental}

Let us analyze the ``accidental'' continuous symmetries of the superpotential eq.~\eqref{eq:Superpot} 
that appear at special points in $(T,U)$ moduli space, cf. ref.~\cite{Nilles:2020gvu} for the 
analogous discussion in the case of the $\mathbbm T^2/\Z3$ orbifold sector. We assume that the four 
twisted matter fields $\Phi^i_{(\nicefrac{-1}{2},\nicefrac{-1}{2})}$ transform identically under 
the enhanced symmetry. To identify continuous symmetries, we define a general $\U{4}$ 
transformation that leaves the K\"ahler potential of $\Phi^i_{(\nicefrac{-1}{2},\nicefrac{-1}{2})}$ 
invariant. Infinitesimally, it reads
\begin{equation}\label{eq:U4Trafo}
\Phi^i_{(\nicefrac{-1}{2},\nicefrac{-1}{2})} ~\xmapsto{\U{4}}~ \left(\Id_4 + \I\, \alpha_a\, \mathrm{T}_a \right)~ \Phi^i_{(\nicefrac{-1}{2},\nicefrac{-1}{2})}\;,
\end{equation}
where summation over $a=1,...,16$ is implied and the $4\times 4$ matrices $\mathrm{T}_a$ denote the 
16 Hermitian generators of the $\U{4}$ Lie algebra. Even though the superpotential has no continuous 
symmetries for general values of the moduli $(T,U)$, we do observe subgroups of $\U{4}$ being 
unbroken at special values of the moduli. We discuss such cases in the following.

Note that from the top-down perspective of string theory, the appearance of continuous symmetries 
is expected. As discussed in appendix~\ref{app:gaugesymmetry}, these ``accidental'' symmetries are 
actually gauge symmetries: at special points in moduli space some winding strings become massless, 
giving rise to the gauge bosons of enhanced gauge symmetries. Consequently, the enhanced symmetries 
that we uncover in this section are exact symmetries to all orders in the superpotential.

\begin{table}[t!]
\centering
\resizebox{\textwidth}{!}{
\begin{tabular}{c|c|c|c|c|c}
\toprule
point in & alignment                                  & unbroken             & basis          & enhanced & charges or\\
moduli   & of coupling                                & Lie                  & change         & gauge    & representations of\\
space    & $\hat{Y}^{(2)}_{\rep{4}_3}(\vev T,\vev U)$ & algebra $\mathrm{t}$ & $M_\mathrm{g}$ & symmetry & matter $\Phi^i_{(\nicefrac{-1}{2},\nicefrac{-1}{2}),\mathrm{g}}$\\
\midrule
\arrayrulecolor{lightgray}
$\vev T=\vev U$ & 
$\begin{pmatrix}
\hat{Y}_1(\vev T, \vev T)\\
\hat{Y}_2(\vev T, \vev T)\\
\hat{Y}_2(\vev T, \vev T)\\
\hat{Y}_4(\vev T, \vev T)
\end{pmatrix}$ & 
$\begin{pmatrix}
0&-\I&0&0\\
\I&0&0&0\\
0&0&0&-\I\\
0&0&\I&0
\end{pmatrix}$ & 
$\dfrac{1}{\sqrt{2}}\,\begin{pmatrix}1&-\I&0&0\\0&0&-\I&1\\0&0&1&-\I\\\I&-1&0&0\end{pmatrix}$ & $\U{1}$ & $\begin{array}{c}+1\\-1\\+1\\-1\end{array}$\\
\cmidrule{1-6} % ---------------------
$\vev T=\vev U=\I$ & 
$c_\I\,\begin{pmatrix}
1\\
\nicefrac{1}{\sqrt{3}}\\
\nicefrac{1}{\sqrt{3}}\\
\nicefrac{1}{3}
\end{pmatrix}$ & 
$\begin{pmatrix}
0&-\I&0&0\\
\I&0&0&0\\
0&0&0&-\I\\
0&0&\I&0
\end{pmatrix}$
& $~\dfrac{1}{2}\,~\begin{pmatrix}-1 & \I & \I & 1\\1 & -\I & \I & 1\\1 & \I & -\I & 1\\-1 & -\I & -\I & 1\end{pmatrix}$ & $\U{1}^2$ & $\begin{array}{cc}(+1,+1)\\(+1,-1)\\(-1,+1)\\(-1,-1)\end{array}$ \\[28pt] 
& $\begin{array}{l}c_\I := \hat{Y}_1(\I, \I)\\ \phantom{c_\I}\approx 0.01706\end{array}$ &
$\begin{pmatrix}
0&0&-\I&0\\
0&0&0&-\I\\
\I&0&0&0\\
0&\I&0&0
\end{pmatrix}$ 
&&&\\
\cmidrule{1-6} % ---------------------
$\vev T=\vev U=\omega$ & 
$c_\omega\,\begin{pmatrix}
1\\
-\I\\
-\I\\
-1
\end{pmatrix}$ & 
$\begin{pmatrix}
0&0&-\I&0\\
0&0&0&\I\\
\I&0&0&0\\
0&-\I&0&0
\end{pmatrix}$ %\;,\;$
& $\dfrac{1}{\sqrt{2}}\,\begin{pmatrix}1&-\I&0&0\\0&0&-\I&1\\0&0&1&-\I\\\I&-1&0&0\end{pmatrix}$ & $\SU{2}$ & $\rep{2}\oplus\rep{2}$\\[28pt]
&$\begin{array}{l}c_\omega := \hat{Y}_1(\omega, \omega)\\ \phantom{c_\omega}\approx 0.01256\end{array}$&
$\begin{pmatrix}
0&0&0&-\I\\
0&0&-\I&0\\
0&\I&0&0\\
\I&0&0&0
\end{pmatrix}$
&&&\\[28pt]
&&
$\begin{pmatrix}
0&-\I&0&0\\
\I&0&0&0\\
0&0&0&-\I\\
0&0&\I&0
\end{pmatrix}$
&&&\\
\arrayrulecolor{black}
\bottomrule
\end{tabular}}
\caption{Gauge symmetry enhancements by Lie algebra elements $\mathrm{t}$ in the case of the 
$\mathbbm T^2/\Z2$ orbifold sector at special points in moduli space, uncovered as ``accidental'' 
symmetries of the superpotential due to the alignment of couplings in flavor space. Note that the 
field basis of twisted matter fields $\Phi^i_{(\nicefrac{-1}{2},\nicefrac{-1}{2}),\mathrm{g}}$ with 
well-defined gauge charges differs from the field basis of localized twisted strings 
$\Phi^i_{(\nicefrac{-1}{2},\nicefrac{-1}{2})}$ using the basis change $M_\mathrm{g}$ in 
eq.~\eqref{eq:FieldBasisChange}.}
\label{tab:AccidentalSymmetry}
\end{table}

In the following, we briefly discuss the results for three special configurations: i) 
$\vev T=\vev U$ for generic $\vev U$, ii) $\vev T=\vev U=\omega$, and iii) $\vev T=\vev U =\I$. In 
each case, we first evaluate the superpotential eq.~\eqref{eq:Superpot} in the respective vev 
configuration by analyzing the alignment of the couplings $\hat{Y}^{(2)}_{\rep{4}_3}(\vev T,\vev U)$ 
in flavor space. Then, we identify the unbroken Lie algebra elements $\mathrm{t}:=\alpha_a\mathrm{T}_a$ 
from the transformation~\eqref{eq:U4Trafo}. Afterwards, we perform a (unitary) basis change
\begin{equation}\label{eq:FieldBasisChange}
\Phi^i_{(\nicefrac{-1}{2},\nicefrac{-1}{2}),\mathrm{g}} ~:=~ M_\mathrm{g}\, \Phi^i_{(\nicefrac{-1}{2},\nicefrac{-1}{2})}\;,
\end{equation}
such that the unbroken Lie algebra elements $\mathrm{t}_\mathrm{g}:=M_\mathrm{g}\, \mathrm{t}\, M_\mathrm{g}^{-1}$ 
are (block-)diagonalized. Finally, we identify the continuous symmetry and the charges (or 
representations) of the twisted matter fields 
$\Phi^i_{(\nicefrac{-1}{2},\nicefrac{-1}{2}),\mathrm{g}}$. The results are summarized in 
table~\ref{tab:AccidentalSymmetry}.

At $\vev T=\vev U$, there appears an enhanced $\U{1}$ symmetry. Note that the traditional flavor 
subgroup $\Z{4} \subset (D_8 \times D_8)/\Z{2}$ generated by $h_1\,h_3$ is a subgroup of this 
$\U{1}$. Hence, one can verify that the traditional flavor symmetry gets enhanced to
\begin{equation}
G_\mathrm{enhanced\ traditional} ~=~ \dfrac{\left(\U{1}\rtimes\Z{2}\right)\times D_8}{\Z{2}} \quad\mathrm{for}\quad \vev T ~=~ \vev U\;,
\end{equation}
where the $\Z2$ quotient identifies $(h_1\,h_3)^2$ from the left factor with $(h_2\,h_4)^2$ from the 
right factor.

At $\vev T = \vev U = \I$ in moduli space, there is an enhanced $\U{1}^2$ symmetry. In this case, 
the traditional flavor subgroup
\begin{equation}
\frac{\Z{4} \times \Z{4}}{\Z{2}} ~\cong~ \Z{4}\times\Z{2} ~\subset~ \frac{D_8 \times D_8}{\Z{2}}
\end{equation}
generated by the order 4 elements $h_1\,h_3$ and $h_2\,h_4$ is a subgroup of this $\U{1}^2$ 
symmetry. Hence, the traditional flavor symmetry is enhanced to
\begin{equation}
G_\mathrm{enhanced\ traditional} ~=~ \dfrac{\left(\U{1}\rtimes\Z2\right)\times\left(\U{1}\rtimes\Z2\right)}{\Z2} \quad\mathrm{for}\quad \vev T ~=~ \vev U ~=~ \I\;,
\end{equation}
where the $\Z2$ quotient identifies $(h_1\,h_3)^2$ with $(h_2\,h_4)^2$, as before.

Finally, in the case $\vev T = \vev U = \omega$, an enhanced $\SU{2}$ symmetry emerges. Note that 
the traditional flavor group $[32,49]$ can be written as
\begin{equation}
[32,49] ~\cong~ \frac{D_8 \times D_8}{\Z{2}} ~\cong~ \frac{Q_8\x Q_8}{\Z2}\;,
\end{equation}
where the first $Q_8\cong [8,4]$ is generated by the elements $h_1\,h_3$ and $h_1\,h_2\,h_4$, the 
second $Q_8$ is generated by $h_1 h_3 h_2$ and $h_1 h_3 h_4$, and the \Z2 identifies the elements 
$-\Id$ of both groups in each product. Now, it turns out that the first $Q_8$ factor is contained 
in $\SU{2}$. Then, we can write the enhanced traditional flavor symmetry at $T=U=\omega$ as 
\begin{equation}
G_\mathrm{enhanced\ traditional} ~=~  \frac{\SU{2}\x Q_8}{\Z{2}} \quad\mathrm{for}\quad \vev T ~=~ \vev U ~=~ \omega\,.
\end{equation}

\newpage

%%%%%%%%%%%%%%%%%%%%%%%%%%%%%%%%%%%%%%%%%%%%%%%%%%%%%%%%%%%%%%%%%%%%%%%%%%%%%%%%%%%%%%%%%%%%%%%%%%%%%%%%%%%%%%%%%%%%%%%%%%%%%%%%%%%%%%%%%%
\section{Conclusions and Outlook}
\label{sec:conclusions}

We performed a detailed analysis of the modular symmetries of the $\mathbbm T^2/\Z2$ orbifold which 
(among others) might be relevant for the (discrete) flavor symmetries of string compactifications 
with an elliptic fibration. The $\mathbbm T^2/\Z2$ case has two unconstrained moduli with 
$\SL{2,\Z{}}_T\times \SL{2,\Z{}}_U$ modular symmetry and allows contact with previous bottom-up 
constructions that have more than one modulus~\cite{deMedeirosVarzielas:2019cyj,King:2019vhv,King:2021fhl,Ding:2020zxw,Ding:2021iqp}. 
In the present paper, we completed the discussion of our earlier work~\cite{Baur:2020jwc} now 
including the automorphy factors of modular symmetry. This leads to an additional $R$-symmetry 
$\Z{4}^R$ (for the given modular weights of matter fields) that plays the role of a so-called 
``shaping symmetry'' and extends the discrete flavor symmetry. In more detail, the traditional 
flavor symmetry of ref.~\cite{Baur:2020jwc} is extended from [32,49] via $\Z{4}^R$ to [64,216], see 
eq.~\eqref{eq:Z2traditionalFlavorWithR}. Together with the finite modular flavor symmetry 
[144,115] (and [288,880] including \CP) discussed in section~\ref{sec:AutomorphyFactors} and 
appendix~\ref{app:TrafoOfTwistedStrings}, this leads to an eclectic flavor symmetry of order 
4608 (and order 9216 with \CP).

This picture reveals the fact that the top-down discussion of modular flavor symmetry constitutes 
an extremely restrictive scenario, which is confirmed in other top-down 
scenarios~\cite{Kobayashi:2018rad,Ohki:2020bpo,Kikuchi:2020frp,Almumin:2021fbk,Hoshiya:2021nux}.
As in the case of the bottom-up discussion, firstly the role of (otherwise freely chosen) flavons 
is played by the moduli $T$ and $U$, and secondly we arrive at a specific finite modular group, 
being $\Gamma_2^T\x\Gamma_2^U = S_3^T\x S_3^U$ for the $\mathbbm T^2/\Z2$ orbifold. In addition, we 
have to consider the restrictions from the automorphy factors with modular weights fixed from 
string theory (in contrast to the bottom-up case where these values can be chosen freely). Moreover, 
in addition to the finite modular symmetry, string theory provides a traditional flavor symmetry, 
which gives severe restrictions for K\"ahler- and superpotential of the theory (discussed in 
section~\ref{sec:EFT}). Finally, the representations of the relevant matter fields of the 
traditional and modular flavor symmetries are determined by the theory. We summarize them 
(along with the corresponding modular weights) in table~\ref{tab:Z2modularWeights}.

Compared to the earlier discussions~\cite{Baur:2019kwi,Baur:2019iai} where one modulus was frozen, 
the two-modulus case allows a full understanding of mirror symmetry (as discussed in 
section~\ref{sec:RSymmetries} and appendix~\ref{app:MirrorForStrings}), including the situation of 
matter fields whose modular weights $n_T$ and $n_U$ differ from each other, see 
eq.~\eqref{eq:RelationBetweenModularWeights}. In this case, mirror symmetry requires the presence 
of matching representations where $n_T$ and $n_U$ are interchanged.

We observe enhancements of the traditional flavor group at specific locations in moduli space. 
These unified flavor groups are discussed in section~\ref{sec:flavorUnification} and summarized in 
figure~\ref{fig:LocalFlavorGroups}. The largest group is located at $T=U=\exp(\nicefrac{\pi\I}{3})$ 
and has 2304 elements (including \CP). We also provide a detailed discussion of the tetrahedral 
$\mathbbm T^2/\Z2$ orbifold, which leads to the group $[192,1509]$ as extension of the traditional 
flavor group. It includes a ``geometrical'' $A_4$ as an $R$-symmetry, where twisted matter fields 
transform as $\rep3\oplus\rep1''$ of this $A_4^R$, as explained in section~\ref{sec:A4symmetry}.

The restrictions on K\"ahler- and superpotential are discussed in section~\ref{sec:EFT}. The 
traditional flavor symmetry is extremely powerful towards the restrictions on the K\"ahler 
potential. As in the $\mathbbm T^2/\Z{3}$ discussed earlier~\cite{Nilles:2020kgo}, the traditional 
flavor group restricts the K\"ahler potential to its trivial diagonal form~\eqref{eq:generalK}: a 
fact that seems to hold in full generality. In contrast, both symmetries are relevant for the form 
of the superpotential given in eq.~\eqref{eq:Superpot}. The traditional flavor symmetry reduces 
the 256 terms in eq.~\eqref{eq:PreSuperpot2} down to 16, and the modular flavor symmetry reduces 
the remaining 16 to 2, see eq.~\eqref{eq:Superpot2}.

A further special feature of string theory is the possible appearance of continuous gauge (flavor) 
symmetries in moduli space. At special points in moduli space, winding modes of the string can 
become massless and are candidates for the gauge bosons, as discussed in section~\ref{sec:Accidental} 
(see table~\ref{tab:AccidentalSymmetry}) and appendix~\ref{app:gaugesymmetry} (with 
table~\ref{tab:GaugeSymmetry}). These symmetries, of course, reflect themselves in the symmetries 
of the superpotential. From a bottom-up perspective they might appear as accidental symmetries, but 
from the top-down point of view they correspond to continuous gauge symmetries of string theory.

Together with our earlier discussion~\cite{Nilles:2020gvu} of the $\mathbbm T^2/\Z{K}$ orbifolds 
with $K=3,4,6$, we now have uncovered the basic properties of the flavor symmetries of 
two-dimensional orbifold compactifications for the case of up to two unconstrained moduli. One 
might expect that some of these properties will generalize from the case of toroidal orbifolds to 
more general string compactifications with an elliptic fibration. Moreover, from the string 
theory point of view, the next step would be the consideration of orbifolds with Wilson lines as 
additional moduli. This would require the embedding of $\SL{2,\Z{}}_T\times \SL{2,\Z{}}_U$ and 
mirror symmetry in the Siegel modular group $\mathrm{Sp}(4,\Z{})$, as discussed in 
ref.~\cite{Baur:2020yjl}, see also refs.~\cite{Ding:2020zxw,Ding:2021iqp}, where bottom-up model 
building based on $\mathrm{Sp}(4,\Z{})$ has been initiated.

%%%%%%%%%%%%%%%%%%%%%%%%%%%%%%%%%%%%%%%%%%%%%%%%%%%%%%%%%%%%%%%%%%%%%%%%%%%%%%%%%%%%%%%%%%%%%%%%%%%%%%%%%%%%%%%%%%%%%%%%%%%%%%%%%%%%%%%%%%
\section*{Acknowledgments}
A.B.\ and P.V. are supported by the Deutsche Forschungsgemeinschaft (SFB1258).

%%%%%%%%%%%%%%%%%%%%%%%%%%%%%%%%%%%%%%%%%%%%%%%%%%%%%%%%%%%%%%%%%%%%%%%%%%%%%%%%%%%%%%%%%%%%%%%%%%%%%%%%%%%%%%%%%%%%%%%%%%%%%%%%%%%%%%%%%%
\appendix

\section{Strings on orbifolds}
\label{app:StringsOnOrbifolds}

Geometrically, two spatial extra dimensions are compactified on a $\mathbbm T^2/\Z{K}$ orbifold by 
identifying points $y$ in $\mathbbm{R}^2$ if they are related by the orbifold 
action~\cite{Dixon:1985jw,Dixon:1986jc,Ibanez:1986tp}
\begin{equation}\label{eq:GeometricOrbifold}
y ~\sim~ \theta^k\,y + e\,n\;,
\end{equation}
where $k\in\{0,\ldots,K-1\}$ enumerates the twisted sectors and the orbifold twist $\theta$ is of 
order $K$, i.e.\ $\theta^K=\Id_2$. In addition, the column vectors $e_i$, $i\in\{1,2\}$, of the 
geometrical vielbein $e$ span the two-torus $\mathbbm T^2$ and $n=(n_1,n_2)^\mathrm{T}\in\Z{}^2$ 
are called winding numbers. We focus on the case $\theta = - \Id_2$ of order $K=2$ that generates a 
$\Z{2}$ point group. The $\mathbbm T^2/\Z2$ orbifold has four inequivalent fixed points where 
twisted strings are localized. We denote the corresponding matter fields by 
$(\phi_{(0,0)},\phi_{(1,0)},\phi_{(0,1)},\phi_{(1,1)})^{\mathrm{T}}$, where $\phi_{(n_1,n_2)}$ is 
localized at the fixed point $y_\mathrm{f}=\nicefrac{1}{2}(n_1e_1+n_2e_2)$ satisfying identically 
the fixed point condition $y_\mathrm{f}\stackrel!=\theta\,y_\mathrm{f}+e\,n$ from 
eq.~\eqref{eq:GeometricOrbifold}.

We focus here on the case without background Wilson lines. In the Narain formulation of the 
heterotic string~\cite{Narain:1985jj,Narain:1986am,Narain:1986qm}, the string coordinate in 
extra-dimensional space $y$ is split into right- and left-moving string modes $y_\mathrm{R}$ and 
$y_\mathrm{L}$, respectively. Then, we define 
\begin{equation}\label{eq:Yleftright}
Y ~:=~ \begin{pmatrix} y_\mathrm{R}\\y_\mathrm{L}\end{pmatrix}
\end{equation}
and eq.~\eqref{eq:GeometricOrbifold} is extended to
\begin{equation}\label{eq:NarainOrbifold}
Y ~\sim~ \Theta^k\,Y + E\,\hat{N}\;, \quad\mathrm{where}\quad \Theta = \begin{pmatrix} \theta_\mathrm{R}&0\\0&\theta_\mathrm{L}\end{pmatrix} \quad\mathrm{and}\quad \hat{N} ~=~ \begin{pmatrix} n\\m\end{pmatrix} ~\in~\Z{}^4\;.
\end{equation}
Here, $k\in\{0,\ldots,K-1\}$ for a Narain twist $\Theta$, with $\theta_\mathrm{R}$, 
$\theta_\mathrm{L}\in\SO{2}$, that is of order $K$, i.e.\ $\Theta^K = \Id_4$. The Narain twist 
generates the Narain point group $P_\mathrm{Narain} \cong \Z{K}$ and the orbifold action 
$Y \mapsto \Theta^kY + E\hat{N}$ defines the so-called Narain space group $S_\mathrm{Narain}$, 
\begin{equation}\label{eq:NarainBC}
Y ~\mapsto~ g\, Y ~:=~ \Theta^k\,Y + E\,\hat{N}\;, \quad\mathrm{where}\quad g ~=~ \left(\Theta^k, E\,\hat{N}\right) ~\in~ S_\mathrm{Narain}\;.
\end{equation}
Each (conjugacy class) $g\in S_\mathrm{Narain}$ defines a closed string and, therefore, we call $g$ 
the constructing element. We focus on symmetric orbifolds by setting 
$\theta_\mathrm{R}=\theta_\mathrm{L}=\theta$ and choose $\theta=-\Id_2$ for the $\mathbbm T^2/\Z2$ 
orbifold. Furthermore, $\hat{N}=(n,m)^\mathrm{T}$ contains winding numbers $n\in\Z{}^2$ and 
Kaluza--Klein numbers $m\in\Z{}^2$, so that the vectors $E\,\hat{N}$ give rise to a four-dimensional 
auxiliary lattice $\Gamma = \{ E\hat{N} ~|~ \hat{N}\in\Z{}^4\}$, called the Narain lattice. The 
$4\times 4$ matrix $E$ is called the Narain vielbein. Due to worldsheet modular invariance of the 
one-loop string partition function, $E$ has to satisfy the condition
\begin{equation}\label{eq:NarainCondition}
E^\mathrm{T} \eta\, E ~=~ \hat{\eta} ~:=~ \begin{pmatrix}0&\Id_2\\ \Id_2&0\end{pmatrix}\;, \quad\mathrm{where}\quad \eta ~:=~ \begin{pmatrix}-\Id_2&0\\ 0&\Id_2\end{pmatrix}\;.
\end{equation}
Therefore, the Narain lattice $\Gamma$ is an even, integer, self-dual lattice of signature $(2,2)$. 
In the absence of Wilson lines, the Narain vielbein $E$ can be parameterized in terms of the 
geometrical vielbein $e$, its inverse transposed $e^{-\mathrm{T}}$, the geometrical metric 
$G=e^\mathrm{T}e$ and the $B$-field background $B$,
\begin{equation}\label{eq:NarainVielbein}
E ~:=~ \begin{pmatrix}
\frac{1}{\sqrt{2\alpha'}}e^{-\mathrm{T}}(G+B) & -\sqrt{\frac{\alpha'}{2}}e^{-\mathrm{T}} \\
\frac{1}{\sqrt{2\alpha'}}e^{-\mathrm{T}}(G-B) &  \sqrt{\frac{\alpha'}{2}}e^{-\mathrm{T}} \\
\end{pmatrix}\;,
\end{equation}
see for example refs.~\cite{GrootNibbelink:2017usl,GrootNibbelink:2020dib} (where we changed the 
convention from $B$ to $-B$). Then, a two-torus compactification can be parameterized by a K\"ahler 
modulus $T$ and a complex structure modulus $U$, defined as
\begin{subequations}\label{eq:TU}
\begin{eqnarray}
T & := & \frac{1}{\alpha'} \left(B_{12} + \I\, \sqrt{\det G}\right)\;,\\
U & := & \frac{1}{G_{11}} \left(G_{12} + \I\, \sqrt{\det G}\right) ~=~ \frac{e_2}{e_1}\;.
\end{eqnarray}
\end{subequations}
In the last equation of $U$, we have taken both two-dimensional column vectors $e_i$ of the 
geometrical vielbein $e$ to be complex numbers, $e_i \in\mathbbm{C}$, so that $\nicefrac{e_2}{e_1}$ 
is defined. Note that $T$ determines the strength of the $B$-field and the area of the 
extra-dimensional two-torus, while $U$ specifies the shape of the two-torus. It is convenient to 
associate a generalized metric $\mathcal{H}$ to the Narain vielbein $E$ and express $\mathcal{H}$ 
in terms of the moduli $T$ and $U$, 
\begin{equation}
\label{eq:ExplicitGeneralizedMetric}
\mathcal{H}(T,U) ~=~ E^\mathrm{T} E ~=~ \frac{1}{\im T\im U}
 \begin{pmatrix}
   |T|^2       & |T|^2 \re U  & -\re T\re U &  \re T \\
   |T|^2 \re U & |T\,U|^2     & -|U|^2\re T &  \re T\re U\\
   -\re T\re U &-|U|^2\re T   &  |U|^2      & -\re U \\
   \re T       & \re T\re U   & -\re U      & 1
 \end{pmatrix},
\end{equation}
see for example ref.~\cite{Nilles:2020gvu}. Furthermore, we define the Narain twist in the lattice 
basis as
\begin{equation}
\hat\Theta ~:=~ E^{-1} \Theta\, E ~\in~ \mathrm{GL}(4,\Z{})\;,
\end{equation}
such that it maps the Narain lattice $\Gamma$ to itself. $\hat\Theta$ generates the Narain point 
group in the lattice basis $\hat{P}_\mathrm{Narain}$. Moreover, due to the left-right structure of 
$\hat\Theta$ given in eq.~\eqref{eq:NarainOrbifold}, $\hat\Theta$ has to satisfy the conditions
\begin{equation}\label{eq:NarainTwistConditions}
\hat\Theta^\mathrm{T} \hat\eta\, \hat\Theta ~=~ \hat\eta \quad\mathrm{and}\quad\hat\Theta^\mathrm{T} \mathcal{H}(T,U) \,\hat\Theta ~=~ \mathcal{H}(T,U)\;.
\end{equation}
In general, the condition that involves the generalized metric $\mathcal{H}(T,U)$ can stabilize $T$ 
and/or $U$. However, for the symmetric $\mathbbm T^2/\Z{2}$ orbifold under consideration, both 
moduli remain unconstrained.

Let us focus in the following on bulk strings, i.e.\ on strings that close under the identification 
eq.~\eqref{eq:NarainOrbifold} with constructing element $(\Id_4, E\hat{N})\in S_\mathrm{Narain}$. 
Then, right- and left-moving momenta $p_\mathrm{R}$ and $p_\mathrm{L}$ of a string have to be 
quantized, because the extra dimensions are compact. As the Narain lattice $\Gamma$ is self-dual, 
$p_\mathrm{R}$ and $p_\mathrm{L}$ must belong to $\Gamma$, too. Hence,
\begin{equation}\label{eq:NarainMomentum}
P ~:=~ \begin{pmatrix}
p_\mathrm{R}\\p_\mathrm{L}
\end{pmatrix} ~=~ E\,\hat{N} ~\in~ \Gamma\;.
\end{equation}
Then, one can see easily using eqs.~\eqref{eq:NarainCondition} 
and~\eqref{eq:ExplicitGeneralizedMetric} that
\begin{equation}\label{eq:pLRsqr}
-\left(p_\mathrm{R}\right)^2 + \left(p_\mathrm{L}\right)^2 ~=~ \hat{N}^\mathrm{T} \hat{\eta}\, \hat{N} ~=~ 2\,n^\mathrm{T} m
\qquad\mathrm{and}\qquad 
\left(p_\mathrm{R}\right)^2 + \left(p_\mathrm{L}\right)^2 ~=~ \hat{N}^\mathrm{T} \mathcal{H}(T,U)\, \hat{N}\;.
\end{equation}
In order to identify (massless) string states from the bulk, one has to consider the right- and 
left-moving mass equations 
\begin{subequations}\label{eq:UntwistedMassEqs}
\begin{eqnarray}
\frac{\alpha'}{2}M_\mathrm{R}^2 & = & q^2 + \left(p_\mathrm{R}\right)^2 + 2\left(N_\mathrm{R}-\frac{1}{2}\right)\;,\label{eq:RightMovingMassEquation}\\
\frac{\alpha'}{2}M_\mathrm{L}^2 & = & p^2 + \left(p_\mathrm{L}\right)^2 + 2\left(N_\mathrm{L}-1\right)\;, \label{eq:LeftMovingMassEquation}
\end{eqnarray}
\end{subequations}
where $N_\mathrm{R}\geq 0$ and $N_\mathrm{L}\geq 0$ count the number of right- and left-moving 
oscillator excitations. In addition, the mass equations~\eqref{eq:UntwistedMassEqs} are subject 
to level-matching $M_\mathrm{R}^2 = M_\mathrm{L}^2$. Note that the total mass $M^2(\hat{N};T,U)$ of 
a bulk string with winding and KK numbers $\hat{N}\in\Z{}^4$ depends on the moduli $T$ and $U$ via
\begin{equation}\label{eq:TotalMass}
M^2(\hat{N};T,U) ~:=~ \frac{\alpha'}{2}\left(M_\mathrm{R}^2 + M_\mathrm{L}^2\right) ~=~ \hat{N}^\mathrm{T} \mathcal{H}(T,U)\, \hat{N} + q^2 + p^2 + 2\left(N_\mathrm{R}+N_\mathrm{L}-\frac{3}{2}\right)\;.
\end{equation}
In eq.~\eqref{eq:RightMovingMassEquation}, $q=(q_0,q_1,q_2,q_3)$ denotes the bosonized momentum of 
the right-moving worldsheet fermions. It is called the $H$-momentum. $q$ has to be an element of 
one of the following weight lattices of $\SO{8}$: either the vector lattice $\rep{8}_\mathrm{v}$ or 
the spinor lattice $\rep{8}_\mathrm{s}$, see for example ref.~\cite{Athanasopoulos:2016aws}. The 
shortest $H$-momenta $q$ satisfy $q^2=1$, i.e.
\begin{equation}\label{eq:VectorMultiplet}
q ~\in~ \bigg\{ \left(\underline{\pm 1,0,0,0}\right)\;,\;\left(\pm\frac{1}{2},\pm\frac{1}{2},\pm\frac{1}{2},\pm\frac{1}{2}\right)\bigg\}\;.
\end{equation}
Here, in the first case ($\rep{8}_\mathrm{v}$), the underline denotes all permutations and, in the 
second case ($\rep{8}_\mathrm{s}$), the number of plus-signs must be even. The first component 
$q_0$ of $q$ defines the four-dimensional chirality. For example, $q_0 = 0$ yields a scalar. Note 
that in the four-dimensional effective quantum field theory, we use the convention that the scalar 
components of left-chiral superfields $\phi$ from the bulk are associated with string states having 
$q\in \{\left(0,\underline{+1,0,0}\right)\}$, such that string states with 
$q\in \{\left(0,\underline{-1,0,0}\right)\}$ give rise to their CPT partners.

%%%%%%%%%%%%%%%%%%%%%%%%%%%%%%%%%%%%%%%%%%%%%%%%%%%%
\subsection{The origin of modular symmetries}
\label{app:OriginOfModular}

The rotational outer automorphisms of the $(2,2)$-dimensional Narain lattice $\Gamma$ are given by 
those transformations $\Sigma$ that satisfy for all $E\hat{N}\in\Gamma$
\begin{equation}\label{eq:BulkTrafo}
\left(\Id_4, E\,\hat{N}\right) ~\mapsto~ \left(\Sigma, 0\right)^{-1}\, \left(\Id_4, E\,\hat{N}\right)\, \left(\Sigma, 0\right) ~=~ \left(\Id_4, E\,\hat{\Sigma}^{-1}\hat{N}\right) ~\stackrel{!}{=}~  \left(\Id_4, E\,\hat{N}'\right)\;,
\end{equation}
where we defined $\hat\Sigma := E^{-1}\Sigma E$ and $\hat{N}'=\hat{\Sigma}^{-1}\hat{N}\in\Z{}^4$ so 
that $E\hat{N}'\in\Gamma$. Hence, $\hat\Sigma\in \mathrm{GL}(4,\Z{})$. Furthermore, we demand that 
$\Sigma$ leaves the metric $\eta$ invariant,
\begin{equation}
\Sigma^\mathrm{T}\eta\, \Sigma  ~\stackrel{!}{=}~ \eta \qquad\Leftrightarrow\qquad \hat\Sigma^{\mathrm{T}}\hat\eta\,\hat\Sigma ~\stackrel{!}{=}~ \hat\eta\;.
\end{equation}
In other words, we demand that $\Sigma$ leaves any Narain scalar product $P^\mathrm{T}\eta P'$ for 
$P,P'\in\Gamma$ invariant. The resulting transformations $\hat\Sigma$ form a group
\begin{equation}\label{eq:O22D}
\mathrm{O}_{\hat{\eta}}(2,2,\Z{}) ~:=~ \big\langle ~\hat\Sigma~\big|~ \hat\Sigma ~\in~\mathrm{GL}(4,\Z{}) \quad\mathrm{with}\quad \hat\Sigma^\mathrm{T}\hat\eta\,\hat\Sigma = \hat\eta~\big\rangle\;,
\end{equation}
the so-called modular group of the Narain lattice $\Gamma$. It is easy to see that 
$\mathrm{O}_{\hat{\eta}}(2,2,\Z{})$ contains two factors of $\SL{2,\Z{}}$, i.e.\ we can define
\begin{equation}\label{eq:SigmaTU}
\hat{\Sigma}_{(\gamma_T,\gamma_U)} ~:=~
\left(\begin{array}{cccc}
\phantom{-}d_T\, a_U &           -d_T\, b_U &           -c_T\, b_U &           -c_T\, a_U \\
          -d_T\, c_U & \phantom{-}d_T\, d_U & \phantom{-}c_T\, d_U & \phantom{-}c_T\, c_U \\
          -b_T\, c_U & \phantom{-}b_T\, d_U & \phantom{-}a_T\, d_U & \phantom{-}a_T\, c_U \\
          -b_T\, a_U & \phantom{-}b_T\, b_U & \phantom{-}a_T\, b_U & \phantom{-}a_T\, a_U
\end{array}\right)\;.
\end{equation}
Then, $\hat{\Sigma}_{(\gamma_T,\gamma_U)}\in \mathrm{O}_{\hat{\eta}}(2,2,\Z{})$ if 
\begin{equation}
\gamma_T ~:=~ \begin{pmatrix}a_T & b_T\\ c_T & d_T \end{pmatrix} ~\in~ \SL{2,\Z{}}_T\quad\mathrm{and}\quad
\gamma_U ~:=~ \begin{pmatrix}a_U & b_U\\ c_U & d_U \end{pmatrix} ~\in~ \SL{2,\Z{}}_U\;.
\end{equation}
As a remark, $\hat{\Sigma}_{(\gamma_T,\gamma_U)}$ satisfies the property of being a representation, 
\begin{equation}\label{eq:SigmaRep}
\hat{\Sigma}_{(\gamma_T, \gamma_U)}\, \hat{\Sigma}_{(\delta_T, \delta_U)} ~=~ \hat{\Sigma}_{(\gamma_T\, \delta_T, \gamma_U\, \delta_U)}\;,
\end{equation}
for all $\gamma_T, \delta_T \in \SL{2,\Z{}}_T$ and $\gamma_U, \delta_U \in \SL{2,\Z{}}_U$. The 
generators $\mathrm{S}$ and $\mathrm{T}$ of the modular group $\SL{2,\Z{}}$ can be represented by 
the $2 \times 2$ matrices
\begin{equation}
\mathrm{S} ~=~ \begin{pmatrix}0 & 1\\ -1 & 0 \end{pmatrix} \quad\mathrm{and}\quad \mathrm{T} ~=~ \begin{pmatrix}1 & 1\\ 0 & 1 \end{pmatrix}\;,
\end{equation}
respectively. Then, we can define
\begin{equation}
\hat{K}_{\mathrm{S}} ~:=~ \hat{\Sigma}_{(\mathrm{S},\Id_2)}\quad, \quad \hat{K}_{\mathrm{T}} ~:=~ \hat{\Sigma}_{(\mathrm{T},\Id_2)} \quad\mathrm{and}\quad
\hat{C}_{\mathrm{S}} ~:=~ \hat{\Sigma}_{(\Id_2,\mathrm{S})}\quad, \quad \hat{C}_{\mathrm{T}} ~:=~ \hat{\Sigma}_{(\Id_2,\mathrm{T})}\;,
\end{equation}
where $\hat{K}_{\mathrm{S}}$ and $\hat{K}_{\mathrm{T}}$ generate the $\SL{2,\Z{}}_T$ factor 
associated with the K\"ahler modulus $T$ and $\hat{C}_{\mathrm{S}}$ and $\hat{C}_{\mathrm{T}}$ 
generate the $\SL{2,\Z{}}_U$ factor associated with the complex structure modulus $U$. The two 
remaining generators of $\mathrm{O}_{\hat{\eta}}(2,2,\Z{})$ are\footnote{Following 
ref.~\cite{Baur:2020yjl}, we have changed the conventions compared to ref.~\cite{Baur:2019iai} by 
redefining $\hat{K}_{\mathrm{S}}$, $\hat{C}_{\mathrm{S}}$ and $\hat{M}$.} 
\begin{equation}
\label{eq:GeneratorsMirrorAndCP}
\hat{M} ~:=~ \begin{pmatrix}0&0&1&0\\0&-1&0&0\\1&0&0&0\\0&0&0&-1\end{pmatrix} \quad \mathrm{and} \quad
\hat{\Sigma}_* ~:=~ \begin{pmatrix}-1&0&0&0\\0&1&0&0\\0&0&-1&0\\0&0&0&1\end{pmatrix}\;, 
\end{equation}
where one can easily show that mirror symmetry $\hat{M}$ interchanges the $\SL{2,\Z{}}$ factors,
\begin{equation}\label{eq:MirrorOnNarainSigma}
\hat{M}\,\hat{\Sigma}_{(\gamma_T,\gamma_U)}\,\hat{M}^{-1} ~=~ \hat{\Sigma}_{(\gamma_U,\gamma_T)}\;.
\end{equation}

Having identified the modular symmetries of a toroidal compactification, the modular symmetries 
of an orbifold are given by the rotational outer automorphisms of the Narain space group 
$S_\mathrm{Narain}$ that preserve the Narain metric $\eta$. They can be understood as those modular 
transformations $\hat\Sigma\in\mathrm{O}_{\hat{\eta}}(D,D,\Z{})$ (with $D=2$ in the present case) 
that are also from the normalizer of the Narain point group,
\begin{equation}
\hat\Sigma\, \hat{P}_\mathrm{Narain}\, \hat\Sigma^{-1} ~=~ \hat{P}_\mathrm{Narain}\;.
\end{equation}
Note that the Narain twist is not an outer automorphism, but an inner automorphism of 
$S_\mathrm{Narain}$. For example, for the $\mathbbm T^2/\Z2$ orbifold we have $\hat{\Theta}=-\Id_4$ 
and $\hat{P}_\mathrm{Narain}\cong\Z{2}$. Hence, $\mathrm{O}_{\hat{\eta}}(2,2,\Z{})/\Z2$ is the 
modular group of the $\mathbbm T^2/\Z2$ orbifold. However, we consider the two-dimensional 
$\mathbbm T^2/\Z2$ orbifold to be contained in a full six-dimensional orbifold. Hence, we assume 
that the underlying six-dimensional torus is factorized as 
$\mathbbm T^6=\mathbbm T^2\oplus\mathbbm T^2\oplus\mathbbm T^2$ and that the Narain twist of the 
$(6,6)$-dimensional Narain lattice takes the form 
\begin{equation}
\hat{\Theta} ~=~ \hat{\Theta}_{(2)} ~\oplus~ \hat{\Theta}_{(K_2)} ~\oplus~ \hat{\Theta}_{(K_3)}\;.
\end{equation}
Here, $\hat{\Theta}_{(K_i)}$ denotes an order $K_i$ Narain twist of the $i$-th $(2,2)$-dimensional 
Narain sublattice for $i\in\{1,2,3\}$, where $K_1=2$ and $\hat{\Theta}_{(2)} = -\Id_4$. Then, we 
can define a so-called sublattice rotation $\hat{\Theta}_{(2)}\oplus\Id_4\oplus\Id_4$ which is an 
outer automorphism of the Narain space group of the full six-dimensional orbifold. Consequently, 
the modular group in the $\mathbbm T^2/\Z2$ orbifold sector is $\mathrm{O}_{\hat{\eta}}(2,2,\Z{})$.

%%%%%%%%%%%%%%%%%%%%%%%%%%%%%%%%%%%%%%%%%%%%%%%%%%%%
\subsection{Transformation of bulk fields under modular symmetries}

In this section, we analyze the action of modular transformations from 
$\mathrm{O}_{\hat{\eta}}(2,2,\Z{})$ on those fields of the effective four-dimensional theory that 
originate from the bulk of the extra dimensions. The transformation of twisted matter 
fields will be discussed later in appendix~\ref{app:TrafoOfTwistedStrings}.

First, we discuss the moduli (i.e.\ the K\"ahler modulus $T$ and the complex structure modulus $U$, 
see eq.~\eqref{eq:TU}). According to eq.~\eqref{eq:BulkTrafo}, a modular transformation 
$\hat\Sigma\in\mathrm{O}_{\hat{\eta}}(2,2,\Z{})$ acts as
\begin{equation}
E ~\xmapsto{~\hat\Sigma~}~ E\,\hat\Sigma^{-1}
\end{equation}
on the Narain vielbein $E$. Consequently, we can use the generalized metric to compute the 
transformation of the moduli,
\begin{equation}\label{eq:TrafoOfGeneralizedMetric}
\mathcal{H}(T,U) ~\xmapsto{~\hat\Sigma~}~ \hat\Sigma^{-\mathrm{T}} \mathcal{H}(T,U) \hat\Sigma^{-1} ~=:~ \mathcal{H}(T',U')\;,
\end{equation}
This can be used to show that $\hat{\Sigma}_{(\gamma_T,\gamma_U)}$ from eq.~\eqref{eq:SigmaTU} acts 
on the moduli as
\begin{equation}
T ~\mapsto~ \frac{a_T\,T+b_T}{c_T\,T+d_T} \quad\mathrm{and}\quad U ~\mapsto~ \frac{a_U\,U+b_U}{c_U\,U+d_U}\;.
\end{equation}
Moreover, using eq.~\eqref{eq:TrafoOfGeneralizedMetric} we can confirm that the mirror 
transformation $\hat{M}$ interchanges the moduli, $T \leftrightarrow U$, while the $\CP$-like 
transformation $\hat\Sigma_*$ acts as
\begin{equation}
T ~\xmapsto{~\hat\Sigma_*}~ -\bar{T} \quad\mathrm{and}\quad U ~\xmapsto{~\hat\Sigma_*}~ -\bar{U}\;.
\end{equation}

As a remark, we can now understand the conditions~\eqref{eq:NarainTwistConditions} on a Narain 
twist as follows: A Narain twist $\hat\Theta\in\hat{P}_\mathrm{Narain}$ must be a modular 
transformation ($\hat\Theta\in\mathrm{O}_{\hat{\eta}}(2,2,\Z{})$) that leaves the moduli invariant 
(compare eq.~\eqref{eq:TrafoOfGeneralizedMetric} to eq.~\eqref{eq:NarainTwistConditions}).

Next, we consider a general (massive) bulk field $\phi_{(\hat{N})}$ labeled by its winding and KK 
numbers $\hat{N}\in\Z{}^4$ that corresponds to a closed string with boundary 
condition~\eqref{eq:NarainBC} given by the constructing element $(\Id_4, E\,\hat{N})$. Its 
total mass $M^2(\hat{N};T,U)$ is moduli dependent via $\hat{N}^\mathrm{T}\mathcal{H}(T,U)\hat{N}$, 
as shown in eq.~\eqref{eq:TotalMass}. Then, the corresponding mass terms in the superpotential read 
schematically
\begin{equation}
\mathcal{W} ~\supset~ \sum_{\hat{N} \in\Z{}^4} M^2(\hat{N};T,U)\, \left(\phi_{(\hat{N})}\right)^2\;.
\end{equation}
Under a (non-\CP-like) modular transformation $\hat\Sigma \in \mathrm{O}_{\hat{\eta}}(2,2,\Z{})$, 
moduli and bulk fields transform as
\begin{equation}
T ~\xmapsto{~\hat\Sigma~}~ T' \quad,\quad U ~\xmapsto{~\hat\Sigma~}~ U'  \quad\mathrm{and}\quad \phi_{(\hat{N})} ~\xmapsto{~\hat\Sigma~}~ \phi'_{(\hat{N})} ~=~ \pm \phi_{(\hat{N}')}\;,
\end{equation}
where we suppress the automorphy factor for $\phi'_{(\hat{N})}$. In addition, we have 
$\hat{N}'=\hat\Sigma^{-1}\hat{N}$ as shown in eq.~\eqref{eq:BulkTrafo} and the factor $\pm 1$ of 
$\phi_{(\hat{N}')}$ will be derived later in eq.~\eqref{eq:BulkVertexTrafo}. Then, due to its 
moduli-dependence, the total string mass $M^2(\hat{N};T,U)$ transforms as
\begin{equation}
M^2(\hat{N};T,U) ~\xmapsto{~\hat\Sigma~}~ M^2(\hat{N};T',U') ~=~ M^2(\hat{N}';T,U)\;,
\end{equation}
using eq.~\eqref{eq:TrafoOfGeneralizedMetric}. Thus, we obtain
\begin{equation}
M^2(\hat{N};T,U)\, \left(\phi_{(\hat{N})}\right)^2 ~\xmapsto{~\hat\Sigma~}~ M^2(\hat{N};T',U')\, \left(\phi'_{(\hat{N})}\right)^2 ~=~ M^2(\hat{N}';T,U)\, \left(\phi_{(\hat{N}')}\right)^2\;,
\end{equation}
as expected for the superpotential $\mathcal{W}$.

%%%%%%%%%%%%%%%%%%%%%%%%%%%%%%%%%%%%%%%%%%%%%%%%%%%%%%%%%%%%%%%%%%%%%%%%%%%%%%%%%%%%%%%%%%%%%%%%%%%%%%%%%%%%%%%%%%%%%%%%%%%%%%%%%%%%%%%%%%
\section[Action of mirror symmetry on strings with nT != nU]{\boldmath Action of mirror symmetry on strings with $n_T \neq n_U$\unboldmath}
\label{app:MirrorForStrings}

In the string construction of the $\mathbbm T^2/\Z2$ orbifold sector (without Wilson lines), a 
string state reads
\begin{equation}\label{eq:StringState}
|p_\mathrm{R};\, q_{\mathrm{sh}}\rangle_\mathrm{R} ~\otimes~ \prod_f \left(\tilde\alpha_f^i\right)^{N^i_f}\, \left(\tilde\alpha_f^{\bar{i}}\right)^{\bar{N}^{\bar{i}}_f}\, |p_\mathrm{L};\, p_{\mathrm{sh}}\rangle_\mathrm{L}\;,
\end{equation}
where $q_{\mathrm{sh}} := q + k v$ is the so-called shifted right-moving $H$-momentum and 
$p_{\mathrm{sh}} := p + k V$ denotes the shifted left-moving gauge momentum. In addition, the 
string state~\eqref{eq:StringState} is excited by $N^i_f \in\mathbbm{N}_0$ left-moving bosonic 
oscillators $\tilde\alpha_f^i$ of ``frequency'' $f<0$ in the complex direction 
$z^i:=y_\mathrm{L}^{2i-1} + \I\, y_\mathrm{L}^{2i}$ for $i\in\{1,2,3\}$ (i.e.\ holomorphic 
oscillators), while $\bar{N}^{\bar{i}}_f\in\mathbbm{N}_0$ counts the (independent) anti-holomorphic 
oscillators in the direction $\bar{z}^{\bar{i}}$. Moreover, $k\in\{0,1\}$ corresponds to an untwisted or 
twisted string, respectively. The right- and left-moving momenta $(p_\mathrm{R}, p_\mathrm{L})$ are 
given in eq.~\eqref{eq:NarainMomentum}, $q$ is from an $\SO{8}$ weight lattice and $p$ specifies 
the gauge charges as $p$ is from the $\E{8}\times\E{8}$ (or $\mathrm{Spin}(32)/\Z{2}$) weight 
lattice. Finally, we assume an orbifold twist $\theta$ with twist vector 
$v = (0,\nicefrac{1}{2},v_2,v_3)$, such that the $\mathbbm T^2/\Z2$ orbifold sector is in the first 
out of three complex extra dimensions, while the so-called shift vector $V$ determines the gauge 
embedding of $v$. Then, the modular weights $(n_T, n_U)$ of the string state 
eq.~\eqref{eq:StringState} are defined as~\cite{Dixon:1989fj,Ibanez:1992hc}
\begin{subequations}\label{eq:DefnTnU}
\begin{align}
\mathrm{if\ } q^1_{\mathrm{sh}} ~\in~\{0,1\}\ \     :\ \ n_T &~:=~ -q^1_{\mathrm{sh}}              &\ ,\quad n_U &~:=~ -q^1_{\mathrm{sh}}\;,\\
\mathrm{if\ } q^1_{\mathrm{sh}} ~\not\in~\{0,1\}\ \ :\ \ n_T &~:=~ -1+q^1_{\mathrm{sh}}-\Delta N^1 &\ ,\quad n_U &~:=~ -1+q^1_{\mathrm{sh}}+\Delta N^1\;,
\end{align}
\end{subequations}
where $\Delta N^i := N^i-\bar{N}^{\bar{i}}\in\mathbbm{N}_0$ is the total number of holomorphic 
minus anti-holomorphic oscillators with internal index $i=1$ or $\bar{i}=\bar{1}$ in the direction 
of the $\mathbbm T^2/\Z2$ orbifold sector, i.e.\ $N^i=\sum_f N^i_f$ and 
$\bar{N}^{\bar{i}}=\sum_f \bar{N}^{\bar{i}}_f$. Hence, 
\begin{equation}\label{eq:nTnURelation}
n_U - n_T ~=~ 0 \ \mathrm{mod}\ 2\;,
\end{equation}
and $n_T$ and $n_U$ coincide if the associated string state carries no oscillator excitations. For 
example, a massless matter field from the bulk ($k=0$) has no oscillators and 
$q\in \{\left(0,\underline{+1,0,0}\right)\}$, so that $n_T=n_U \in\{0,-1\}$, while a massless 
twisted matter field ($k=1$) without oscillators has $q^1_{\mathrm{sh}}=\frac{1}{2}$ and
$n_T=n_U=\nicefrac{-1}{2}$, see table~\ref{tab:Z2modularWeights}. In addition, there are twisted 
string states (massless or massive) that are excited by oscillators. According to 
eq.~\eqref{eq:DefnTnU}, the modular weights are increased/decreased by adding oscillator 
excitations,
\begin{subequations}
\begin{eqnarray}
\mathrm{add\ holomorphic\ oscillator:}             & n_T \mapsto n_T-1\;, &\quad n_U \mapsto n_U+1 \\
\mathrm{add\ anti\text{-}holomorphic\ oscillator:} & n_T \mapsto n_T+1\;, &\quad n_U \mapsto n_U-1\;.
\end{eqnarray}
\end{subequations}
In both cases, the resulting string states transform identically under the $\Z{2}$ orbifold 
projection (this is a special property of $\Z{2}$ orbifolds and not true for $\Z{K}$ orbifolds with 
$K \neq 2$). Hence, for each matter field $\Phi_{(n_T, n_U)}$ with $n_T\neq n_U$, there exists a 
partner with exactly the same mass and identical quantum numbers except for interchanged weights, 
i.e.\ $\Phi_{(n_T, n_U)}$ has a partner $\Phi_{(n_U, n_T)}$ if $n_T\neq n_U$, cf.\ 
table~\ref{tab:Z2modularWeights}.

Mirror symmetry interchanges holomorphic and anti-holomorphic left-moving oscillators. In order to 
see this, we rewrite $\hat{M}$ (given in eq.~\eqref{eq:GeneratorsMirrorAndCP}) into the left-right 
coordinate basis $(y_\mathrm{R}, y_\mathrm{L})$ at $T=U$ in moduli space. This results in 
\begin{equation}
M ~:=~ E\, \hat{M}\,E^{-1} ~=~ \begin{pmatrix} -1&0&0&0\\ 0&-1&0&0\\ 0&0&1&0\\ 0&0&0&-1\end{pmatrix}, \quad\mathrm{such\ that\ } \begin{pmatrix} y_\mathrm{R}^1\\y_\mathrm{R}^2\\y_\mathrm{L}^1\\y_\mathrm{L}^2\end{pmatrix}~\xmapsto{~M~}~ M\,\begin{pmatrix} y_\mathrm{R}^1\\y_\mathrm{R}^2\\y_\mathrm{L}^1\\y_\mathrm{L}^2\end{pmatrix}\;.
\end{equation}
Recall that a general transformation $\Sigma:=E\,\hat\Sigma\,E^{-1}$ acts on the coordinate $Y$ 
eq.~\eqref{eq:Yleftright} as $Y\xmapsto{~\Sigma~}\Sigma\,Y$~\cite[app. A.2]{Baur:2019iai}. 
Hence, the mirror transformation $M$ acts on the complex left-moving string coordinate $z^1$ in the 
direction of the $\mathbbm T^2/\Z2$ orbifold sector as
\begin{equation}
z^1 ~:=~ y_\mathrm{L}^1 + \I\, y_\mathrm{L}^2 ~\xmapsto{~M~}~ y_\mathrm{L}^1 - \I\, y_\mathrm{L}^2 ~=~ \bar{z}^{\bar{1}}\;.
\end{equation}
Hence, a mirror transformation interchanges holomorphic and anti-holomorphic oscillators resulting 
in eq.~\eqref{eq:MirrorOnTwistedMatter}.

%%%%%%%%%%%%%%%%%%%%%%%%%%%%%%%%%%%%%%%%%%%%%%%%%%%%%%%%%%%%%%%%%%%%%%%%%%%%%%%%%%%%%%%%%%%%%%%%%%%%%%%%%%%%%%%%%%%%%%%%%%%%%%%%%%%%%%%%%%
\section{Gauge symmetry enhancement}
\label{app:gaugesymmetry}

It is a well-known feature of string theory that at special points $(T, U)$ in moduli space, 
additional gauge symmetries arise whose gauge bosons are associated with massless winding strings. 
These massless strings become massive by moving in moduli space away from the special points. 
Hence, the enhanced gauge symmetry gets broken spontaneously by the moduli vevs. In order to 
identify the enhanced gauge symmetries, we look for additional massless strings from the orbifold 
bulk that become massless only at certain points in moduli space. We do this in two steps: first, 
we construct the massless string states on the torus $\mathbbm T^2$ and then move on to the 
$\mathbbm T^2/\Z2$ orbifold by projecting the massless torus states onto $\Z{2}$-invariant states.

In general, a massless string has to satisfy $M_\mathrm{R}^2 = M_\mathrm{L}^2 = 0$. Then, from 
eq.~\eqref{eq:RightMovingMassEquation} together with $q^2=1$ it follows that $N_\mathrm{R}=0$ and 
$p_\mathrm{R} = 0$. Hence, for $p_\mathrm{R}=0$ eqs.~\eqref{eq:NarainMomentum} and ~\eqref{eq:pLRsqr} yield
\begin{equation}\label{eq:pRZero}
p_\mathrm{L} ~=~ \sqrt{\frac{2}{\alpha'}}\,e\,n \quad\mathrm{and}\quad \left(p_\mathrm{L}\right)^2 ~=~ \hat{N}^\mathrm{T} \mathcal{H}(T,U)\, \hat{N} ~=~ 2\, n^\mathrm{T} m\;.
\end{equation}
For generic points $(T,U)$ in moduli space, eq.~\eqref{eq:pRZero} is satisfied only for 
$\hat{N}=(0^4)$, i.e.\ massless strings carry in general neither KK numbers $m$ nor winding numbers 
$n$ along the compactified dimensions. However, for special points $(T, U)$ in moduli space, 
additional massless strings can originate from specific solutions of the left-moving mass 
equation~\eqref{eq:LeftMovingMassEquation}, 
\begin{equation} 
N_\mathrm{L}~=~0\,,\qquad p ~=~ 0\,, \qquad \left(p_\mathrm{L}\right)^2 ~=~ 2 \quad\Rightarrow\quad M_\mathrm{L}^2 ~=~ 0\;.
\end{equation} 
Consequently, we can find additional massless strings if the following conditions are satisfied
\begin{equation}\label{eq:ConditionsForMasslessStrings}
\hat{N}^\mathrm{T} \mathcal{H}(T, U)\, \hat{N} ~=~ 2 \quad\mathrm{and}\quad n^\mathrm{T} m ~=~ 1 \quad\mathrm{for}\quad \hat{N} ~=~ \begin{pmatrix} n\\m \end{pmatrix} ~\in~ \Z{}^4\;.
\end{equation}
We denote the set of all solutions $\{\hat{N}\}$ of eq.~\eqref{eq:ConditionsForMasslessStrings} at 
$(T,U)$ in moduli space by $N_\mathrm{g}(T,U)\subset\Z{}^4$. 

Note that if $\hat{N}\in N_\mathrm{g}(T,U)$, also $-\hat{N}\in N_\mathrm{g}(T,U)$ is a solution to 
eq.~\eqref{eq:ConditionsForMasslessStrings}. This statement can be generalized as follows: Assume 
that there is a transformation $\hat\Sigma\in\mathrm{O}_{\hat{\eta}}(2,2,\Z{})$, such that a 
specific point $(T, U)$ in moduli space is invariant under $\hat\Sigma$ (using the transformation 
of the generalized metric given in eq.~\eqref{eq:TrafoOfGeneralizedMetric}). Now, take a massless 
string solution $\hat{N} \in N_\mathrm{g}(T,U)$. Then, define $\hat{N}' := \hat\Sigma\,\hat{N}$ 
and consider
\begin{equation}
\hat{N}'^{\mathrm{T}} \hat\eta\, \hat{N}' ~=~ \hat{N}^\mathrm{T} \hat\Sigma^\mathrm{T} \hat\eta\, \hat\Sigma\, \hat{N} ~=~ \hat{N}^\mathrm{T} \hat\eta\, \hat{N} ~=~ 2 \quad\Rightarrow\quad {n'}^\mathrm{T} m' ~=~ 1\;,
\end{equation}
and
\begin{equation}
\hat{N}'^{\mathrm{T}} \mathcal{H}(T, U)\, \hat{N}' ~=~ \hat{N}^\mathrm{T} \hat\Sigma^\mathrm{T} \mathcal{H}(T, U)\, \hat\Sigma\, \hat{N} ~=~ \hat{N}^\mathrm{T} \mathcal{H}(T, U)\, \hat{N} ~=~ 2\;.
\end{equation}
Hence, we see from eq.~\eqref{eq:ConditionsForMasslessStrings} that also $\hat{N}'\in N_\mathrm{g}(T,U)$ 
corresponds to a massless string. The set of transformations 
$\hat\Sigma\in\mathrm{O}_{\hat{\eta}}(2,2,\Z{})$ that leave the specific point $(T, U)$ 
in moduli space invariant is defined as the stabilizer subgroup $H_{(T, U)}$ of 
$\mathrm{O}_{\hat{\eta}}(2,2,\Z{})$ at $(T, U)$. 

The set of solutions $N_\mathrm{g}(T,U)$ gives rise to additional massless string states with nontrivial 
left-moving momenta $p_\mathrm{L}$,
\begin{equation}\label{eq:StringStateLadder}
|0;\, q_\mathrm{g}\rangle_\mathrm{R} ~\otimes~ |p_\mathrm{L};\, 0\rangle_\mathrm{L} \quad\mathrm{with}\quad p_\mathrm{L} ~=~ \sqrt{\frac{2}{\alpha'}}\,e\,n \quad\mathrm{for}\quad \hat{N} ~=~ \begin{pmatrix} n\\m\end{pmatrix} ~\in~ N_\mathrm{g}(T,U)\;,
\end{equation}
see eq.~\eqref{eq:StringState} (with $k=0$ for the bulk). As we are especially interested in the 
massless gauge bosons we choose $q_\mathrm{g} := (\pm 1,0,0,0)$ in eq.~\eqref{eq:VectorMultiplet}.
In addition, there are two massless gauge bosons with $p_\mathrm{R}=p_\mathrm{L}=(0^2)$, 
$p=(0^{16})$, $q_\mathrm{g} = (\pm 1,0,0,0)$ and $N_\mathrm{L}=1$. The associated string states 
read
\begin{equation}\label{eq:Cartan}
|0;\, q_\mathrm{g}\rangle_\mathrm{R} ~\otimes~ \tilde\alpha_{-1}^1|0;\, 0\rangle_\mathrm{L} \quad\mathrm{and}\quad |0;\, q_\mathrm{g}\rangle_\mathrm{R} ~\otimes~ \tilde\alpha_{-1}^{\bar{1}}|0;\, 0\rangle_\mathrm{L}
\end{equation}
where the indices $i=1$ and $\bar{i}=\bar{1}$ lie in the two-torus that will be orbifolded by the 
$\Z{2}$ action.

Note that the string states~\eqref{eq:Cartan} correspond to the Cartan generators, while the string 
states~\eqref{eq:StringStateLadder} correspond to raising operators (with $+\hat{N}\in N_\mathrm{g}(T,U)$) 
and lowering operators (with $-\hat{N}\in N_\mathrm{g}(T,U)$) of some non-Abelian, enhanced gauge 
symmetry. The root lattice of this symmetry group is spanned by the left-moving momenta 
$p_\mathrm{L}$ that correspond to the solutions $\hat{N}\in N_\mathrm{g}(T,U)$ using 
eqs.~\eqref{eq:NarainVielbein} and~\eqref{eq:NarainMomentum}. Thus, the stabilizer subgroup 
$H_{(T, U)}$ of $\mathrm{O}_{\hat{\eta}}(2,2,\Z{})$ at $(T, U)$ in moduli space gives rise to the 
rotational symmetries of the lattice spanned by $N_\mathrm{g}(T,U)$ and, as such, contains the Weyl 
group $W$ of the resulting gauge symmetry.

Under the $\mathbbm T^2/\Z2$ orbifold, the gauge bosons of the Cartan generators~\eqref{eq:Cartan} 
are projected out (since $\tilde\alpha_{-1}^1 \rightarrow -\tilde\alpha_{-1}^1$ and 
$\tilde\alpha_{-1}^{\bar{1}} \rightarrow -\tilde\alpha_{-1}^{\bar{1}}$ under the $\Z{2}$ orbifold), 
while the gauge bosons of the raising and lowering operators get combined to $\Z{2}$-invariant 
linear combinations
\begin{equation}\label{eq:Z2OrbifoldInvariantString}
|0;\, q_\mathrm{g}\rangle_\mathrm{R} ~\otimes~ |+p_\mathrm{L};\, 0\rangle_\mathrm{L} ~+~ |0;\, q_\mathrm{g}\rangle_\mathrm{R} ~\otimes~ |-p_\mathrm{L};\, 0\rangle_\mathrm{L}\;,
\end{equation}
where $\pm p_\mathrm{L}$ is given by $\pm\hat{N}\in N_\mathrm{g}(T,U)$, respectively.

We analyze three special points in moduli space:\footnote{Note that here, in contrast with
section~\ref{sec:flavorUnification}, we choose $T=U=\omega$ instead of $T=U=e^{\nicefrac{\pi\I}{3}}$. 
However, they correspond to equivalent points.} i) $T=U$, ii) $T=U=\I$ and iii) $T=U=\omega$ and 
summarize the results in table~\ref{tab:GaugeSymmetry}. Consequently, the enhanced continuous 
symmetries identified in section~\ref{sec:Accidental} are actually gauge symmetries.

\begin{table}[t!]
\centering
\resizebox{\textwidth}{!}{
\begin{tabular}{lllll}
\toprule
point in     & massless strings on $\mathbbm T^2$ & stabilizer   & gauge              & gauge  \\
moduli       & $N_\mathrm{g}(T,U)$                & $H_{(T, U)}$ & symmetry           & symmetry\\
space        &                                    &              & for $\mathbbm T^2$ & for $\mathbbm T^2/\Z2$ \\
\midrule
\arrayrulecolor{lightgray}
$T=U$        & $\begin{array}{l}(1, 0, 1, 0)^\mathrm{T},\, (-1, 0,-1, 0)^\mathrm{T}\end{array}$ & $\begin{array}{c}\langle-\Id_4, \hat{M}\rangle\\\cong~\Z{2}\times\Z{2}\end{array}$ & $\SU{2}\times\U{1}$ & $\U{1}$\\
\cmidrule{1-5}
$T=U=\I$     & $\begin{array}{l}( 1, 0, 1, 0)^\mathrm{T},\, (-1, 0,-1, 0)^\mathrm{T},\,\\ ( 0, 1, 0, 1)^\mathrm{T},\,(0,-1, 0,-1)^\mathrm{T}\end{array}$ & $\begin{array}{c}\langle\hat{M},\hat{C}_\mathrm{S},\hat\Sigma_*\rangle\\\cong[32,27]\cong\Z2\ltimes\Z2^4\end{array}$ & $\SU{2}^2$ & $\U{1}^2$\\
\cmidrule{1-5}
$T=U=\omega$ & $\begin{array}{l}( 1, 0, 1, 0)^\mathrm{T},\, (-1, 0,-1,0)^\mathrm{T},\,\\ (0, 1,-1, 1)^\mathrm{T},\, (0,-1, 1,-1)^\mathrm{T},\,\\ ( 1, 1, 0, 1)^\mathrm{T},\, (-1,-1, 0,-1)^\mathrm{T}\end{array}$ & $\begin{array}{l}\langle-\Id_4,\hat{M},\hat{C}_\mathrm{S}\hat{C}_\mathrm{T},\hat\Sigma_*\hat{C}_\mathrm{T}\hat{K}_\mathrm{T}\rangle\\ \cong[72,46]\\ \cong S_3 \x S_3\x\Z{2}\end{array}$ & $\SU{3}$ & $\SU{2}$\\
\arrayrulecolor{black}
\bottomrule
\end{tabular}}
\caption{Gauge symmetry enhancements of the $\mathbbm T^2/\Z2$ orbifold sector at special points in 
moduli space. We use $\omega:=e^{\nicefrac{2\pi\I}{3}}$.}
\label{tab:GaugeSymmetry}
\end{table}

%%%%%%%%%%%%%%%%%%%%%%%%%%%%%%%%%%%%%%%%%%%%%%%%%%%%%%%%%%%%%%%%%%%%%%%%%%%%%%%%%%%%%%%%%%%%%%%%%%%%%%%%%%%%%%%%%%%%%%%%%%%%%%%%%%%%%%%%%%
\section[Vertex operators of the Z2 Narain orbifold]{\boldmath Vertex operators of the \Z2 Narain orbifold \unboldmath}
\label{app:TrafoOfTwistedStrings}

The spectrum of the $\mathbbm{T}^{2}/\Z2$ orbifold sector includes untwisted strings, associated 
with constructing elements $(\Id,\hat{N}) \in \hat{S}_{\mathrm{Narain}}$, and twisted strings 
constructed by elements $(\hat{\Theta},\hat{N})\in\hat{S}_{\mathrm{Narain}}$. In this appendix, we 
study how the symmetries of the theory act on these strings by inspecting the transformations of 
their corresponding vertex operators.

%%%%%%%%%%%%%%%%%%%%%%%%%%%%%%%%%%%%%%%%
\subsection{Untwisted vertex operators}
The zero-mode vertex operator corresponding to a bosonic string on a toroidal background with 
winding and Kaluza-Klein numbers $\hat{N}=(n,m)^\mathrm{T}\in\Z{}^4$ is given 
by~\cite[eq.~(3.41)]{Freidel:2017wst}
\begin{equation}\label{eq:VOclosedString}
V(\hat{N}) ~=~ \e^{-\nicefrac{\pi\I}{4}\, \hat{N}^\mathrm{T}\hat{\eta}\hat{N}} \e^{2\pi\I\, \hat{N}^\mathrm{T}\hat{\eta}\,\rep{Y}}\;,
\end{equation}
where the string coordinate operator $\rep{Y}$ results from promoting $E^{-1}Y$ to an operator (see 
eq.~\eqref{eq:Yleftright}). $\rep{Y}$ satisfies the commutation relations\footnote{The matrix 
components of the commutators~\eqref{eq:CanonicalCommutatorsClosedString} are such that 
$\com{\rep{Y}}{\rep{Y}^\mathrm{T}}_{IJ}=\com{\rep{Y}_I}{\rep{Y}_J}=\I \hat{\omega}_{IJ}/ 4\pi$.} 
(derived from the action of the sigma model)
\begin{equation}
\label{eq:CanonicalCommutatorsClosedString}
\left[ \rep{Y} , \rep{Y}^\mathrm{T}\right] 
~=~\dfrac{\I}{4\pi}
\; \hat{\omega},
\qquad\text{where}~~
\hat{\omega}~=~\begin{pmatrix}0& \Id_2 \\ -\Id_2 & 0\end{pmatrix}
\end{equation}
is the symplectic structure in the Narain basis. The nonzero value of the 
commutator~\eqref{eq:CanonicalCommutatorsClosedString} is a result of intrinsic non-commutative 
effects of closed strings~\cite{Freidel:2017wst}. The zero-mode vertex 
operators~\eqref{eq:VOclosedString} in combination with the 
commutator~\eqref{eq:CanonicalCommutatorsClosedString} are subject to the so-called Weyl 
quantization relation
\begin{equation}\label{eq:WeylConditionOriginal}
V(\hat{N}_1)\,V(\hat{N}_2)~=~\e^{\nicefrac{\pi\I}{2}\,\hat{N}_1^\mathrm{T} \left( \hat{\eta}+\hat{\omega}\right)\,\hat{N}_2}\,V(\hat{N}_1+\hat{N}_2)\;.
\end{equation}
According to ref.~\cite{Sakamoto:1989ig}, this relation is instrumental to evaluate the time 
ordering of operators as required in the computation of scattering amplitudes. The quantization 
relation~\eqref{eq:WeylConditionOriginal} must hold independently of whether the vertex operators 
have been affected by modular transformations. As we shall shortly see (cf.\ 
eqs.~\eqref{eq:PhasesGenerators}), this helps determine the phases required for the modular 
generators to act consistently on twisted vertex operators~\cite{Erler:1991an}.

Using eq.~\eqref{eq:VOclosedString}, one finds that the \Z2 orbifold-invariant untwisted vertex 
operators are given by
\begin{equation}
\label{eq:OrbInvVO}
  V(\hat{N})^{\mathrm{orb.}} ~:=~ \frac{1}{\sqrt{2}} \left(V(\hat{N}) + V(-\hat{N})\right)\;.
\end{equation}
These untwisted vertex operators can be arranged into 16 classes (corresponding to the 16 
conjugacy classes $[(\Id_4,\hat{N}^0)]$),
\begin{equation}
\label{eq:ClassesOfUntwistedStrings}
V^{\hat{N}^0} ~:=~ V^{(n^0,m^0)^\mathrm{T}} 
~=~ \sum_{\hat{N}\in \Z{}^{4}}\e^{\pi\I\, \hat{N}^\mathrm{T}\hat{\eta}\hat{N}^0}~ V(\hat{N}^0+2\hat{N})^{\mathrm{orb.}}\;.
\end{equation}
They are characterized by a representative winding number $n^0$ and a representative KK number 
$m^0$, also called charges and collected in $\hat N^0:=(n^0,m^0)^\mathrm{T}$, with 
$n^0,m^0 \in \{(0,0)^\mathrm{T},(0,1)^\mathrm{T},(1,0)^\mathrm{T},(1,1)^\mathrm{T}\}$. 
Note that the phases in eq.~\eqref{eq:ClassesOfUntwistedStrings} let one establish a relation 
between the representative vertex operator $V^{\hat{N}^0}$ of a class and any other member of the 
class through
\begin{equation}\label{eq:FlippingPhase}
V^{\hat{N}^0+2\hat{N}'} ~=~ \e^{-\pi\I\, \hat{N}^{\prime\mathrm{T}} \hat{\eta}\, \hat{N}^0} V^{\hat{N}^0}\;,
\qquad\text{with}\quad \hat{N}'\in\Z{}^4\;.
\end{equation}
The phases further ensure that each class holds uniform properties, particularly in couplings and 
under outer automorphisms.

Let us now focus on the transformations of untwisted vertex operators $V^{\hat N^0}$ under the 
action of general outer automorphisms of the Narain orbifold space group,
\begin{equation}
\label{eq:Outers}
\mathrm{Out}(\hat{S}_\mathrm{Narain})~=~ 
\left\langle (\hat{\Sigma},0), \hat{h}_i:=(\Id_4,T_i) ~\big|~ 
\hat{\Sigma}\in \mathrm{O}_{\hat{\eta}}(2,2,\Z{})\,,~ T_i\in\tfrac12\Z{}^4\right\rangle\;.
\end{equation}
As discussed in appendix~\ref{app:OriginOfModular}, the rotational outer automorphisms defined by 
$\hat\Sigma\in \mathrm O_{\hat\eta}(2,2,\Z{})=\langle\hat{K}_\mathrm{S},\hat{K}_\mathrm{T},\hat{C}_\mathrm{S},\hat{C}_\mathrm{T},\hat{M},\hat\Sigma_*\rangle$ 
can be interpreted as modular transformations. In addition, the translational outer automorphisms 
$\hat{h}_i$, $i=1,2,3,4$, are defined by the shift vectors $T_i$ whose components in the lattice 
basis are $T_i{}^j=\frac12\delta_i{}^j$, cf.\ ref.~\cite[app.~A]{Baur:2020jwc}.

To determine the transformation of $V^{\hat N^0}$ under a translation $\hat h_i$, we observe that
\begin{equation}\label{eq:VOtranslation}
V(\hat{N}) ~\xmapsto{~\hat{h}_i~}~ \e^{2\pi\I\, T_i^\mathrm{T} \hat{\eta}\, \hat{N}} \; V(\hat{N})\;.
\end{equation}
This implies that $V(\hat{N})$ acquires a \Z2 phase, which is identical for $V(-\hat{N})$. 
Consequently, the orbifold invariant vertex operator~\eqref{eq:OrbInvVO} inherits the same phase. 
It thus follows that the untwisted vertex operator class $V^{\hat N^0}$ gets a \Z2 phase too, 
\begin{equation}\label{eq:ClassTranslation}
V^{\hat{N}^0} ~\xmapsto{~\hat{h}_i~}~ \e^{2\pi\I\, T_i^\mathrm{T} \hat{\eta}\, \hat{N}^0} \,V^{\hat{N}^0}\,.
\end{equation}

Under a rotational outer automorphism $\hat\Sigma$, $\hat N$ transforms to $\hat\Sigma^{-1}\hat N$. 
Then, we expect that the vertex operator $V(\hat N)$ transforms according to
\begin{equation}\label{eq:VOrotationAnsatz}
V(\hat{N}) ~\xmapsto{~\hat{\Sigma}~}~ \varphi_{\hat{\Sigma}}(\hat{N}) \, V(\hat{\Sigma}^{-1}\hat{N})\;.
\end{equation}
Here, we propose, due to the nontrivial commutation 
relations~\eqref{eq:CanonicalCommutatorsClosedString}, a phase $\varphi_{\hat{\Sigma}}(\hat{N})$ 
that is given by the ansatz
\begin{equation}\label{eq:PhiSigmaAnsatz}
\varphi_{\hat{\Sigma}}(\hat{N}) ~=~ \e^{\pi\I\, \hat{N}^\mathrm{T} A_{\hat{\Sigma}} \hat{N} 
                                 + \pi\I\,C_{\hat{\Sigma}}^\mathrm{T}\hat{N}}\;.
\end{equation}
The $4\x4$ matrix $A_{\hat{\Sigma}}$ (with only half-integral off-diagonal entries) and the 
vector $C_{\hat{\Sigma}}\in\Z{}^4$ will be determined next. Note that, with these conditions, 
$\varphi_{\hat{\Sigma}}(\hat{N})$ can only be a \Z2 phase. By demanding that the Weyl quantization 
relation~\eqref{eq:WeylConditionOriginal} be preserved by $\hat\Sigma$ and using the abbreviation 
$\hat{\mu}=\frac{1}{2}(\hat{\eta}+\hat{\omega})$, one arrives at 
\begin{equation}\label{eq:PhaseAmatrixForm}
A_{\hat{\Sigma}} ~=~ \dfrac{1}{2}\left( \hat{\mu} - \hat{\Sigma}^{-T} \hat{\mu}\,\hat{\Sigma}^{-1}\right) \mod 2\;.
\end{equation}
In contrast, $C_{\hat{\Sigma}}$ cannot be constrained by the quantization condition. However, the 
effect of $C_{\hat{\Sigma}}$ is equivalent to the one of a translation $\hat{h}_i$ in the Narain 
lattice, given in eq.~\eqref{eq:ClassTranslation}. These translations generate the traditional 
flavor symmetry, which is unbroken independently of the moduli. Therefore, the traditional flavor 
symmetry allows for a free choice of the vector $C_{\hat{\Sigma}}$. We choose 
$C_{\hat{K}_\mathrm{T}}=(1,1,0,0)^\mathrm{T}$, $C_{\hat{M}}=(1,1,1,1)^\mathrm{T}$, and 
$C_{\hat\Sigma}=0$ for $\hat\Sigma\notin\{\hat{K}_\mathrm{T},\hat{M}\}$ such that the 
transformations~\eqref{eq:VOrotationAnsatz} generate only the finite modular group with a minimal 
amount of traditional flavor transformations. Thus, we are led to the phases 
\begin{subequations}
\label{eq:PhasesGenerators}
\begin{align}
  \varphi_{\hat{K}_{\mathrm{S}}}(\hat{N}) & = \e^{\pi\I\, \left( m_1n_1 +m_2n_2\right) }\;,\qquad 
& \varphi_{\hat{K}_{\mathrm{T}}}(\hat{N}) & = \e^{\pi\I\, \left( n_1n_2 + n_1 + n_2\right) }\;,\\
  \varphi_{\hat{C}_{\mathrm{S}}}(\hat{N}) & = 1\;, 
& \varphi_{\hat{C}_{\mathrm{T}}}(\hat{N}) & = 1\;,\\
  \varphi_{\hat{M}}(\hat{N})              & = \e^{\pi\I\, \left( m_1n_1 + n_1 + n_2 + m_1 + m_2\right) }\;,
& &
\end{align}
\end{subequations}
where $\hat N=(n,m)^\mathrm{T}=(n_1,n_2,m_1,m_2)^\mathrm{T}$. We do not determine the phase 
corresponding to the \CP-like generator $\hat\Sigma_*$ by the previous procedure because the result 
would be trivial. Instead, we fix its value by demanding that the transformations of the untwisted 
vertex operators $V^{\hat N^0}$ be compatible with the transformation 
$\phi_{n}\xmapsto{\hat\Sigma_*}\bar\phi_{n}$ of twisted states in the operator product expansions 
(OPEs) of twisted fields discussed in section~\ref{subsec:OPE}. We then find 
\begin{equation}
\label{eq:PhaseSigma*}
  \varphi_{\hat{\Sigma}_{*}}(\hat{N})~:=~ \e^{\pi\I\, \left( m_1n_1 + m_2n_2\right) }\;.
\end{equation}
Noticing that the \Z2 phases~\eqref{eq:PhasesGenerators} and~\eqref{eq:PhaseSigma*}
coincide for $\hat N$ and $-\hat N$, we find that the orbifold invariant vertex 
operators $V(\hat N)^{\mathrm{orb.}}$ transform just as
\begin{equation}\label{eq:BulkVertexTrafo}
V(\hat{N})^{\mathrm{orb.}} ~\xmapsto{~\hat\Sigma~}~ \varphi_{\hat{\Sigma}}(\hat{N})\ V(\hat\Sigma^{-1} \hat{N})^{\mathrm{orb.}}\;.
\end{equation}
Therefore, a rotational outer automorphism $\hat\Sigma$ acts as
\begin{equation}\label{eq:ClassRotation}
V^{\hat{N}^{0}} ~\xmapsto{~\hat{\Sigma}~}~ \varphi_{\hat{\Sigma}}(\hat{N}^{0}) \, V^{\hat{\Sigma}^{-1}\hat{N}^{0}}\;.
\end{equation}
Since $\hat{\Sigma}^{-1}\hat{N}^{0}$ does not always take the form $(n^{0'},m^{0'})^\mathrm{T}$
with $n^{0'},m^{0'}\in\{(0,0)^\mathrm{T},(0,1)^\mathrm{T},(1,0)^\mathrm{T},(1,1)^\mathrm{T}\}$, 
expressing $V^{\hat{\Sigma}^{-1}\hat{N}^{0}}$ in terms of the vertex operator 
classes~\eqref{eq:ClassesOfUntwistedStrings} through eq.~\eqref{eq:FlippingPhase}
introduces an extra \Z2 phase in the transformation~\eqref{eq:ClassRotation}.

%%%%%%%%%%%%%%%%%%%%%%%%%%%%%%%%%%%%%%%%
\subsection{Operator product expansions of twisted vertex operators}
\label{subsec:OPE}

Even though vertex operators of twisted string states are more involved than untwisted vertex 
operators, OPEs of two twisted states in two-dimensional orbifolds are known. 
Let us consider the twisted vertex operators $\phi_{n^a}$ and $\phi_{n^b}$ of twisted strings 
localized at orbifold fixed points given by the winding numbers $n^a$, 
$n^b\in\{(0,0)^\mathrm{T},(0,1)^\mathrm{T},(1,0)^\mathrm{T},(1,1)^\mathrm{T}\}$. Up to a constant 
overall factor, they satisfy the OPE~\cite{Lauer:1990tm}
\begin{equation}\label{eq:OPEgeneral}
\bar{\phi}_{n^a}\,\phi_{n^b} ~=~ \sum_{m^0} C(n^a,n^b\,;n^0,m^0)\,V^{\hat N^0}\;,
\qquad \text{where} \quad n^0 ~=~ \left( n^b-n^a\right)\mod2\;.
\end{equation}
$V^{\hat N^0}$ with $\hat{N}^0 =(n^0,m^0)^\mathrm{T}$ are the untwisted vertex operator 
classes~\eqref{eq:ClassesOfUntwistedStrings}, the bar on $\phi_{n^a}$ denotes conjugation, and 
$C(n^a,n^b\,;n^0,m^0)\in\mathbbm{C}$ are known as coupling constants. For the 
$\mathbbm{T}^{2}/\Z2$ orbifold, they read
\begin{equation}
  C(n^a,n^b\,;n^0,m^0) ~=~ \e^{\pi\I\,\left( n^b\right)^\mathrm{T}m^0}~\e^{\nicefrac{\pi\I}{2}\,\left( n^0\right)^\mathrm{T}m^0}\,.
\end{equation}

By inverting eq.~\eqref{eq:OPEgeneral}, we can express the classes of untwisted string states 
$V^{\hat N^0}$ in terms of combinations of different OPEs of twisted states $\bar{\phi}_{n^a}\phi_{n^b}$. 
Explicitly, one finds
\begin{subequations}\label{eq:TwistedOPEReverse}
	\begin{eqnarray}
	V^{(0,0,0,0)^\mathrm{T}} &=\, \phantom{-\I}\, \bar{\phi}_{(0,0)}\phi_{(0,0)}\, + \phantom{\I}\, \bar{\phi}_{(1,0)}\phi_{(1,0)}\, + \phantom{\I}\, \bar{\phi}_{(0,1)}\phi_{(0,1)}\, + \phantom{\I}\, \bar{\phi}_{(1,1)}\phi_{(1,1)}\,,\\
	V^{(0,0,1,0)^\mathrm{T}} &=\, \phantom{-\I}\, \bar{\phi}_{(0,0)}\phi_{(0,0)}\, - \phantom{\I}\, \bar{\phi}_{(1,0)}\phi_{(1,0)}\, + \phantom{\I}\, \bar{\phi}_{(0,1)}\phi_{(0,1)}\, - \phantom{\I}\, \bar{\phi}_{(1,1)}\phi_{(1,1)}\,,\\
	V^{(0,0,0,1)^\mathrm{T}} &=\, \phantom{-\I}\, \bar{\phi}_{(0,0)}\phi_{(0,0)}\, + \phantom{\I}\, \bar{\phi}_{(1,0)}\phi_{(1,0)}\, - \phantom{\I}\, \bar{\phi}_{(0,1)}\phi_{(0,1)}\, - \phantom{\I}\, \bar{\phi}_{(1,1)}\phi_{(1,1)}\,,\\
	V^{(0,0,1,1)^\mathrm{T}} &=\, \phantom{-\I}\, \bar{\phi}_{(0,0)}\phi_{(0,0)}\, - \phantom{\I}\, \bar{\phi}_{(1,0)}\phi_{(1,0)}\, - \phantom{\I}\, \bar{\phi}_{(0,1)}\phi_{(0,1)}\, + \phantom{\I}\, \bar{\phi}_{(1,1)}\phi_{(1,1)}\,,%\\[6pt]
	\end{eqnarray}
	\begin{eqnarray}
	V^{(1,0,0,0)^\mathrm{T}} &=\, \phantom{-\I}\, \bar{\phi}_{(0,0)}\phi_{(1,0)}\, + \phantom{\I}\, \bar{\phi}_{(1,0)}\phi_{(0,0)}\, + \phantom{\I}\, \bar{\phi}_{(0,1)}\phi_{(1,1)}\, + \phantom{\I}\, \bar{\phi}_{(1,1)}\phi_{(0,1)}\,,\\
	V^{(1,0,1,0)^\mathrm{T}} &=\, \phantom{-}\I\, \bar{\phi}_{(0,0)}\phi_{(1,0)}\, - \I\, \bar{\phi}_{(1,0)}\phi_{(0,0)}\, + \I\, \bar{\phi}_{(0,1)}\phi_{(1,1)}\, - \I\, \bar{\phi}_{(1,1)}\phi_{(0,1)}\,,\\
	V^{(1,0,0,1)^\mathrm{T}} &=\, \phantom{-\I}\, \bar{\phi}_{(0,0)}\phi_{(1,0)}\, + \phantom{\I}\, \bar{\phi}_{(1,0)}\phi_{(0,0)}\, - \phantom{\I}\, \bar{\phi}_{(0,1)}\phi_{(1,1)}\, - \phantom{\I}\, \bar{\phi}_{(1,1)}\phi_{(0,1)}\,,\\
	V^{(1,0,1,1)^\mathrm{T}} &=\, \phantom{-}\I\, \bar{\phi}_{(0,0)}\phi_{(1,0)}\, - \I\, \bar{\phi}_{(1,0)}\phi_{(0,0)}\, - \I\, \bar{\phi}_{(0,1)}\phi_{(1,1)}\, + \I\, \bar{\phi}_{(1,1)}\phi_{(0,1)}\,,%\\[6pt]
	\end{eqnarray}
	\begin{eqnarray}
	V^{(0,1,0,0)^\mathrm{T}} &=\, \phantom{-\I}\, \bar{\phi}_{(0,0)}\phi_{(0,1)}\, + \phantom{\I}\, \bar{\phi}_{(1,0)}\phi_{(1,1)}\, + \phantom{\I}\, \bar{\phi}_{(0,1)}\phi_{(0,0)}\, + \phantom{\I}\, \bar{\phi}_{(1,1)}\phi_{(1,0)}\,,\\
	V^{(0,1,1,0)^\mathrm{T}} &=\, \phantom{-\I}\, \bar{\phi}_{(0,0)}\phi_{(0,1)}\, - \phantom{\I}\, \bar{\phi}_{(1,0)}\phi_{(1,1)}\, + \phantom{\I}\, \bar{\phi}_{(0,1)}\phi_{(0,0)}\, - \phantom{\I}\, \bar{\phi}_{(1,1)}\phi_{(1,0)}\,,\\
	V^{(0,1,0,1)^\mathrm{T}} &=\, \phantom{-}\I\, \bar{\phi}_{(0,0)}\phi_{(0,1)}\, + \I\, \bar{\phi}_{(1,0)}\phi_{(1,1)}\, - \I\, \bar{\phi}_{(0,1)}\phi_{(0,0)}\, - \I\, \bar{\phi}_{(1,1)}\phi_{(1,0)}\,,\\
	V^{(0,1,1,1)^\mathrm{T}} &=\, \phantom{-}\I\, \bar{\phi}_{(0,0)}\phi_{(0,1)}\, - \I\, \bar{\phi}_{(1,0)}\phi_{(1,1)}\, - \I\, \bar{\phi}_{(0,1)}\phi_{(0,0)}\, + \I\, \bar{\phi}_{(1,1)}\phi_{(1,0)}\,,%\\[6pt]
	\end{eqnarray}
	\begin{eqnarray}
	V^{(1,1,0,0)^\mathrm{T}} &=\, \phantom{-\I}\, \bar{\phi}_{(0,0)}\phi_{(1,1)}\, + \phantom{\I}\, \bar{\phi}_{(1,0)}\phi_{(0,1)}\, + \phantom{\I}\, \bar{\phi}_{(0,1)}\phi_{(1,0)}\, + \phantom{\I}\, \bar{\phi}_{(1,1)}\phi_{(0,0)}\,,\\
	V^{(1,1,1,0)^\mathrm{T}} &=\, \phantom{-}\I\, \bar{\phi}_{(0,0)}\phi_{(1,1)}\, - \I\, \bar{\phi}_{(1,0)}\phi_{(0,1)}\, + \I\, \bar{\phi}_{(0,1)}\phi_{(1,0)}\, - \I\, \bar{\phi}_{(1,1)}\phi_{(0,0)}\,,\\
	V^{(1,1,0,1)^\mathrm{T}} &=\, \phantom{-}\I\, \bar{\phi}_{(0,0)}\phi_{(1,1)}\, + \I\, \bar{\phi}_{(1,0)}\phi_{(0,1)}\, - \I\, \bar{\phi}_{(0,1)}\phi_{(1,0)}\, - \I\, \bar{\phi}_{(1,1)}\phi_{(0,0)}\,,\\
	V^{(1,1,1,1)^\mathrm{T}} &=\, -\phantom{\I}\, \bar{\phi}_{(0,0)}\phi_{(1,1)}\, + \phantom{\I}\, \bar{\phi}_{(1,0)}\phi_{(0,1)}\, + \phantom{\I}\, \bar{\phi}_{(0,1)}\phi_{(1,0)}\, - \phantom{\I}\, \bar{\phi}_{(1,1)}\phi_{(0,0)}\,.
	\end{eqnarray}\label{eq:OPEs}%
\end{subequations}
These expressions together with the transformation properties of untwisted operators, 
eqs.~\eqref{eq:ClassTranslation} and~\eqref{eq:ClassRotation}, can lead to the 
corresponding transformations of the twisted vertex operators $\phi_{n}$, as we now discuss.

%%%%%%%%%%%%%%%%%%%%%%%%%%%%%%%%%%%%%%%%
\subsection{Transformations of twisted vertex operators}

With the help of the explicit relations~\eqref{eq:OPEs} between the OPEs of twisted string states 
and the untwisted states, and the transformations~\eqref{eq:ClassTranslation} 
and~\eqref{eq:ClassRotation} of the latter, we can deduce the action of 
$\mathrm{Out}(\hat{S}_\mathrm{Narain})$ on $\bar{\phi}_{n^a}\,\phi_{n^b}$. One can then infer the 
action of those transformations on the single twisted operators, arranged in a twisted multiplet 
$(\phi_{(0,0)},\phi_{(1,0)},\phi_{(0,1)},\phi_{(1,1)})^\mathrm{T}$. Note that, since no oscillator 
excitation is present in these twisted string states, this multiplet must correspond to the 
components of the twisted matter field $\Phi_{(\nicefrac{-1}{2},\nicefrac{-1}{2})}$, i.e.\ with 
$n_T=n_U=\nicefrac{-1}{2}$. In these terms, the transformations of twisted states can be encoded in 
transformation matrices $\rho_{\rep r}(\hat\Sigma)$ or $\rho_{\rep r}(h_i)$, which denote 
$r$-dimensional representations of the outer automorphisms $\hat\Sigma$ and $\hat h_i$. Our goal 
here is to present those transformation matrices.

%%%%%%%%%%%%%%%%%%%%%%%%%%%%%%%%%%%%%%%%
\subsubsection{Traditional flavor group}
\label{sec:appDTraditionalGroup}

Let us first inspect the action of the translational outer automorphisms $\hat h_i$, defined in 
eq.~\eqref{eq:Outers}. Applying the transformations~\eqref{eq:ClassTranslation} on the untwisted 
operators~\eqref{eq:TwistedOPEReverse} and interpreting for the twisted operators that build 
$\Phi_{(-\nicefrac{1}{2},-\nicefrac{1}{2})}$, we find that the twisted multiplet transforms under a 
translational outer automorphism as
\begin{equation}
\Phi_{(\nicefrac{-1}{2},\nicefrac{-1}{2})} ~\xmapsto{~\hat h_i~}~ \rho_{\rep{4}}(h_i)\,\Phi_{(\nicefrac{-1}{2},\nicefrac{-1}{2})}\;,
\end{equation}
where the representation matrices are given by
\begin{subequations}\label{eq:TraditionalRhoOfTwistedStrings}
   \begin{align}
	\rho_{\rep{4}}(h_1) &= \begin{pmatrix} 0&1&0&0\\ 1&0&0&0\\ 0&0&0&1\\ 0&0&1&0\end{pmatrix}\;, & 
	\rho_{\rep{4}}(h_2) &= \begin{pmatrix} 0&0&1&0\\ 0&0&0&1\\ 1&0&0&0\\ 0&1&0&0\end{pmatrix}\;,\label{eq:h1andh2}\\
	\rho_{\rep{4}}(h_3) &= \begin{pmatrix} 1&0&0&0\\0&-1&0&0\\0&0& 1&0\\0&0&0&-1\end{pmatrix}\;, &
	\rho_{\rep{4}}(h_4) &= \begin{pmatrix} 1&0&0&0\\0& 1&0&0\\0&0&-1&0\\0&0&0&-1\end{pmatrix}\;.\label{eq:h3andh4}
   \end{align}
\end{subequations}
They generate the traditional flavor group
\begin{equation}\label{eq:TFG}
\left(D_8 \times D_8\right)/\Z{2} ~\cong~ [32,49]\;.
\end{equation}
The irreducible representations of this group are: one four-dimensional representation $\rep{4}$ 
and 16 one-dimensional representations $\rep{1}_{\alpha\beta\gamma\delta}$, such that
\begin{equation}
\label{eq:traditionalSinglets}
\rep{r}~=~\rep{1}_{\alpha\beta\gamma\delta} \quad\mathrm{defined\ by}\quad \rho_{\rep{r}}(h_1) ~=~ \alpha \;,\; \rho_{\rep{r}}(h_2) ~=~ \beta \;,\; \rho_{\rep{r}}(h_3) ~=~ \gamma \;,\; \rho_{\rep{r}}(h_4) ~=~ \delta\;,
\end{equation}
where $\alpha, \beta, \gamma, \delta \in\{-1,+1\}$ and $\rep{1}_0:=\rep{1}_{++++}$ is the trivial 
singlet.

The traditional flavor group eq.~\eqref{eq:TFG} contains the $\Z{2}\times\Z{2}\times\Z{2}$ space 
and point group (PG) selection rules~\cite{Ramos-Sanchez:2018edc}. For example, twisted matter 
strings transform as follows
\begin{subequations}\label{eq:SGandPG}
\begin{eqnarray}
\Z{2}^{\mathrm{PG}} & : \quad \phi_{(n_1,n_2)} & \xmapsto{(h_1h_3)^2}~ -\phi_{(n_1,n_2)}\;,\label{eq:PG}\\
\Z{2}^{e_1}         & : \quad \phi_{(n_1,n_2)} & \xmapsto{~h_3~}~ (-1)^{n_1}\,\phi_{(n_1,n_2)}\;,\\
\Z{2}^{e_2}         & : \quad \phi_{(n_1,n_2)} & \xmapsto{~h_4~}~ (-1)^{n_2}\,\phi_{(n_1,n_2)}\;,
\end{eqnarray}
\end{subequations}
while the bulk strings $\Phi_{(0,0)}$ and $\Phi_{(-1,-1)}$ transform as $\rep{1}_0$. Hence, they 
are invariant under the transformations~\eqref{eq:SGandPG}.

Since twisted strings transform in the (real) representation $\rep4$ of $[32,49]$, the OPEs 
$\bar\phi_{n^a}\phi_{n^b}$ can be associated with the tensor product
\begin{equation}
\rep{4}~\otimes~\rep{4} ~=~ \bigoplus\limits_{\alpha,\beta,\gamma,\delta~\in~\{+,-\}}\rep{1}_{\alpha\beta\gamma\delta}\;.
\end{equation}
Hence, eq.~\eqref{eq:OPEs} implies that the 16 classes of untwisted vertex operators $V^{\hat N^0}$ 
build the 16 one-dimensional representations $\rep{1}_{\alpha\beta\gamma\delta}$ of the traditional 
flavor symmetry. Further, the representation $\rep{4}\otimes\rep{4}$ is not faithful. Thus, taking 
only the untwisted vertex operators $V^{\hat N^0}$ into account, the traditional flavor symmetry is 
only $[16,14]\cong\left(\Z{2}\right)^4 $. This can be confirmed by the explicit one-dimensional 
representations~\eqref{eq:traditionalSinglets}.

Finally, since oscillator excitations are not affected by the transformations associated with 
$\hat h_i$, all other twisted matter fields $\Phi_{(n_T,n_U)}$ (see 
table~\ref{tab:Z2modularWeights}) must transform in the same $4$-dimensional representation defined 
by eq.~\eqref{eq:TraditionalRhoOfTwistedStrings}.

%%%%%%%%%%%%%%%%%%%%%%%%%%%%%%%%%%%%%%%%
\subsubsection{Modular flavor group}
\label{sec:appDModularGroup}

Let us consider the twisted matter field $\Phi_{(\nicefrac{-1}{2},\nicefrac{-1}{2})}$, which builds 
a $\rep4$-plet of the traditional flavor group, as seen in the previous section. Its 
transformations under rotational automorphisms $\hat\Sigma$ are governed by 
eq.~\eqref{eq:generalModTrafoOnFields}, where the automorphy factors 
$j^{(n_T,n_U)}(\hat\Sigma,T,U)$ are given by eqs.~\eqref{eq:jfromSP4Z} 
or~\eqref{eq:AutomorphyFactorSLxSL}, depending on $\hat\Sigma$. Further, the 4-dimensional matrix 
representation of these transformations acting on the multiplet 
$\Phi_{(\nicefrac{-1}{2},\nicefrac{-1}{2})}$ are identified using the OPEs~\eqref{eq:OPEs}, 
resulting in
\begin{subequations}\label{eq:ModularRhoOfTwistedStrings}
  \begin{align}
     \rho_{\rep{4}_1}(\hat{K}_\mathrm{S}) &= \frac{1}{2}\begin{pmatrix}1&1&1&1\\1&1&-1&-1\\1&-1&1&-1\\1&-1&-1&1\\ \end{pmatrix}\;,
   & \rho_{\rep{4}_1}(\hat{K}_\mathrm{T}) &= \begin{pmatrix}-1&0&0&0\\0&1&0&0\\0&0&1&0\\0&0&0&1\\ \end{pmatrix}\;, \label{eq:ModularRhoOfTwistedStringsKaehler} \\
     \rho_{\rep{4}_1}(\hat{C}_\mathrm{S}) &= \begin{pmatrix}1&0&0&0\\0&0&1&0\\0&1&0&0\\0&0&0&1\end{pmatrix}\;,
   & \rho_{\rep{4}_1}(\hat{C}_\mathrm{T}) &= \begin{pmatrix}1&0&0&0\\0&1&0&0\\0&0&0&1\\0&0&1&0\end{pmatrix}\;, \label{eq:ModularRhoOfTwistedStringsCS}\\
     \rho_{\rep{4}_1}(\hat{M})            &= \frac{1}{\sqrt{2}}\begin{pmatrix}0&0&-1&1\\0&0&1&1\\1&-1&0&0\\-1&-1&0&0\end{pmatrix}\;.
   & & \label{eq:ModularRhoMirror}
  \end{align}
\end{subequations}
They build the representation $\rep{4}_1$ of the finite modular flavor group 
$(S_3^T \times S_3^U )\rtimes \Z{4}^{\hat{M}}\cong[144,115]$ without \CP (see 
appendix~\ref{app:FMGCharacterTable}).

Note that, as in the traditional flavor group, the tensor product $\rep{4}_1\otimes\rep{4}_1$ of 
twisted vertex operators does not behave as a faithful representation. Hence, we learn that the 
classes of untwisted vertex operators $V^{\hat N^0}$ transform only under the subgroup $[72,40]$ of 
the full finite modular symmetry.

The finite modular flavor group can be extended by \CP. Since the \CP-like transformation 
$\hat\Sigma_*$ interchanges matter fields with their conjugates, the 4-dimensional representation 
of twisted matter strings transforms as 
$\Phi_{(\nicefrac{-1}{2},\nicefrac{-1}{2})} \leftrightarrow \bar{\Phi}_{(\nicefrac{-1}{2},\nicefrac{-1}{2})}$. 
So, the representation of $\hat\Sigma_*$ acts on 
$(\Phi_{(\nicefrac{-1}{2},\nicefrac{-1}{2})}, \bar{\Phi}_{(\nicefrac{-1}{2},\nicefrac{-1}{2})})^\mathrm{T}$. 
Furthermore, to determine the corresponding automorphy factor we follow the discussion in 
section~\ref{sec:AutomorphyFactors} and consider the associated $\mathrm{GSp}(4,\Z{})$ element 
$M_*=\mathrm{diag}(-1,-1,1,1)$~\cite[eq.~(39)]{Baur:2020yjl}. This implies $A=-\Id_2$, $B=0$, 
$C=0$ and $D=\Id_2$ in the context of eq.~\eqref{eq:jfromSP4Z}. It follows that the automorphy 
factor of \CP is trivial. Hence, we find that $\hat\Sigma_*$ acts on the twisted multiplet as 
\begin{equation}\label{eq:ModularRhoCP}
\begin{pmatrix}\Phi_{(\nicefrac{-1}{2},\nicefrac{-1}{2})}\\ \bar{\Phi}_{(\nicefrac{-1}{2},\nicefrac{-1}{2})}\end{pmatrix}
~\xmapsto{~\hat{\Sigma}_*~}~
\rho_{\rep4_1\oplus\rep4_1} (\hat{\Sigma}_*)\begin{pmatrix}\Phi_{(\nicefrac{-1}{2},\nicefrac{-1}{2})}\\ \bar{\Phi}_{(\nicefrac{-1}{2},\nicefrac{-1}{2})}\end{pmatrix}
~=~ \begin{pmatrix}0& \Id_4 \\ \Id_4 & 0\end{pmatrix} \begin{pmatrix}\Phi_{(\nicefrac{-1}{2},\nicefrac{-1}{2})}\\ \bar{\Phi}_{(\nicefrac{-1}{2},\nicefrac{-1}{2})}\end{pmatrix}\;.
\end{equation}
This enhances the finite modular flavor group to 
$\left[( S_3^T \x S_3^U) \rtimes \Z4^{\hat{M}} \right]\x\Z2^\CP\cong[288,880]$.

%%%%%%%%%%%%%%%%%%%%%%%%%%%%%%%%%%%%%%%%%%%%%%%%%%%%%%%%%%%%%%%%%%%%%%%%%%%%%%%%%%%%%%%%%%%%%%%%%%%%%%%%%%%%%%%%%%%%%%%%%%%%%%%%%%%%%%%%%%
\section{Details on the superpotential}

%%%%%%%%%%%%%%%%%%%%%%%%%%%%%%%%%%%%%%%%%%%%%%%%%%%%
\subsection{Representation of modular forms}
\label{app:FMGIrreps}

The four-dimensional multiplet of modular forms $\hat{Y}^{(2)}_{\rep{4}_3}(T,U)$ has been given in 
eq.~\eqref{eq:ModularFormWeight22}. It transforms under modular transformations with a $\rep{4}_3$ 
representation of the finite modular group $(S_3^T\times S_3^U)\rtimes\Z4^{\hat M}\cong [144,115]$. 
In detail, for a modular transformation $\hat{\Sigma}$ we find
\begin{equation}
\label{eq:YTrafo}
\hat{Y}^{(2)}_{\rep{4}_3}(T,U) ~\xmapsto{~\hat{\Sigma}~}~ j^{(2)}(\hat{\Sigma},T,U)~ \rho_{\rep{4}_3}(\hat{\Sigma})\,\hat{Y}^{(2)}_{\rep{4}_3}(T,U)\;,
\end{equation}
where $j^{(2)}(\hat{\Sigma},T,U)$ is the automorphy factor, cf.\ section \ref{sec:AutomorphyFactors}, 
and the representations $\rho_{\rep{4}_3}(\hat{\Sigma})$ read 
\begin{subequations}\label{eq:repOfY}
\begin{align}
\rho_{\rep{4}_3}(\hat{K}_{\mathrm{S}}) &=
\begin{pmatrix}
-\frac{1}{2} & -\frac{\sqrt{3}}{2} & 0 & 0 \\[2pt]
-\frac{\sqrt{3}}{2} & \frac{1}{2} & 0 & 0 \\[2pt]
0 & 0 & -\frac{1}{2} & -\frac{\sqrt{3}}{2} \\[2pt]
0 & 0 & -\frac{\sqrt{3}}{2} & \frac{1}{2}
\end{pmatrix}\;,
&
\rho_{\rep{4}_3}(\hat{K}_{\mathrm{T}}) =
\begin{pmatrix}
1 & 0 & 0 & 0 \\
0 & -1 & 0 & 0 \\
0 & 0 & 1 & 0 \\
0 & 0 & 0 & -1 \\
\end{pmatrix}\;,
\\[4pt]
\rho_{\rep{4}_3}(\hat{C}_{\mathrm{S}}) &=
\begin{pmatrix}
-\frac{1}{2} & 0 & -\frac{\sqrt{3}}{2} & 0 \\[2pt]
0 & -\frac{1}{2} & 0 & -\frac{\sqrt{3}}{2} \\[2pt]
-\frac{\sqrt{3}}{2} & 0 & \frac{1}{2} & 0 \\[2pt]
0 & -\frac{\sqrt{3}}{2} & 0 & \frac{1}{2}
\end{pmatrix}\;,
&
\rho_{\rep{4}_3}(\hat{C}_{\mathrm{T}}) =
\begin{pmatrix}
1 & 0 & 0 & 0 \\
0 & 1 & 0 & 0 \\
0 & 0 & -1 & 0 \\
0 & 0 & 0 & -1 \\
\end{pmatrix}\;,
\\[4pt]
\rho_{\rep{4}_3}(\hat{M}) &=
\begin{pmatrix}
1 & 0 & 0 & 0 \\
0 & 0 & 1 & 0 \\
0 & 1 & 0 & 0 \\
0 & 0 & 0 & 1 \\
\end{pmatrix}\;.
&&
\end{align}
\label{eq:rho4_3}
\end{subequations}
Note however that this is an unfaithful representation, i.e.\ $\rho_{\rep{4}_3}(\hat{\Sigma})$ only 
spans the group $(S_3^T\times S_3^U) \rtimes\Z2\cong[72,40]$.

%%%%%%%%%%%%%%%%%%%%%%%%%%%%%%%%%%%%%%%%%%%%%%%%%%%%
\subsection{Components of the superpotential}
\label{app:superpotential}

There are 16 terms in the product of four twisted matter fields 
$\Phi^i_{(\nicefrac{-1}{2},\nicefrac{-1}{2})}=(\phi^i_{(0,0)}, \phi^i_{(1,0)}, \phi^i_{(0,1)}, \phi^i_{(1,1)})^\mathrm{T}$, 
$i\in\{1,2,3,4\}$, that are invariant under the traditional flavor symmetry 
$\left(D_8 \times D_8\right)/\Z{2} \cong [32,49]$. They read
\begin{align}
\mathcal{I}_1 ~&  =~ \phi^{1}_{(0,0)} \phi^{2}_{(0,0)} \phi^{3}_{(0,0)} \phi^{4}_{(0,0)} + \phi^{1}_{(1,0)} \phi^{2}_{(1,0)} \phi^{3}_{(1,0)} \phi^{4}_{(1,0)} \nonumber\\
& ~~+ \phi^{1}_{(0,1)} \phi^{2}_{(0,1)} \phi^{3}_{(0,1)} \phi^{4}_{(0,1)} + \phi^{1}_{(1,1)} \phi^{2}_{(1,1)} \phi^{3}_{(1,1)} \phi^{4}_{(1,1)}\;,
\end{align}
and \enlargethispage{\baselineskip}
\begin{subequations}
\begin{align}
	\mathcal{I}_2 ~&=~ \Delta_1\left(\phi_{(0,0)},\phi_{(1,0)},\phi_{(0,1)},\phi_{(1,1)}\right)\;,
	& \mathcal{I}_{11} ~&=~ \Xi\left(\phi_{(0,0)},\phi_{(1,0)},\phi_{(0,1)},\phi_{(1,1)}\right)\;, \\
	\mathcal{I}_3 ~&=~ \Delta_2\left(\phi_{(0,0)},\phi_{(1,0)},\phi_{(0,1)},\phi_{(1,1)}\right)\;, 
	& \mathcal{I}_{12} ~&=~ \Xi\left(\phi_{(0,0)},\phi_{(0,1)},\phi_{(1,0)},\phi_{(1,1)}\right)\;, \\
	\mathcal{I}_4 ~&=~ \Delta_3\left(\phi_{(0,0)},\phi_{(1,0)},\phi_{(0,1)},\phi_{(1,1)}\right)\;, 
	&\mathcal{I}_{13} ~&=~ \Xi\left(\phi_{(0,0)},\phi_{(1,0)},\phi_{(1,1)},\phi_{(0,1)}\right)\;, \\
	\mathcal{I}_5 ~&=~ \Delta_1\left(\phi_{(0,0)},\phi_{(0,1)},\phi_{(1,0)},\phi_{(1,1)}\right)\;, 
	& \mathcal{I}_{14} ~&=~ \Xi\left(\phi_{(0,0)},\phi_{(1,1)},\phi_{(1,0)},\phi_{(0,1)}\right)\;,\\
	\mathcal{I}_6 ~&=~ \Delta_2\left(\phi_{(0,0)},\phi_{(0,1)},\phi_{(1,0)},\phi_{(1,1)}\right)\;, 
	& \mathcal{I}_{15} ~&=~ \Xi\left(\phi_{(0,0)},\phi_{(0,1)},\phi_{(1,1)},\phi_{(1,0)}\right)\;, \\
	\mathcal{I}_7 ~&=~ \Delta_3\left(\phi_{(0,0)},\phi_{(0,1)},\phi_{(1,0)},\phi_{(1,1)}\right)\;, 
	& \mathcal{I}_{16} ~&=~ \Xi\left(\phi_{(0,0)},\phi_{(1,1)},\phi_{(0,1)},\phi_{(1,0)}\right)\;, \\
	\mathcal{I}_8 ~&=~ \Delta_1\left(\phi_{(0,0)},\phi_{(1,1)},\phi_{(1,0)},\phi_{(0,1)}\right)\;, \\
	\mathcal{I}_9 ~&=~ \Delta_2\left(\phi_{(0,0)},\phi_{(1,1)},\phi_{(1,0)},\phi_{(0,1)}\right)\;, \\
	\mathcal{I}_{10} ~&=~ \Delta_3\left(\phi_{(0,0)},\phi_{(1,1)},\phi_{(1,0)},\phi_{(0,1)}\right)\;,
\end{align}
\end{subequations}
where we have used the following abbreviations:
\begin{subequations}
	\begin{align}
	\Xi(A,B,C,D) ~&=~ A^1\,B^2\,C^3\,D^4 + A^2\,B^1\,C^4\,D^3 + A^3\,B^4\,C^1\,D^2 + A^4\,B^3\,C^2\,D^1\;, \\
	\Delta_i(A,B,C,D) ~&=~ \tilde\Delta_i(A,B) + \tilde\Delta_i(C,D)\;,
	\end{align}
\end{subequations}
with
\begin{subequations}
	\begin{align}
	\tilde\Delta_1(A,B) ~&=~ A^1\,A^2\,B^3\,B^4 + A^3\,A^4\,B^1\,B^2 \;, \\
	\tilde\Delta_2(A,B) ~&=~ A^1\,A^3\,B^2\,B^4 + A^2\,A^4\,B^1\,B^3 \;, \\
	\tilde\Delta_3(A,B) ~&=~ A^1\,A^4\,B^2\,B^3 + A^2\,A^3\,B^1\,B^4 \;.
	\end{align}
\end{subequations}

As described in eq.\ (\ref{eq:ITrafosUnderFMG}), the vector $(\mathcal{I}_1, \dots , \mathcal{I}_{16})^\mathrm{T}$ transforms under the modular symmetry $\left[144,115\right]$ with $R(\hat{\Sigma})$. These $16\times 16$ matrices are generated by 
\begin{equation}
R(\hat{C}_{\mathrm{S}}) =
\begin{pmatrix}
1 & 0 & 0 & 0 & 0 & 0 & 0 & 0 \\
0 & 0 & \Id_3 & 0 & 0 & 0 & 0 & 0 \\
0 & \Id_3 & 0 & 0 & 0 & 0 & 0 & 0 \\
0 & 0 & 0 & \Id_3 & 0 & 0 & 0 & 0 \\
0 & 0 & 0 & 0 & 0 & 1 & 0 & 0 \\
0 & 0 & 0 & 0 & 1 & 0 & 0 & 0 \\
0 & 0 & 0 & 0 & 0 & 0 & 0 & \Id_2 \\
0 & 0 & 0 & 0 & 0 & 0 & \Id_2 & 0 \\
\end{pmatrix}
\;,~
R(\hat{C}_{\mathrm{T}}) =
\begin{pmatrix}
1 & 0 & 0 & 0 & 0 & 0 & 0 & 0 \\
0 & \Id_3 & 0 & 0 & 0 & 0 & 0 & 0 \\
0 & 0 & 0 & \Id_3 & 0 & 0 & 0 & 0 \\
0 & 0 & \Id_3 & 0 & 0 & 0 & 0 & 0 \\
0 & 0 & 0 & 0 & 0 & \Id_2 & 0 & 0 \\
0 & 0 & 0 & 0 & \Id_2 & 0 & 0 & 0 \\
0 & 0 & 0 & 0 & 0 & 0 & 0 & 1 \\
0 & 0 & 0 & 0 & 0 & 0 & 1 & 0 \\
\end{pmatrix}\;.
\end{equation}
In addition, $R(\hat{M})$ acts on each four-dimensional subspace defined by
\begin{equation}
\begin{pmatrix}
\mathcal{I}_1 \\ \mathcal{I}_2 \\ \mathcal{I}_3 \\ \mathcal{I}_4
\end{pmatrix}\;,~
\begin{pmatrix}
\mathcal{I}_5 \\ \mathcal{I}_8 \\ \mathcal{I}_{11} \\ \mathcal{I}_{13}
\end{pmatrix}\;,~
\begin{pmatrix}
\mathcal{I}_6 \\ \mathcal{I}_9 \\ \mathcal{I}_{12} \\ \mathcal{I}_{14}
\end{pmatrix}\quad \mathrm{and} \quad
\begin{pmatrix}
\mathcal{I}_7 \\ \mathcal{I}_{10} \\ \mathcal{I}_{15} \\ \mathcal{I}_{16}
\end{pmatrix}
\end{equation}
with the $4 \times 4$ matrix
\begin{equation}
\dfrac{1}{2}\,
\begin{pmatrix}
	1 & 1 & 1 & 1 \\
	1 & 1 & -1 & -1 \\
	1 & -1 & 1 & -1 \\
	1 & -1 & -1 & 1
\end{pmatrix}\;.
\end{equation}

%%%%%%%%%%%%%%%%%%%%%%%%%%%%%%%%%%%%%%%%%%%%%%%%%%%%
\subsection[(S3xS3):Z4 character table]{\boldmath $(S_3^T \times S_3^U) \rtimes\Z4^{\hat M}$ character table \unboldmath}
\label{app:FMGCharacterTable}

\begin{table}[t!]
	\centering
	\resizebox{\textwidth}{!}{
		\begin{tabular}{c|rrrrrrrrrrrrrrrrrr}
			\toprule
			order & 1 & 2 & 2 & 2 & 2 & 2 & 3 & 3 & 4 & 4 & 4 & 4 & 6 & 6 & 6 & 6 & 12 & 12\\
			size & 1 & 1 & 6 & 6 & 9 & 9 & 4 & 4 & 6 & 6 & 18 & 18 & 4 & 4 & 12 & 12 & 12 & 12 \\
			name & $C_1$ & $C_2$ & $C_3$ & $C_4$ & $C_5$ & $C_6$ & $C_7$ & $C_8$ & $C_9$ & $C_{10}$ & $C_{11}$ & $C_{12}$ & $C_{13}$ & $C_{14}$ & $C_{15}$ & $C_{16}$ & $C_{17}$ & $C_{18}$ \\
			\midrule
			$\rep{1}_0$  &$1$&$1$&$1$&$1$&$1$&$1$&$1$&$1$&$1$&$1$&$1$&$1$&$1$&$1$&$1$&$1$&$1$&$1$\\
			$\rep{1}_{1}$&$1$&$1$&$-1$&$-1$&$1$&$1$&$1$&$1$&$1$&$1$&$-1$&$-1$&$1$&$1$&$-1$&$-1$&$1$&$1$\\
			$\rep{1}_{2}$&$1$&$1$&$-1$&$-1$&$1$&$1$&$1$&$1$&$-1$&$-1$&$1$&$1$&$1$&$1$&$-1$&$-1$&$-1$&$-1$\\
			$\rep{1}_{3}$&$1$&$1$&$1$&$1$&$1$&$1$&$1$&$1$&$-1$&$-1$&$-1$&$-1$&$1$&$1$&$1$&$1$&$-1$&$-1$\\
			$\rep{1}_{4}$&$1$&$-1$&$1$&$-1$&$1$&$-1$&$1$&$1$&$-\I$&$\I$&$-\I$&$\I$&$-1$&$-1$&$1$&$-1$&$-\I$&$\I$\\
			$\rep{1}_{5}$&$1$&$-1$&$-1$&$1$&$1$&$-1$&$1$&$1$&$\I$&$-\I$&$-\I$&$\I$&$-1$&$-1$&$-1$&$1$&$\I$&$-\I$\\
			$\rep{1}_{6}$&$1$&$-1$&$-1$&$1$&$1$&$-1$&$1$&$1$&$-\I$&$\I$&$\I$&$-\I$&$-1$&$-1$&$-1$&$1$&$-\I$&$\I$\\
			$\rep{1}_{7}$&$1$&$-1$&$1$&$-1$&$1$&$-1$&$1$&$1$&$\I$&$-\I$&$\I$&$-\I$&$-1$&$-1$&$1$&$-1$&$\I$&$-\I$\\[4pt]
			$\rep{2}_{1}$&$2$&$2$&$0$&$0$&$-2$&$-2$&$2$&$2$&$0$&$0$&$0$&$0$&$2$&$2$&$0$&$0$&$0$&$0$\\
			$\rep{2}_{2}$&$2$&$-2$&$0$&$0$&$-2$&$2$&$2$&$2$&$0$&$0$&$0$&$0$&$-2$&$-2$&$0$&$0$&$0$&$0$\\[4pt]
			$\rep{4}_{1}$&$4$&$-4$&$2$&$-2$&$0$&$0$&$1$&$-2$&$0$&$0$&$0$&$0$&$2$&$-1$&$-1$&$1$&$0$&$0$\\
			$\rep{4}_{2}$&$4$&$-4$&$-2$&$2$&$0$&$0$&$1$&$-2$&$0$&$0$&$0$&$0$&$2$&$-1$&$1$&$-1$&$0$&$0$\\
			$\rep{4}_{3}$&$4$&$4$&$0$&$0$&$0$&$0$&$-2$&$1$&$2$&$2$&$0$&$0$&$1$&$-2$&$0$&$0$&$-1$&$-1$\\
			$\rep{4}_{4}$&$4$&$4$&$0$&$0$&$0$&$0$&$-2$&$1$&$-2$&$-2$&$0$&$0$&$1$&$-2$&$0$&$0$&$1$&$1$\\
			$\rep{4}_{5}$&$4$&$4$&$2$&$2$&$0$&$0$&$1$&$-2$&$0$&$0$&$0$&$0$&$-2$&$1$&$-1$&$-1$&$0$&$0$\\
			$\rep{4}_{6}$&$4$&$4$&$-2$&$-2$&$0$&$0$&$1$&$-2$&$0$&$0$&$0$&$0$&$-2$&$1$&$1$&$1$&$0$&$0$\\
			$\rep{4}_{7}$&$4$&$-4$&$0$&$0$&$0$&$0$&$-2$&$1$&$2\I$&$-2\I$&$0$&$0$&$-1$&$2$&$0$&$0$&$-\I$&$\I$\\
			$\rep{4}_{8}$&$4$&$-4$&$0$&$0$&$0$&$0$&$-2$&$1$&$-2\I$&$2\I$&$0$&$0$&$-1$&$2$&$0$&$0$&$\I$&$-\I$\\
			\bottomrule
\end{tabular}}
\caption{Character table of $(S_3^T\times S_3^U)\rtimes\Z4^{\hat M}\cong [144,115]$, 
	where `size' means the number of elements in a conjugacy class 
	and `order' denotes the order of its elements.}
	\label{tab:144115char}
\end{table}

A presentation for the group $(S_3^T \times S_3^U) \rtimes\Z4^{\hat M} \cong [144,115]$ is given by
\begin{subequations}
\begin{align}
	\left\langle \mathsf{\hat{C}_\mathrm{S}},~\mathsf{\hat{C}_\mathrm{T}},~\mathsf{\hat{M}}
	~\big|~ \right. \mathsf{\hat{C}_\mathrm{S}}^2=\mathsf{\hat{C}_\mathrm{T}}^2=\mathsf{\hat{M}}^4= \left(\mathsf{\hat{C}_\mathrm{S}\hat{C}_\mathrm{T}}\right)^3= \left(\mathsf{\hat{C}_\mathrm{S}\hat{M}}^2\right)^2 = \left(\mathsf{\hat{C}_\mathrm{T}\hat{M}}^2\right)^2 = \Id  ~~\mathrm{and}~~~~~ \\ 
	~~~\left.\mathsf{\hat{M}\,\hat{C}}_i\mathsf{\,\hat{M}\,\hat{C}}_j= \mathsf{\hat{C}}_j\mathsf{\,\hat{M}\,\hat{C}}_i\mathsf{\,\hat{M}}~~ \mathrm{where}~~i,j\in\{\mathsf{\mathrm{S,T}}\} \right\rangle\,.
\end{align}
\end{subequations}
Furthermore, we define $\mathsf{\hat{K}_\mathrm{S}}=\mathsf{\hat{M}}^3\mathsf{\hat{C}_\mathrm{S}\hat{M}}$ and 
$\mathsf{\hat{K}_\mathrm{T}}=\mathsf{\hat{M}}^3\mathsf{\hat{C}_\mathrm{T}\hat{M}}$,
such that the names of the abstract generators $\mathsf{\hat{C}_\mathrm{S},\hat{C}_\mathrm{T},\hat{K}_\mathrm{S},\hat{K}_\mathrm{T}}$ 
and $\mathsf{\hat{M}}$ allow for an intuitive association with the modular transformations 
$\hat{C}_\mathrm{S},\hat{C}_\mathrm{T},\hat{K}_\mathrm{S},\hat{K}_\mathrm{T}$ and $\hat{M}$.
The group has 18 conjugacy classes
\begin{subequations}
\begin{align}
		C_1~&=~ [\Id]\;, &
		C_2~&=~ [\mathsf{\hat{M}^2}]\;, &
		C_3~&=~ [\mathsf{\hat{C}_\mathrm{S}}]\;, \\
		C_4~&=~ [\mathsf{\hat{M}^2\,\hat{C}_\mathrm{S}}]\;, &
		C_5~&=~ [\mathsf{\hat{C}_\mathrm{S}\,\hat{K}_\mathrm{S}}]\;, &
		C_6~&=~ [\mathsf{\hat{M}^2\,\hat{C}_\mathrm{S}\,\hat{K}_\mathrm{S}}]\;, \\
		C_7~&=~ [\mathsf{\hat{C}_\mathrm{S}\hat{C}_\mathrm{T}}]\;, &
		C_8~&=~ [\mathsf{\hat{C}_\mathrm{S}\hat{C}_\mathrm{T}\,\hat{K}_\mathrm{S}\hat{K}_\mathrm{T}}]\;, &
		C_9~&=~ [\mathsf{\hat{M}}]\;, \\
		C_{10}~&=~ [\mathsf{\hat{M}^3}]\;, &
		C_{11}~&=~ [\mathsf{\hat{M}\,\hat{C}_\mathrm{S}}]\;, &
		C_{12}~&=~ [\mathsf{\hat{M}^3\,\hat{C}_\mathrm{S}}]\;, \\
		C_{13}~&=~ [\mathsf{\hat{M}^2\,\hat{C}_\mathrm{S}\hat{C}_\mathrm{T}\,\hat{K}_\mathrm{S}\hat{K}_\mathrm{T}}]\;, &
		C_{14}~&=~ [\mathsf{\hat{M}^2\,\hat{C}_\mathrm{S}\hat{C}_\mathrm{T}}]\;, &
		C_{15}~&=~ [\mathsf{\hat{C}_\mathrm{S}\hat{C}_\mathrm{T}\,\hat{K}_\mathrm{S}}]\;, \\
		C_{16}~&=~ [\mathsf{\hat{M}^2\,\hat{C}_\mathrm{S}\hat{C}_\mathrm{T}\,\hat{K}_\mathrm{S}}]\;, &
		C_{17}~&=~ [\mathsf{\hat{M}\,\hat{C}_\mathrm{S}\hat{C}_\mathrm{T}}]\;, &
		C_{18}~&=~ [\mathsf{\hat{M}^3\,\hat{C}_\mathrm{S}\hat{C}_\mathrm{T}}]\;.
\end{align}
\end{subequations}
The character table is given in table~\ref{tab:144115char}. Note that out of the four-dimensional 
representations, the $\rep{4}_i$-plets with $i=3,\ldots,6$ are not faithful representations 
and span only the group $(S_3^T \times S_3^U) \rtimes\Z2^{\hat M} \cong [72,40]$.

%%%%%%%%%%%%%%%%%%%%%%%%%%%%%%%%%%%%%%%%%%%%%%%%%%%%
\subsection[A4 character table]{\boldmath $A_4$ character table\unboldmath}
\label{app:A4}

We denote as $\mathsf{a},\mathsf{b}$ and $\mathsf{c}$ the abstract $A_4$ generators associated with 
$h_1$, $h_2$ and $(\hat C_\mathrm{T}\hat C_\mathrm{S})^2$, respectively, see 
section~\ref{sec:A4symmetry}. In these terms, $A_4$ is defined by the presentation
\begin{equation}
A_4 ~=~ \left\langle \mathsf{a}, \mathsf{b}, \mathsf{c} ~\big|~
                     \mathsf{a}^2=\mathsf{b}^2=\mathsf{c}^3=(\mathsf{a}\mathsf{b})^2=\Id,~ 
                     \mathsf{b}\mathsf{c}\mathsf{a}=\mathsf{b}\mathsf{a}\mathsf{c}\mathsf{b}=\mathsf{c}\right\rangle\,,
\end{equation}
and has four conjugacy classes:
\begin{equation}
C_1~=~ [\Id]\;,\quad
C_2~=~ [\mathsf{c}^2]\;,\quad
C_3~=~ [\mathsf{c}]\;, \quad
C_4~=~ [\mathsf{a}]\;.
\end{equation}
The character table of $A_4$ is shown in table~\ref{tab:A4char}, where we also present the order 
and number (size) of the elements in each conjugacy class.
\begin{table}[h!]
\centering
\begin{tabular}{c|cccc}
\toprule
order & 1 & 3 & 3 & 2\\
size  & 1 & 4 & 4 & 3\\
name  & $C_1$ & $C_2$ & $C_3$ & $C_4$ \\
\midrule
$\rep{1}$   & $1$ & $1$ & $1$ & $1$\\
$\rep{1}'$  & $1$ & $\omega$ & $\omega^2$ & $1$ \\
$\rep{1}''$ & $1$ & $\omega^2$ & $\omega$ & $1$ \\
$\rep{3}$   & $3$ & $0$ & $0$ & $-1$ \\
\bottomrule
\end{tabular}
\caption{Character table of $A_4$. We use the definition $\omega:=\exp(\nicefrac{2\pi\I}{3})$.}
\label{tab:A4char}
\end{table}

%\bibliography{Orbifold}
%\bibliographystyle{OurBibTeX}
\providecommand{\bysame}{\leavevmode\hbox to3em{\hrulefill}\thinspace}

\end{document}